\documentclass[12pt]{JHEP3}
\pdfoutput=1
\usepackage{amsmath,epsfig}
\usepackage{amssymb,amsfonts}
\usepackage{latexsym}
\usepackage[latin1]{inputenc}
\usepackage{float}
\usepackage{subeqnarray}
\usepackage{xcolor}

\usepackage{graphicx}
\usepackage{longtable}
\usepackage{enumerate}
\usepackage{braket}
\usepackage{empheq}
\relax
\renewcommand{\theequation}{\arabic{section}.\arabic{equation}}
\def\be{\begin{equation}}
\def\ee{\end{equation}}
\def\bea{\begin{eqnarray}}
\def\eea{\end{eqnarray}}
\newcommand\fverb{\setbox\pippobox=\hbox\bgroup\verb}
\newcommand\fverbdo{\egroup\medskip\noindent%
                        \fbox{\unhbox\pippobox}\ }
\newcommand\fverbit{\egroup\item[\fbox{\unhbox\pippobox}]}

\newcommand{\bear}{\begin{eqnarray}}

\newcommand{\eear}{\end{eqnarray}}
\newcommand{\boxedeq}[2]{\begin{empheq}[box={\fboxsep=6pt\fbox}]{align}\label{#1}#2\end{empheq}}

\newcommand{\bsea}{\begin{subeqnarray}}
\newcommand{\esea}{\end{subeqnarray}}
\newbox\pippobox

\newcommand{\abs}[1]{|{#1}|}

\newcommand{\ud}{\mathrm{d}}

\def\6{\partial}

\def\a{\alpha}
\def\g{\gamma}

\def\le{\left}
\def\ri{\right}

\def\pa{\partial}

\def\e{\epsilon}
\def\m{\mu}
\def\n{\nu}
\def\r{\rho}
\def\s{\sigma}

\def\sp{,\;\;}

\def\z{\zeta}
\def\sq
\def\a{\alpha}
\def\b{\beta}
\def\l{\lambda}

\def\k{\chi}
\def\hri#1#2{\href{http://arxiv.org/abs/#1}{[ArXiv:#1]#2}}

\def\h{\eta}
\def\e{\epsilon}

\def\d{\delta}

\title{Criticality and Transport in Magnetized Holographic Systems}
\author{{\large N. Angelinos\\
		~\\
		\href{http://hep.physics.uoc.gr/}
		{Crete Center for Theoretical Physics}, Department of Physics, University of Crete
		71003 Heraklion, Greece}}

\preprint{CCTP-2018-14}

\abstract{In this master's thesis the Einstein-Maxwell-Dilaton theory is used to model
	the dynamics of 2+1-dimensional, strongly coupled, large-$N$ quantum field theories with intrinsic T-violation, at finite density and temperature, in the presence of a magnetic field. We include axion fields in order to introduce momentum relaxation.
	We find analytic expressions for the DC conductivity and present numerical results for the AC conductivity. We also classify the IR-asymptotic hyperscaling violating solutions of the theory. }

 %%%%%%%TeX, LaTeX, % %%%%%%%%NesTeX} %%%%%%%%%\dedicated{Dedicated to\ldots\\if you want.}

\begin{document}

\section{Introduction}

In Holography, a strongly coupled quantum field theory (QFT) at the large-$N$ limit can be studied in terms of higher-dimensional classical gravity, \cite{Mald}, \cite{witten}. This higher-dimensional space-time is commonly called the ``bulk" and the QFT ``lives" on its boundary. Every observable in the QFT is related to a bulk quantity by a precise formula, \cite{Witten:1998zw}, \cite{Gubser:1998bc}. This, in particular, allows for the calculation of transport coefficients. Some of the earliest successes include the calculation of the shear viscosity, \cite{Policastro:2001yc}, and the universal shear viscosity to entropy density ratio, \cite{Kovtun:2004de}. More recent works have focused on the electric conductivity of two-dimensional systems, \cite{Hartnoll:2009sz}, \cite{Herzog:2009xv}.

There is a wide variety of interesting two-dimensional materials, which are not well-understood theoretically. Some examples are the cuprates, which exhibit high-$T_c$ superconductivity, \cite{htcsc}, many other materials with strange metal characteristics, \cite{Iqbal:2011ae}, as well as heavy fermion systems, \cite{hf}.

Early studies of the conductivity have been based on the minimal Einstein-Maxwell action. The electric AC conductivity of a finite-temperature two-dimensional system at zero density is independent of frequency and temperature, \cite{Herzog:2007ij}, \cite{Iqbal:2008by}.
At finite charge density, the dependence becomes non-trivial, however the DC conductivity diverges, \cite{Hartnoll:2009sz}, as the momentum is conserved due to the translational symmetry.

The AC conductivity of a system with charge density $\rho$ in the presence of a magnetic field, $B$, has been studied in \cite{Hartnoll:2007ih} at low frequencies. There is a peak in the longitudinal conductivity at a frequency which was identified with the relativistic cyclotron frequency, \cite{Hartnoll:2007ih}, \cite{Hartnoll:2007ip}.
Meanwhile, the DC conductivity is constrained to the form, \cite{hartnoll},
\be \sigma_{xx}=0\sp \sigma_{xy}={\rho\over B},\ee
due to the absence of momentum relaxation. Further studies of conductivity in the presence of magnetic field in a strange-metal related context using holographic methods can be found in \cite{KLi,KKP}.

Early attempts to introduce momentum relaxation included the use of a spatially dependent boundary condition (``lattice"), \cite{Horowitz:2012gs}, \cite{Horowitz:2012ky}, \cite{ll2}, \cite{ll3}, \cite{ll4}, \cite{ll5}. In such models, the DC conductivity is finite, with a Drude peak at small frequencies.
Including a graviton mass term which breaks the translational, but not rotational, symmetry has been used, \cite{vegh}, \cite{mg3}, \cite{mg4}, \cite{mg5}, \cite{zhou}, \cite{Baggioli:2014roa}, with similar results to the explicit lattice. It turns out that the two approaches are equivalent at zero momentum, \cite{blaketongvegh}.
Other models have considered $3$-dimensional theories in an anisotropic space with helical symmetry, based on the Bianchi VII$_0$ symmetry, \cite{Donos:2014oha}, \cite{Donos:2014gya}, \cite{Erdmenger:2015qqa}.

Another way to break the translational symmetry is to introduce scalars which depend on the spatial coordinates, \cite{Donos:2013eha}, \cite{donos}, \cite{Gouteraux:2014hca}, \cite{Kim:2014bza}, \cite{Andrade:2013gsa}, \cite{Donos:2014cya}.
 Magnetic systems with momentum relaxation have been studied in \cite{Kim:2015wba}, \cite{Blake:2015ina}, \cite{blake}.

In the present thesis the Einstein-Maxwell-Dilaton (EMD) theory is used to study
the dynamics of 2+1-dimensional theories at finite temperature and density, in the presence of a magnetic field. We include the T-violating term $F\wedge F$, as well as ``axion" fields in order to introduce momentum relaxation.

 The structure is as follows. Section \ref{holo} is an introduction to Holography and is safe to skip for experienced readers. In section \ref{setup} we introduce the EMD action, equations of motion and the class of backgrounds which we will study, followed by the calculation of the DC conductivity in section \ref{hcond}. In section \ref{EMDax} we include axion fields in the action in order to break the translational symmetry of the background and we calculate the DC conductivity in section \ref{condax}.
In section \ref{ACcond} we compute the AC conductivity numerically in four simple constant-scalar backgrounds. Analytic formulas are presented for certain limits.
In section \ref{HVsol} we study the IR-asymptotic solutions with hyperscaling violating geometries. Finally, in section \ref{concl} we summarize our results.

\section{The Holographic Correspondence\label{holo}}

An early indicator for Holography was Bekenstein's bound, which implies that the entropy of a black hole scales with its surface area, rather than its volume, \cite{bek}. Bekenstein's bound has been controversial, since it makes QFT seemingly incompatible with gravity. Nowadays it is believed that a successful theory of quantum gravity must satisfy this bound.

The first incarnation of Holography was 't Hooft's large-$N$ gauge theory/string theory duality, \cite{hooft}. This duality is based on the observation that the perturbative structure of string theory is similar to that of the Yang-Mills theory when the number of colors (N) tends to infinity. A large-N gauge theory at strong coupling can be described by an effective string theory at weak coupling. A brief overview of this duality is presented in \ref{gsd}.

The most succesful realization of Holography is Maldacena's conjecture \cite{Mald} in 1997, according to which a superstring theory on AdS$_5\times S^5$ is equivalent to $\mathcal{N}=4$ super Yang-Mills. Since the latter is a CFT, this duality was named the AdS/CFT correspondence. The CFT lives in a space-time homeomorphic to the boundary of AdS$_5$ and every operator can be translated to a field propagating in the AdS bulk, \cite{witten}. After two decades of research Maldacena's conjecture seems to be valid in a far more general context. An overview of the AdS/CFT correspondence is presented in section \ref{AdSCFT}.

In section \ref{gkpw} we introduce the central formula in Holography, which connects boundary observables with bulk data. In  \ref{hren} we describe how the process of renormalization is realized in Holography. We illustrate the above using the simplest example: a scalar field in AdS space-time.

In sections \ref{ft0}, \ref{cpmf} we explain how theories at finite temperature and chemical potential are realized in Holography and how the thermodynamics of the boundary are related to the bulk geometry.

In section \ref{2pf} we motivate the simplest prescription for calculating two-point functions directly in real-time.

In section \ref{dict} we summarize how every feature of the bulk is reflected on the boundary and vice-versa. This is commonly known as the ``Holographic dictionary".

Finally, in section \ref{LRCT} we briefly review the linear response theory, which we will use in order to calculate the electric conductivity.

\subsection{The Gauge/String Duality\label{gsd}}

The gauge/string duality, first suggested by 't Hooft, \cite{hooft}, is based on the observation that the perturbative expansion of string theory is similar to the large-$N$ expansion of a $U(N)$ gauge theory.

\subsubsection{String Theory}

A relativistic string is a one-dimensional object which sweeps a two-dimensional surface (world-sheet) as it moves in space-time. In analogy with the relativistic particle, the equations of motion for the string are derived by minimizing the surface of its world-sheet between two string configurations
\be S_{string}=-T\int dA \label{a2},\ee
where
\be T=(2\pi \ell_s^2)^{-1}\label{a22}\ee
is the tension of the string and $dA$ is the surface element of its worldsheet. The parameter $\ell_s$ is the characteristic length of the theory (string length).
 The world-sheet is a tube for a closed string and a strip for an open string.

Let $\xi^a (a=0,1)$ parametrize the world-sheet and let $G_{\m\n}$ be the metric of the target space. The metric induced on the world-sheet is:
\be\tilde G_{ab}=G_{\m\n}{\partial X^\m\over \partial \xi^a}{\partial X^\n\over \partial \xi^b}.\label{a3}\ee
The string action (\ref{a2}) can be written as:
\be S_{string}=-T\int \ud^2 \xi \sqrt{-\det \tilde G_{ab}}.\label{a4} \ee
This is known as the Nambu-Goto action.

The Polyakov action
\be S_P=-{T\over 2}\int \ud^2 \xi \sqrt{-h}h^{ab}G_{\mu\nu}\partial_a X^\mu\partial_b X^\nu\label{a5}\ee
yields the same equations of motion as the Nambu-Goto action, but is easier to quantize,s as it is linear in the matter fields. Here $h_{ab}$ is the world-sheet metric and is an additional, independent dynamical variable.

Varying (\ref{a5}) with respect to the world-sheet metric, we obtain the equation of motion for $h_{ab}$
\be {1\over 2}h_{ab}h^{cd}\partial_c X^\mu\partial_d X^\nu G_{\mu\nu}=\partial_a X^\mu\partial_b X^\nu G_{\mu\nu}.\label{aa}\ee
The above equation is solved by
\be h_{ab}=e^{2\phi}G_{\mu\nu}\partial_a X^\mu \partial_b X^\nu\label{a0}.\ee
Upon substitution, the field $\phi$ drops out of the equation. This is because of the classical Weyl symmetry of the Polyakov action, \cite{st}.

Using (\ref{aa}) we may calculate
\be h=\det h_{ab}=e^{4\phi}\det G_{\mu\nu}\partial_a X^\mu \partial_b X^\nu=e^{4\phi} \det \tilde G_{ab},\label{aaa1}\ee
where $\tilde G_{ab}$ is defined in (\ref{a3}) and, also,
\be h^{ab}G_{\mu\nu}\partial_a X^\mu\partial_b X^\nu=2e^{-2\phi}.\label{aaa2}\ee
Substituting (\ref{aaa1}) and (\ref{aaa2}) into (\ref{a5}), the fields $\phi$ cancel out and
we obtain the Nambu-Goto action (\ref{a4}).

The equations of motion for a string can be solved in flat target space-time for both Neumann and Dirichlet boundary conditions and then quantized by standard methods, \cite{st}.

\begin{figure}[H]
	\centering
	\includegraphics[width=1\textwidth]{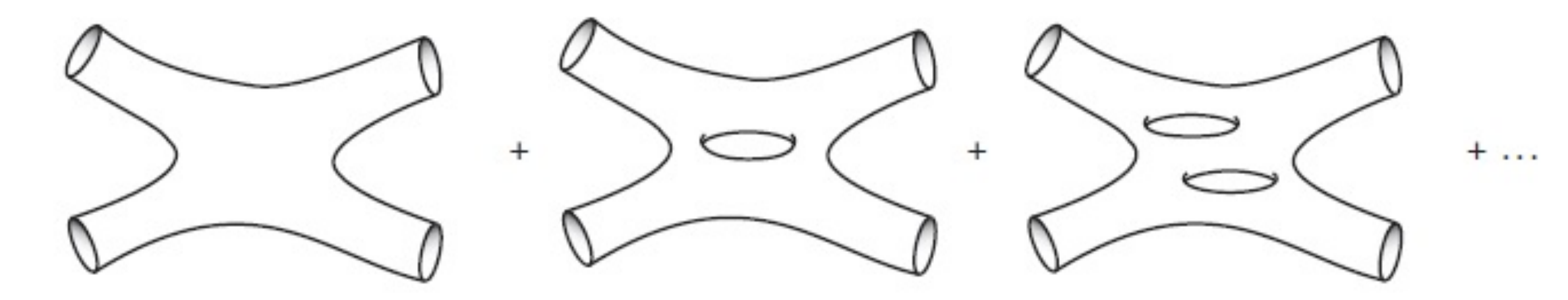}
	\caption{The perturbative expansion of a closed string world-sheet. This figure was taken from \cite{AdS/CFT}}.
	\label{strings}
\end{figure}

In closed string theory the basic interaction is a string splitting into two (or the inverse), \cite{st}. The perturbative expansion of two closed strings merging and splitting is shown in figure \ref{strings}. The world-sheet of the one-loop interaction is a surface of genus $1$ (the genus value counts how many "holes" a surface has). The higher-loop diagrams correspond to surfaces of higher genus. The world-sheets are characterized by the topological number, $h$, which is the value of its genus. Therefore the perturbation series in string theory is a topological expansion. By weighing each triple vertex with the dimensionless string coupling constant $g_s$ the string perturbative expansion is of the form, \cite{st}:
\be\sum_{h=0}^{\infty} g_s^{2h-2} F_h(\ell_s^2). \label{s1}\ee

\subsubsection{Large-N Gauge Theories}
Consider the Lagrangian of a U(N) gauge theory:
\be\mathcal{L}=-{1\over g_{YM}^2}Tr[F_{\m\n}F^{\m\n}], \label{l1} \ee
where $A_\mu$ are $N\times N$ matrices and the field strength tensor is
\be F_{\m\n}=\partial_\m A_\n-\partial_\n A_\m+[A_\m,A_\n].\ee

In order to take the large-$N$ limit, we first have to know how to scale the coupling constant $g_{YM}$. In quantum field theory the beta function encodes the dependence of the coupling parameter $g_{YM}$ on the energy scale $\m$. From the one-loop beta function for a non-Abelian U(N) gauge theory we obtain the RG flow equation, \cite{st}:
\be \m {\ud g_{YM}\over \ud \m}=-{11\over 3}N {g_{YM}^3\over (4\pi)^2}+\mathcal{O}(g_{YM}^5)\ee
We can find the appropriate scaling by demanding that the leading terms are of the same order. Therefore we can see that the combination
\be \lambda=g_{YM}^2 N \label{tH}\ee
which is called the 't Hooft coupling, \cite{hooft}, must be kept constant as N goes to infinity.
The Lagrangian (\ref{l1}) can be rewritten as
\be\mathcal{L}=-{N\over\l}Tr[F_{\m\n}F^{\m\n}]\ee

The vacuum diagrams can be written in the double-line notation, which replaces each line in the Feynman diagrams with two lines of opposite orientations. Every propagator contributes a factor of $\l/N$ and every vertex a factor of $N/\l$. In addition every loop contributes a power of N (because of the summation over N colors). A diagram with E propagators (edges), V vertices and
F loops (faces) has a coefficient proportional to:
\be \left({\l\over N}\right)^E \left({N\over \l}\right)^V N^F=N^\chi \l^{E-V}\label{l3}\ee
where $\chi=V-E+F$ is the Euler number of the surface. For a closed compact surface with $h$ handles $\chi=2-2h$. Therefore such a diagram has a coefficient of order $\mathcal{O}(N^{2-2h})$.

\begin{figure}[H]
	\centering
	\includegraphics[width=0.6\textwidth]{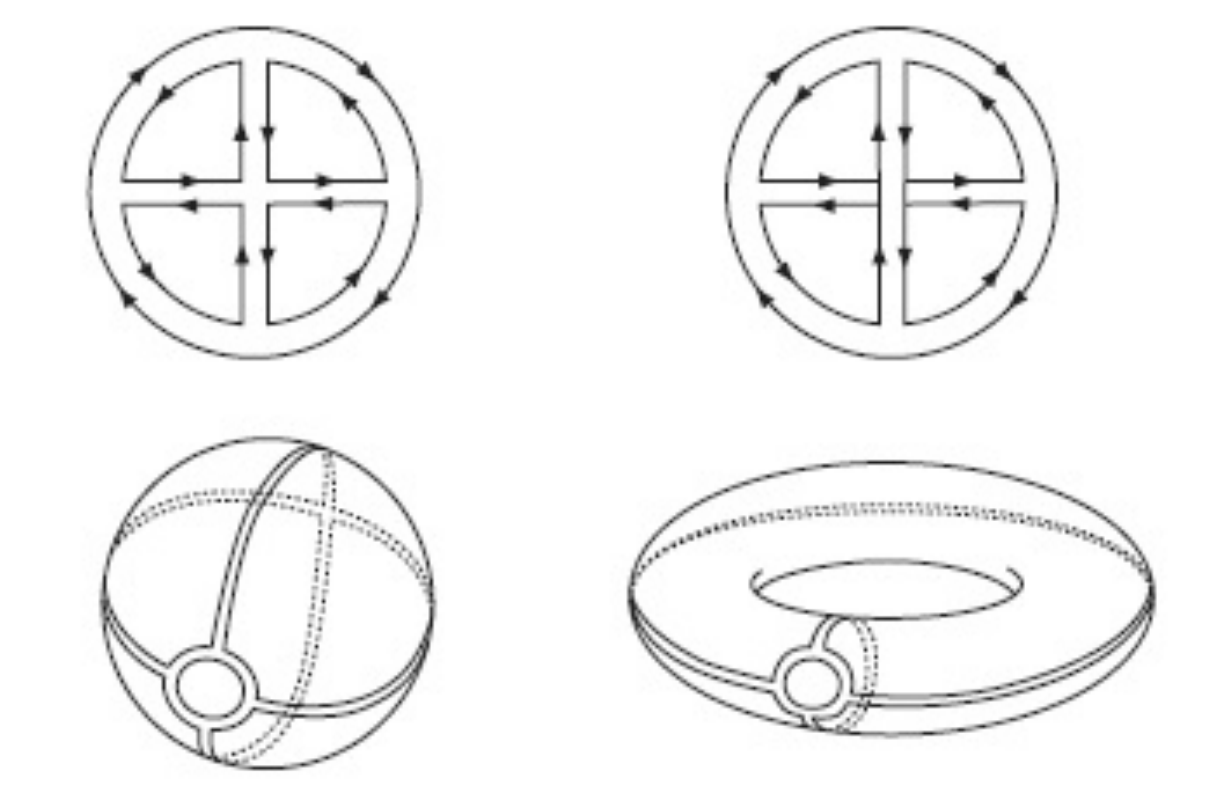}
	\caption{On the left: the zeroth order diagram is homeomorphic to a sphere. On the right: the first order diagram is homeomorphic to a torus. This figure was taken from \cite{AdS/CFT}}.
	\label{diagrams}
\end{figure}

 The large-$N$ expansion is a topological expansion. The dominant diagrams are the ones with the minimum number of handles ($h=0$), which are homeomorphic to a sphere. A higher-order diagram is homeomorphic to a surface of genus $h$ and is suppressed by an additional factor of $1/N^{2h}$.

The standard perturbative expansion for any correlator can be written at large N as, \cite{st}:
\be  \sum_{h=0}^\infty N^{2-2h} Z_h(\l)=\sum_{h=0}^\infty N^{2-2h}\sum_{i=0}^\infty c_{i,h}\l^i\label{l4}\ee
which suggests a connection with the topological expansion of string theory (\ref{s1}) if we identify the string coupling constant $g_s$ with:
\be g_s\sim 1/N\label{l5}\ee
An important observation is that in the large-$N$ limit the dual string theory is weakly coupled.

\subsection{The AdS/CFT Correspondence\label{AdSCFT}}

The AdS/CFT correspondence is a classic example for Holography, \cite{Mald}. This duality connects a superstring theory in AdS$_5\times S^5$ space-time to the $\mathcal{N}=4$ Yang Mills gauge theory in $3+1$ dimensions.

String theory contains objects extended in more than one spatial dimensions, called branes, \cite{st}. The most important ones for the AdS/CFT correspondence are the D$_p$-branes, which are defined as (p+1)-dimensional hypersurfaces on which open strings can end with Dirichlet boundary conditions. On one hand the brane dynamics can be described pertubatively in terms of open strings. On the other hand, the D-branes interact gravitationally and are supergravity solutions. In this section we will review how the AdS/CFT conjecture arises by comparing these two different descriptions of the same system of D$_p$-branes, \cite{Mald}.

\subsubsection{Charged D-branes}

A $D_p$-branes can have two types of excitations. The first type is the motion and deformation of their shapes which can be parametrized by their coordinates $\phi^i$ in the (9-p)-dimensional transverse space. These degrees of freedom are scalar fields on the brane's world-volume. The second type are internal excitations caused by the charged end of a string. In this case the D$_p$-brane has an Abelian gauge field A$_\m, (\m=0,1,...,p)$ living on its world-volume.
The action that describes the brane dynamics is the Dirac-Born-Infeld (DBI) action, \cite{AdS/CFT}:
\be S_{DBI}=-T_{D_p}\int \ud^{p+1}x\sqrt{-\det({g_{\m\n}}+2\pi \ell_s^2 F_{\m\n})}\label{DBI}\ee
where $T_{D_p}=(2\pi)^{-p}g_s^{-1}\ell_s^{-p-1}$ is the tension of the brane, $\ell_s$ the string length and $g_{\m\n}$ is the induced metric on the brane's world-volume which depends on the scalar fields $\phi^i$.

Consider now a $D_3$-brane in flat target space. We can write the induced metric on the brane using the 6 scalar fields $\phi^i$ as $g_{\m\n}=\eta_{\m\n}+(2\pi \ell_s^2)^2 \partial_\m \phi^i \partial_\n\phi^i$. The DBI action (\ref{DBI}) for a $D_3$-brane can be written as:
\be S_{brane}=-T_{D_3}\int \ud^{4}x\sqrt{-\det\left[\eta_{\m\n}+(2\pi \ell_s^2\partial_\m \phi^i) (2\pi \ell_s^2\partial_\n \phi^i)+2\pi \ell_s^2 F_{\m\n}\right]}\label{b1}\ee
We notice that every field $\phi, F$ is accompanied by a factor of $2\pi \ell_s^2$. We can expand in powers of $2\pi \ell_s^2$, which is equivalent to expanding in powers of the gravitational constant $\kappa$ and keeping the string coupling $g_s$ constant since $\kappa\sim g_s \ell_s^4$, \cite{st}. The leading order terms are (ignoring the constant zeroth order term:
\be S_{brane}=-{1\over 2\pi g_s}\int \ud^4 x \left({1\over 4} F_{\m\n}F^{\m\n}+{1\over 2}\partial_\m \phi^i\partial_\n \phi^i+\dots\right) \label{DBI2}\ee
where we used \be T_{D_3}=((2\pi)^3 g_s \ell_s^4)^{-1}\label{b2}\ee The rest of the terms are suppressed by additional factors of $\ell_s^2$. Therefore at the low-energy limit we have a U(1) theory living on the world-volume of the $D_3$-brane.

A system of $N$ coincident, parallel D$_3$-branes ($p=3$) generates a non-Abelian U(N) gauge theory. The branes live in a 10-dimensional bulk space and are located at the same point of the transverse 6-dimensional space. Except for the open strings, which are excitations of the branes, the theory also contains closed strings, which are excitations of the bulk space. We can write the low-energy action of this theory schematically:
\be S=S_{bulk}+S_{brane}+S_{interactions}\label{b3}\ee
The first term $S_{bulk}$ describes the dynamics of the bulk space in terms of closed strings, which are described by IIB superstring theory. The second term $S_{brane}$ describe the dynamics of the branes in terms of open strings. The third term $S_{interactions}$ contains the interactions between open and closed strings.

At the low-energy limit, the superstring theory living on the bulk reduces to free IIB supergravity. By expanding S$_{bulk}$ around the free point in powers of the gravitational constant $\kappa=8\pi G_N$ as $g_{\m\n}=\h_{\m\n}+\kappa h_{\m\n}$ we obtain, \cite{st}:
\be S_{bulk}\sim {1\over 2\kappa^2}\int \ud^{10} x \sqrt{-det(g_{\m\n})}R+\dots\sim \int\ud^{10} x \left[(\partial h)^2 + \kappa h (\partial h)^2+\dots\right]\label{b4} \ee
where we have not indicated explicitly all the bulk fields for simplicity. All the interaction terms are proportional to positive powers of $\kappa$, therefore at very low energies (small $\kappa$) they become very weak compared to the kinetic terms and can be ignored.

The second term S$_{brane}$ governs the brane dynamics in terms of open string. A system of $N$ branes generates a U(N) theory containing a vector boson and 6 scalars transforming as adjoints of U(N). In particular such a system is equivalent to $\mathcal{N}=4, U(N)$ super Yang-Mills (SYM) theory in the low-energy limit, \cite{st}:
\be S_{brane}\sim-{1\over 2\pi g_s}\int \ud^4 x  Tr\left[{1\over 4}F_{\m\n}F^{\m\n}\right]+\dots \label{DBI3}\ee
where we kept only the gauge field terms for simplicity.

The third term describing the interactions between open-closed degrees of freedom is subleading in the low energy limit. We expand S$_{interactions}$ in powers of $\kappa$, \cite{st}:
\be S_{interactions}\sim \int \ud^4 x\sqrt{-det(g_{\m\n})} Tr\left[F^2\right]+\dots\sim \kappa \int \ud^4 x h_{\m\n} Tr\left[F_{\m\n}^2-{\delta_{\m\n}\over 4}F^2\right]+\dots\label{b5} \ee
where again the only terms indicated are the kinetic terms of the gauge field for simplicity.

We conclude that in the low-energy limit, this theory is described by two decoupled theories: free IIB supergravity on the bulk and $\mathcal{N}=4, U(N)$ super Yang-Mills theory on the branes.

\subsubsection{D-branes as Supergravity Solutions}
$D_p$-branes are solutions of supergravity in 10-dimensions. The exact solution for $N$ $D_3$-branes is given by, \cite{st}:
\be ds^2=H^{-1/2}(-dt^2+d\vec{x}^2)+H^{1/2}(dr^2+r^2 d\Omega_5^2)\label{m1}\ee
The 3 spatial coordinates $\vec{x}$ are parallel to the branes, while $dr^2+r^2 d\Omega_5^2$ is the metric of the 6-dimensional transverse space. In particular $r$ is the distance from the branes, and the metric changes as we move along $r$ because of the warp factor:
\be H=1+{L^4\over r^4}\sp L^4=4\pi g_s l_s^4 N\label{w}\ee

The energy measured depends on $r$ due to gravitational redshift. If at some point $r$ we measure energy $E_r$, an observer at infinity would measure $E_\infty=H^{-{1\over 4}}E_r$. Therefore an object moving near the branes $r\to 0$ (which means $H^{-{1\over 4}}\to 0$) would appear to have very low energy to an observer at infinity.

From the point of view of an observer located at infinity there are two types of low energy excitations: massless low energy modes that propagate in the whole bulk and all modes that approach the horizon ($r=0$), since any finite energy is redshifted to zero. In the low-energy limit the two types of excitations decouple from each other, \cite{st}.

Excitations of the first kind propagate away from the branes where the space is flat. For very large r, $H\approx 1$ and the metric (\ref{m1}) becomes flat. Therefore this set of modes is described by free supergravity.

Excitations of the second kind remain confined near the horizon, because there is a potential barrier they have to climb in order to get away, \cite{st}. For very small $r$, $H^{1/2}\approx L^2/r^2$ and the metric (\ref{m1}) becomes:
\be ds^2={r^2\over L^2}(-dt^2+d\vec{x}^2)+{L^2\over r^2}(dr^2+r^2 d\Omega_5^2)\label{b6}\ee
Changing the radial coordinate to $u=L^2/r$ we obtain:
\be ds^2={L^2\over u^2}(du^2-dt^2+d\vec{x}^2)+L^2 d\Omega_5^2 \label{AdSS}\ee
which describes the product space $AdS_5\times S^5$. Therefore the low-energy limit of this system is described by two decoupled theories: IIB free supergravity and IIB supergravity in $AdS_5\times S^5$.

In both cases, the low-energy description of the system of $N$ $D_3$-branes reduces to the sum of two non-interacting theories. One of those theories is free IIB supergravity in both cases. It is natural to conjecture that the two remaining theories, gravity in $AdS_5\times S^5$ and $\mathcal{N}=4, U(N)$ SYM, are equivalent.

\subsubsection{Validity of the Correspondence}
We now examine the region of validity of the two dual descriptions. We would first like to find the connection between the dimensionless parameters of the two theories. Starting from (\ref{DBI3}) and keeping only the gauge field terms:
\be \mathcal{L}_{YM}={1\over 2\pi g_s}  Tr\left[{1\over 4}F_{\m\n}F^{\m\n}\right]={c\over 2\pi g_s} {1\over 4} F^a_{\m\n}F^{a,\m\n}\sp F_{\m\n}=F^a_{\m\n}T^a\label{b7}\ee
where $T^a$ are the generators of the non-Abelian group and c is their normalization constant: $Tr\left[T^a T^b\right]=c \delta^{ab}$. A popular choice that we are going to use is $c=1/2$, \cite{st}. Therefore the Yang-Mills coupling is:
\be g_{YM}^2=4\pi g_s \label{gYM}\ee
Combining (\ref{gYM}) with the expression for the radius of the AdS space-time from (\ref{w}):
\be \left({L\over \ell_s}\right)^4=N g_{YM}^2=\lambda \label{f1}\ee
where $\lambda$ is the 't Hooft coupling defined in (\ref{tH}). This is one of the formulas we were looking for.

The 10-dimensional Newton constant is given by:
\be 16\pi G_{10}=(2\pi)^7 g_s^2 \ell_s^8\label{b8}\ee
Combining this with (\ref{gYM}) and (\ref{tH}) we obtain:
\be {G_{10}\over \ell_s^8}={\pi^4 \lambda^2\over 2 N^2}\label{g10}\ee
We can now find the region of validity of the two descriptions.
From (\ref{g10}) we see that $G_{10}\sim 1/N^2$, which means that quantum effects are suppressed for large N. If the CFT is strongly coupled ($\lambda\gg 1$) then, according to (\ref{f1}), $L\gg \ell_s$, which means that the string theory is weakly curved and can be approximated by supergravity. Therefore the large-$N$ limit of the sYM theory is described well by the two-derivative action of IIB supergravity in $AdS_5\times S^5$.

\subsubsection{The Holographic Coordinate as an Energy Scale}

As we send $\ell_s\to 0$ in the brane theory, the energy an observer measures at infinity is:
\be E_\infty=H^{-{1\over 4}}E_r\sim E_r{r\over \ell_s}\label{b9} \ee
We must keep the energy in the near-horizon region fixed in string units $\ell_s E_r$, \cite{st}, as well as the energy in the near-boundary region $E_\infty$ since this is the energy measured in the CFT. From:
\be E_\infty\sim E_r{r\over l_s}=(E_r \ell_s){r\over \ell_s^2}\label{b10}\ee
we see that $U={r\over \ell_s^2}$ must be kept fixed as we are taking the $r\to 0$ limit. The radial coordinate is therefore proportional to the energy scale of the dual CFT near the horizon. We can change the radial coordinate to $U={r\over \ell_s^2}$ in the near-horizon metric, \cite{Mald}:
\be ds^2=\ell_s^2 \left[{U^2\over \sqrt{4\pi g_s N}}(-dt^2 + d\vec{x}^2)+ \sqrt{4\pi g_s N}\left({dU^2\over U^2}+d\Omega_5^2\right)\right]\label{b11}\ee
The coordinates $t,\vec{x}$ are the space-time coordinates of the CFT. The extra coordinate, $U$, of the AdS part behaves like the energy scale of the CFT.

This argument involved the decoupling limit. There are, however, other arguments that indicate that the radial direction behaves like an energy scale for the gauge theory with the UV located near the boundary. We write the $AdS_5$ metric in Poincar\'e coordinates:
\be ds^2={1\over u^2}(du^2-dt^2+d\vec{x}^2)\label{b12}\ee
The metric is invariant under SO(1,1):
\be (u,t,\vec{x})\to (au,at,a\vec{x})\label{b13}\ee
The boundary is located at $u=0$ in these coordinates. If we scale up the coordinates of the gauge theory on the boundary $(t,\vec{x})$, which means going down in energy, we must also scale up u, which means moving away from the boundary at $u=0$. Therefore the high-energy limit of the gauge theory (UV) corresponds to $u = 0$ (the boundary) while the low-energy limit (IR) corresponds to $u = \infty$ (the Poincar\'e horizon), \cite{st}.

\subsubsection{Bekenstein's Bound in AdS/CFT}

In this section we calculate the degrees of freedom (entropy) of the theory on the boundary and the bulk theory.

We begin by cutting out the 5-sphere part of the metric (\ref{AdSS}). It turns out that by reducing the $S^5$ part, every massless field in the original theory corresponds to an infinite tower of massive fields on AdS$_5$ with ever increasing mass, \cite{st}, \cite{AdS/CFT}. The only interesting detail for our purpose is the value of the 5-dimensional Newton's constant $G_5$ after the reduction on $S^5$. The relationship between $G_{10}$ and $G_5$ can be found by considering the reduction of the Einstein-Hilbert term, \cite{AdS/CFT}:
\be {1\over 16\pi G_{10}} \int \ud^5 x \ud ^5 \Omega \sqrt{-\det(g_{10})}R_{10}={V\over 16\pi G_{10}} \int \ud^5 x \sqrt{-\det(g_{5})}R_{5}\label{b14}\ee
where $V=\pi^3 L^5$ is the volume of an $S^5$ of radius L. It follows that
\be G_5=G_{10}/V={(2\pi)^7l_s^8 g_s^2\over (16\pi) \pi^3 L^5}={\pi L^3\over 2 N^2}\label{b15}\ee
where we also used (\ref{w}).

Now that we removed the sphere we consider the AdS$_5$ metric in global coordinates:
\be ds^2=L^2(-\cosh^2(\rho)d\tau^2+d\rho^2 +\sinh^2(\rho)d\Omega_3^2) \label{m5}\ee
Changing the radial coordinate to $u=\tanh({\rho\over 2})$, (\ref{m5}) becomes:
\be ds^2=L^2\left(-\left({1+u^2\over 1-u^2}\right)^2d\tau^2+{4\over (1-u^2)^2}(du^2+u^2 d\Omega_3^2)\right) \label{m4}\ee
In this coordinate system the boundary is located at $u=1$ and the interior at $0\leq u<1$.

We introduce a cutoff close to the boundary at $u^2=1-\e$ with $\e$ very small. This corresponds to a UV cutoff in the dual gauge theory according to the previous section. The gauge theory lives on an $S^3$ of radius L. The distance between two points on the cutoff sphere scales as $\log(|x_1-x_2|/\epsilon)$. Therefore we may view $\e\ll 1$ as a small distance cutoff in the gauge theory. Since the gauge theory has a small distance cutoff $L\e$ (in units of length), there are $1/\e^3$ fundamental cells of radius L in the 3-sphere. Since the gauge theory has order $N^2$ degrees of freedom and the entropy is proportional to the regularized volume of the boundary we obtain:
\be S_{CFT}\sim {N^2\over \e^3}\label{b16}\ee

On the $AdS_5$ side the area of the sphere at the regularized boundary can be read directly from the metric (\ref{m4}):
\be A= 8  {L^3 u^3\over (1-u^2)^3}\bigg|_{u^2=1-\e}\sim {L^3\over \e^3}\label{a}\ee
The gravitational entropy is given by the Bekenstein bound:
\be S_{AdS}\sim{A\over G_5}\sim {N^2\over \e^3}\label{b17}\ee
where $G_5$ is the AdS$_5$ Newton constant we calculated earlier. The two descriptions have the same degrees of freedom. Therefore the AdS/CFT correspondence successfully realized holography in string theory.

We can also calculate the volume of $AdS_5$ up to the shifted boundary $u^2=1-\e$:
\be V_4=L^4 \int {2u^3\over (1-u^2)^4}\ud u \ud^3\Omega_3=L^4 \Omega_3 \int_{0}^{u^2=1-\e} {2u^3\over (1-u^2)^4}\ud u\sim {L^4\over \e^3}\label{b18}\ee
where $\Omega_3$ is the volume of the 3-sphere. Comparing with (\ref{a}) we notice that the volume of $AdS_5$ scales with the same power of $\e$ as its area close to the boundary. From that aspect holography seems trivial. However the non-triviality of our previous analysis stems from the fact that area and volume scale with different powers of L.

\subsection{The GKPW Formula\label{gkpw}}

The GKPW formula, named after Gubser-Klebanov-Polyakov, \cite{Gubser:1998bc}, and Witten, \cite{witten}, relates the generating functionals of the bulk and boundary theories. The ability to calculate boundary correlators from bulk data is, in fact, what makes Holography useful.
It can be written schematically as
\be \mathcal{Z}_{QFT}[\phi_0]=\mathcal{Z}_{string}[r^{\Delta-d}\phi(x,r)|_{r= 0}= \phi_0(x)].\label{gkpw0}\ee

The left-hand side of (\ref{gkpw0}) represents the generating functional of the $d$-dimensional QFT in the presence of an external source $\phi_0$
\be \mathcal{Z}_{QFT}[\phi_0]=\braket{\exp\left(i\int \mathcal O\phi_0\right)}\label{g2}\ee
 where $\mathcal{O}$ is the scalar operator of the QFT coupled to the source. If $\mathcal{Z}_{QFT}[\phi_0]$ is known, it is straightforward to compute the correlators of $\mathcal O$ by taking functional derivatives with respect to $\phi_0$. However, calculating the generating functional in strongly coupled theories is, in most cases, very difficult.

 The right-hand side of (\ref{gkpw0}) represents the generating functional of the $d+1$-dimensional string theory. The expression in the brackets is the boundary condition that is imposed at the AdS boundary.

We approximate this side by the classical gravity limit, in which the action is calculated on-shell. This limit is necessary in order to construct invariant (boundary) observables. It is equivalent to taking large-$N$ and strong coupling limit in the dual QFT. The large-$N$ limit suppresses string loops, while the strong coupling limit suppresses stringy corrections, \cite{st}. This is what we calculate in Holography in order to compute correlators for the boundary theory. We wrote the $d+1$-dimensional AdS metric in Poincar\'e coordinates:
 \be ds^2={L^2\over r^2}(dr^2+\eta_{\mu\nu}dx^\mu dx^\nu)\label{ads0}\ee
 where $L$ is the AdS radius, $r$ is the Holographic coordinate, $\eta$ is the $d$-dimensional Minkowski metric and $\mu,\nu$ run over the boundary indices. The boundary of the space-time is located at $r=0$.

From the QFT point of view, $\phi_0$ is a non-dynamical field. It may be just a notational trick (a ``source") to compute correlators of the operator $\mathcal O$ or it may represent something more real, such as an external electric field in an experiment. From the bulk point of view, $\phi$ is a dynamical field governed by its own equations of motion. The two sides are connected by the boundary condition $r^{\Delta-d}\phi(x,r)|_{r= 0}= \phi_0(x)$, where the constant $\Delta$ will be explained shortly.

The GKPW relation is formulated in Euclidean time. If taken literally in Lorentzian signature, problems may arise. For example, if the bulk contains a black hole there is a difference in the boundary conditions at the horizon. We will discuss this topic in section \ref{2pf}.

To illustrate the GKPW formula we consider the simplest example: a scalar field in the AdS$_{d+1}$ bulk:
\be S=-{1\over 2}\int \ud^{d+1}x\sqrt{-g}\left[(\partial\phi)^2+m^2\phi^2\right]\label{g3}.\ee
The action can be written equivalently as
\be S=-{1\over 2}\int \ud^{d+1}x\left[\partial_\mu(\sqrt{-g}\phi\partial^\mu\phi)-\phi(\partial_\mu(\sqrt{-g}\partial^\mu \phi)-\sqrt{-g}m^2\phi)\right]\label{g4}.\ee
The second term vanishes by the equations of motion
\be \partial_\mu(\sqrt{-g}\partial^\mu\phi)-\sqrt{-g}m^2\phi=0\label{g5}\ee
and the on-shell action reduces to the boundary term
\be S_{os}={1\over 2}\int \ud^{d}x \sqrt{-g}g^{rr}\phi\partial_r\phi|_{r\to 0}.\label{sosa}\ee
For the AdS geometry (\ref{ads0}) we have
\be \sqrt{-g}=L^{d+1}r^{-d-1}\sp g^{rr}=r^2/L^2.\label{g6}\ee
We Fourier transform the field on the boundary coordinates \be \phi(r,x)=\int{\ud^d k\over (2\pi)^d} e^{ikx}\phi(r,k)\ee and the equation of motion reads:
\be r^2\phi''+(1-d)r\phi'-(m^2L^2+k^2r^2)\phi=0,\label{g7}\ee
where primes denote derivatives with respect to $r$.
Near the boundary ($r=0$) we use the ansatz $\phi\sim r^{\Delta}$ and keep the leading terms to obtain
\be \Delta(\Delta-d)=m^2L^2\label{g8}\ee
which has solutions\footnote{Note that a scalar field in AdS space-time can be tachyonic, as long as $m^2\ge -d^2/(4L^2)$. This is known as the Breitenlohner-Freedman bound. Values of $m^2$ below this bound lead to instabilities.}
\be \Delta_\pm={d\over 2}\pm\sqrt{{d\over 4}^2+m^2L^2}.\label{g9}\ee
Since $\Delta_-<\Delta_+$ we have to set the boundary condition on the $\Delta_-$ mode
\be r^{\Delta-d}\phi|_{r= 0}(r,x)=\phi_0(x)\label{g11}\ee
and the constant $\Delta$ is identified with
\be \Delta=d-\Delta_-=\Delta_+.\label{g12}\ee
Therefore the near-boundary behavior of a scalar $\phi$ in an asymptotically AdS spacetime is
\be \phi(r,x)=r^{\Delta_-}(\phi_0(x)+\cdots)+r^{\Delta_+}(\phi_1(x)+\cdots)\label{ssol}.\ee
Since the scalar equation is of second order and we set only one boundary condition, $\phi_1(x)$ remains arbitrary. The other boundary condition is set in the interior of the bulk, by demanding that the field is regular.

The mode $\phi_1$ has a very simple physical interpretation --- it is the expectation value of the operator $\mathcal{O}$ in the presence of the source $\phi_0$:
\be \phi_1(x)\sim \braket{\mathcal{O}(x)}_s,\label{g13}\ee
where $\braket{\phantom{O}}_s$ denotes an expectation value in presence of the source $\phi_0$.
This is straightforward to show for a massless scalar $m^2=0$. In this case $\Delta_-=0\sp \Delta_+=d$. Since the equation is linear, we can write the solution (\ref{ssol}) as
\be \phi=\phi^{(0)}(1+\phi^{(1)} {r}^d )+\cdots,\label{g14}\ee
where $\phi^{(1)}$ is independent of $\phi^{(0)}$.
Substituting the above expansion into the action (\ref{sosa}) we obtain
\be S_{os}[\phi^{(0)}]=L^{d-1}{d\over 2}\int \ud^d x(\phi^{(0)})^2\phi^{(1)}\label{g15}.\ee
According to the GKPW relation
\be \braket{\mathcal{O}}_s={\delta S[\phi^{(0)}]\over \delta \phi^{(0)}}=L^{d-1}{d}\phi^{(0)}\phi^{(1)},\label{g16}\ee
therefore
\be \phi=\phi^{(0)}+{L\over d} \braket{\mathcal O}_s {r^d\over L^d} +\cdots.\label{svev}\ee

For a massive scalar ($m^2\ne 0$) the on-shell action diverges. The process of handling such divergences is known as \textit{holographic renormalization}. This is the subject of the next section.

\subsection{Holographic Renormalization\label{hren}}

Renormalization in the context of Holography is done in an analogous way to QFT, \cite{Skenderis:2002wp}. First, the action is regularized by putting the boundary at a cutoff radius $\epsilon$. This corresponds to a UV cutoff in the dual QFT. Boundary counterterms are added to the action, such that the terms that diverge as $\epsilon\to 0$ cancel out. These counterterms must be finite in number, they must cancel all divergences and may contain only boundary fields. The calculation of correlators is performed at the cutoff $\epsilon$ and then the $\epsilon\to 0$ limit is taken. We will continue the example of the scalar field to illustrate this process.

Attempting to calculate the on-shell action (\ref{sosa}) for a massive scalar we find that it diverges.
In order to regularize it we put the boundary at $r=\epsilon$.
 We write the solution (\ref{ssol}) as
\be \phi=\phi^{(0)}\left(r^{\Delta_-}+\phi^{(1)} r^{\Delta_+} \right)+\cdots.\label{hr1}\ee
and calculate the leading term of the regularized on-shell action
\be S_{reg}(\epsilon)={L^{d-1}\over 2}\int \ud^d x {\phi'\phi|_{r=\epsilon}\over \epsilon^{d-1}}={L^{d-1}\over 2}\int \ud^d x\left({\Delta_-}\epsilon^{2\Delta_--d}(\phi^{(0)})^2+\cdots\right).\label{hr2} \ee
Clearly this diverges as $\epsilon\to 0$, since
\be 2\Delta_--d=-\sqrt{d^2+4m^2L^2}<0.\label{hr3}\ee
The appropriate counterterm to cancel this divergence is, \cite{Natsuume:2014sfa},
\be S_{CT}(\epsilon)=-{\Delta_-\over 2L}\int \ud^d x \sqrt{-\gamma}\phi^2\bigg|_{r=\epsilon}\label{hr4}\ee
where $\gamma_{ab}$ is the boundary metric. The $S_{CT}$ above checks all the boxes, since it contains finitely many terms, is written only in terms of boundary fields and cancels all divergences. The subtracted bulk action is
\be S_{sub}(\epsilon)=S_{reg}(\epsilon)+S_{CT}(\epsilon)\label{hr5}\ee
and the renormalized action, defined as
\be S_{ren}=\lim\limits_{\epsilon\to 0} S_{sub}(\epsilon),\label{hr6}\ee
is finite on-shell:
\be S_{ren}[\phi^{(0)}]=L^{d-1}{{2\Delta_+-d}\over 2}\int \ud^d x(\phi^{(0)})^2\phi^{(1)}.\label{hr7}\ee
According to the GKPW relation
\be \braket{\mathcal{O}}_s={\delta S_{ren}[\phi^{(0)}]\over \delta \phi^{(0)}}=L^{d-1}(2\Delta_+-d)\phi^{(0)}\phi^{(1)},\label{hr8}\ee
therefore
\be \phi=\phi^{(0)}r^{\Delta_-}+{L^{1-d}\over 2\Delta_+-d} \braket{\mathcal{O}}_s r^{\Delta_+} +\cdots\label{iref1}\ee
which reduces to (\ref{svev}) in the massless case.

\subsection{Finite Temperature\label{ft0}}

Holography can be extended to study theories at finite temperature. A field theory at finite temperature is dual to a space-time with a black hole, \cite{Witten:1998zw}.
We may think of a conformal theory as the UV limit of a theory deformed by finite temperature. Restoration of conformal symmetry at the UV implies that the space-time is asymptotically AdS. The IR of the geometry is dominated by the existence of the black hole. The UV physics are affected by the IR geometry through the regularity condition of the bulk fields at the horizon.

The simplest asymptotically AdS black hole is the AdS-Schwarzschild space-time
\be ds^2={L^2\over r^2}\left(-f(r)dt^2+{dr^2\over f(r)}+dx_idx^i\right)\sp i=1,2,\dots,d-1\label{Sch0}\ee
where
\be f(r)=1-{r^d\over r_h^d}.\label{ft1}\ee
Near the boundary $r\to 0$ the metric is AdS, as required. The IR part of the geometry contains a black hole with its horizon at $r_h$. The above metric is a solution to the Einstein equations with negative cosmological constant.

We follow the notes on \cite{Hartnoll:2009sz} in order to show that the dual field theory actually has finite temperature.
Rotating to Euclidean time $\tau=it$, the metric (\ref{Sch0}) becomes
\be ds^2={L^2\over r^2}\left(f(r)d\tau^2+{dr^2\over f(r)}+dx_idx^i\right).\label{ft2}\ee
We study the geometry near the horizon by changing into the near-horizon coordinate $\epsilon=r_h-r$, with $\epsilon$ very small and positive
\be ds^2={L^2d\epsilon\over r_h^3}d\tau^2+{L^2\over r_hd\epsilon}{d\epsilon^2}+{L^2\over r_h^2}dx_idx^i.\label{ft3}\ee
The coordinates $x_i$ decouple, so we focus on the $\epsilon-\tau$ part. In the coordinates $\rho=2L\sqrt{\epsilon/(r_hd)}$ and $\phi=\tau d/(2r_h)$ we have
\be ds_2^2=\rho^2 d\phi+d\rho^2.\label{ft4}\ee
This is similar to the $\mathbb R^2$ metric in polar coordinates, but $\phi$ is not periodic. This is actually the metric of a cone, with the conical singularity\footnote{In simple words, a closed curve of constant $\rho$ has circumference different than $2\pi\rho$ as $\rho\to 0$.} at $\rho=0$. In order to eliminate the singularity we need to require that $\phi$ has period $2\pi$. Identifying $\phi\sim \phi+2\pi$, the Euclidean time also becomes periodic
\be \tau\sim \tau+{4\pi r_h\over d}.\label{ft5}\ee

We now examine how this affects the dual field theory. Near the boundary the metric behaves as
\be g_{\mu\nu}(r)|_{r\to 0}={L^2\over r^2}g_{\mu\nu}^{(0)}+\cdots.\label{ft6}\ee
The pullback of $g_{\mu\nu}^{(0)}$ to the boundary is interpreted as the metric of the boundary theory. There is an ambiguity in this definition, because a conformally invariant theory is not affected by the overall conformal factor of the metric. The metric of the field theory is, therefore
\be ds^2=d\tau^2+dx_idx^i.\label{ft7}\ee
This is a field theory with periodically identified Euclidean time. This is an attribute of a statistical system at finite temperature equal to the inverse periodicity
\be T={d\over 4\pi r_h}.\label{tempsch0}\ee

The thermodynamics of the theory can be studied by evaluating the partition function.
 We start from the Einstein action in Euclidean time with the additional boundary terms
\be S_E=-M^{d-1}\int d^{d+1}\sqrt{-g}\left(R+{d(d-1)\over L^2}\right)+M^{d-1}\int_{\partial V} d^{d}\sqrt{-\g}\left(-2K+{2(d-1)\over L}\right).\label{ft8}\ee
The first of the boundary terms is the Gibbons-Hawking boundary term, which is needed on geometries with a boundary in order for the variational problem of the metric to be well-defined. The fields that appear are the induced metric on the boundary $\gamma$ and the trace of the extrinsic curvature $K=\g^{\mu\nu}\nabla_\mu n_{\nu}$, where $n_\mu$ is an outward-pointing unit vector, normal to the boundary. The last term is a constant counterterm to cancel divergences at the boundary. The above action evaluated on the solution (\ref{Sch0}) is
\be S_E=-{M^{d-1}L^{d-1}}{V_{d-1}\over T r_h^d}=-{M^{d-1}L^{d-1}}{(4\pi)^dV_{d-1}\over  d^d}T^{d-1},\label{ft9}\ee
where $V_{d-1}$ is the spatial volume of the boundary.
 The Helmholtz free energy is
\be F=-T\log Z=TS_E=-{M^{d-1}L^{d-1}}{(4\pi)^dV_{d-1}\over  d^d}T^{d}\label{ft10}\ee
and the entropy is
\be \mathcal S=-{\partial F\over \partial T}={M^{d-1}L^{d-1}}{(4\pi)^dV_{d-1}\over  d^{d-1}}T^{d-1}.\label{ft11}\ee
The entropy density, using (\ref{tempsch0}) to express the temperature in terms of $r_h$, is
\be s={\mathcal S\over V_{d-1}}=M^{d-1}4\pi {L^{d-1}\over r_h^{d-1}},\label{ft12}\ee
which is proportional to the area of the black hole's horizon.

\subsection{Chemical Potential and Magnetic Field \label{cpmf}}

A common additional structure, important for condensed matter theories, is a global U(1) symmetry. In Holography, a global symmetry corresponds to a gauged symmetry in the bulk, \cite{Hartnoll:2009sz}. The minimal bulk action with a U(1) gauge field is the Einstein-Maxwell action
\be S=M^{d-1}\int \ud^{d+1}x\sqrt{-g}\left(R+{d(d-1)\over L^2}-{1\over 4}F^2\right),\label{c1}\ee
where $F_{\mu\nu}=\pa_{\mu}A_{\n}-\pa_{\nu}A_{\mu}$ is the strength tensor of $A_\mu$, the fields are dimensionless and $M$ has mass dimension $1$ for normalization of the action.

 The near-boundary ($r\to 0$) behavior of the bulk gauge field is
\be MA_\mu(r)=A_\mu^{(0)}+\cdots.\label{c2}\ee
The field theory's gauge field is the pullback of $A_\mu^{(0)}$ to the boundary.
The component $\mu=A_t^{(0)}$ is a chemical potential, sourcing a charge density $q$ in the field theory.
The solution with finite chemical potential is the AdS-Reissner-Nordstrom black hole \cite{Hartnoll:2009sz}
\be A=\mu\left(1-\left({r\over r_h}\right)^{d-2}\right)dt\sp ds^2={L^2\over r^2}\left(-f(r)dt^2+{dr^2\over f(r)}+dx_idx^i\right)\label{c3}\ee
with
\be f(r)=1-\left(1+{r_h^2\mu^2\over \gamma^2}\right)\left({r\over r_h}\right)^d+{r_h^2\mu^2\over \g^2}\left({r\over r_h}\right)^{2(d-1)},\label{c4}\ee
where
\be \gamma^2={d-1\over d-2}L^2\label{c5}.\ee
$A_t$ has to vanish at the horizon so that the one-form $A$ is well-defined there \cite{Koba}. This determines the constant term in $A_t$.

The temperature of this theory is
\be T={1\over 4\pi r_h}\left(d-{(d-2)r_h^2\mu^2\over \g^2}\right)\label{temp0}.\ee
The thermodynamic potential is obtained by calculating the renormalized Euclidean action on the solution, as in the previous section. No additional counterterms are needed, as the Maxwell sector of the action does not diverge at the boundary. In the grand canonical ensemble, $\mu$ is fixed and the grand potential is, \cite{Hartnoll:2009sz},
\be \Omega=-T\log Z=-{M^{d-1}L^{d-1}\over r_h^d}\left(1+{r_h^2\mu^2\over \g^2}\right)=\mathcal F\left({T\over \mu}\right)V_{d-1} T^d,\label{c6}\ee
where $\mathcal F$ is obtained by solving (\ref{temp0}) for $r_h$.

In $d=3$ dimensions we can also add a magnetic field without breaking the rotational symmetry
\be A=\mu\left(1-{r\over r_h}\right)dt+hx dy.\label{c7}\ee
The blackening factor now becomes,  \cite{Hartnoll:2009sz},
\be f(r)=1-\left(1+{r_h^2\mu^2+r_h^4h^2\over \gamma^2}\right)\left({r\over r_h}\right)^3+{r_h^2\mu^2+r_h^4h^2\over \g^2}\left({r\over r_h}\right)^{4}\label{c8}\ee
and the temperature is
\be  T={1\over 4\pi r_h}\left(3-{r_h^2\mu^2+r_h^4h^2\over \g^2}\right)\label{temp10}.\ee
The grand potential becomes
\be \Omega=-T\log Z=-{M^{2}L^{2}\over r_h^d}\left(1+{r_h^2\mu^2\over \g^2}-{3r_h^4h^2\over \g^2}\right).\label{c9}\ee
Two important quantities are the charge density,  \cite{Hartnoll:2009sz},
\be q=-{1\over V_{2}}{\partial\Omega\over\partial\mu}={M^2L^2}{4\mu\over r_h\g^2}\label{c10}\ee
and the magnetization density,  \cite{Hartnoll:2009sz},
\be m=-{1\over V_2}{\partial\Omega\over\partial h}=-{M^2L^2}{4r_hh\over\g^2}\label{c11}.\ee

\subsection{Real-time Two-point Functions\label{2pf}}

As we mentioned in \ref{gkpw}, the GKPW relation was formulated in Euclidean time and it needs to be appropriately modified for a Lorentzian signature. In principle, a Euclidean correlator can be analytically continued to obtain the Feynman propagator and subsequently calculate the rest of the Green's functions. However, when the bulk equations can be solved only in certain limits, it is challenging to analytically continue the Green's functions.

 A prescription for calculating two-point functions directly in real-time was formulated in \cite{Son}. A problem with this prescription is that it does not work for higher-point functions and that it seemingly cannot be derived by variations of the on-shell action. In \cite{HS} a more general prescription was formulated. The correlators can be calculated by taking derivatives of an appropriately modified on-shell action. In the case of two-point functions, the two prescriptions are in agreement. A geometric rephrasing of the Schwinger-Keldysh  prescription can be found in \cite{Skenderis:2008dh}, with examples of its use in \cite{Skenderis:2008dg}.
 Since we are only interested in two-point functions, we will use the prescription in \cite{Son}. We will briefly describe the idea behind it.

Consider fluctuations of a scalar in a finite-temperature AdS background
\be S=-{1\over 2}\int \ud^{d+1}x\sqrt{-g}\left[(\partial\phi)^2+m^2\phi^2\right],\label{d1}\ee
with
\be ds^2={L^2\over r^2}\left(-f(r)dt^2+{dr^2\over f(r)}+dx_idx^i\right)\label{d2}.\ee
The fluctuation equation for the scalar in real time is
\be {1\over \sqrt{-g}}\partial_r (\sqrt{-g}g^{rr}\partial_r\phi)-\partial^t\partial_t\phi+\partial^i\partial_i\phi-m^2\phi=0\label{rt0}\ee
and rotating to Euclidean time $\tau=it$ we obtain
\be {1\over \sqrt{-g}}\partial_r (\sqrt{-g}g^{rr}\partial_r\phi)+\partial^\tau\partial_\tau\phi+\partial^i\partial_i\phi-m^2\phi=0.\label{euc0}\ee
We Fourier transform the field as follows
\be \phi(r,x)=\int {\ud^dx\over (2\pi)^d} e^{i\omega_E\tau+ikx}\pi(r,k)\phi_0(k),\label{d3}\ee
where $\phi_0$ is determined by the boundary condition
\be \phi(0,x)=\int {\ud^dx\over (2\pi)^d} e^{i\omega_E\tau+ikx}\phi_0(k),\label{bc0}\ee
so that $\pi=1$ at the boundary.
The Euclidean equation (\ref{euc0}) in Fourier space is
\be \pi''+\left({f'\over f}-{d+1\over r}\right)\pi'+{1\over f}\left(-{\omega_E^2\over f}-k^2-{m^2L^2\over r^2}\right)\pi=0\label{d4},\ee
where primes denote derivatives with respect to $r$.
To find the leading mode near the horizon we use the ansatz
\be \pi(r)\simeq f(r)^{\nu}\label{d5}\ee
to obtain
\be \nu^2={\omega^2\over f'(r_h)}\label{d6}\ee
hence
\be \nu_\pm=\pm \omega/f'(r_h)\label{d7}.\ee
Regularity of the solution at the horizon requires us to choose the positive mode $\nu_-$. This, along with the boundary condition (\ref{bc0}) at the AdS boundary, uniquely fixes the solution.

We now consider the real-time equation. We write (\ref{rt0}) in Fourier space
\be \pi''+\left({f'\over f}-{d+1\over r}\right)\pi'+{1\over f}\left({\omega^2\over f}-k^2-{m^2L^2\over r^2}\right)\pi=0.\label{refeq0}\ee
To find the leading mode near the horizon we use the ansatz
\be \pi(r)\simeq f(r)^{\nu}\label{d8}\ee
to obtain
\be \nu^2=-{\omega^2\over f'(r_h)}\label{d9}\ee
hence
\be \nu=\pm i \omega/f'(r_h)\label{d10}.\ee
Now there is no clear choice. The modes $\n_-$ or $\n_+$ correspond to an in-going or out-going wave respectively. This ambiguity is a reflection of the multiple Green's functions that can be defined in real time. From a physical perspective we should require that no fields escape from the horizon, hence we should choose the in-going boundary condition. It was conjectured in \cite{Son} that the in-going boundary condition corresponds to the retarded Green's function in the field theory. This, however, does not resolve all the problems.

Having solved the equation with the in-going condition at the horizon, the on-shell action can be written in the form
\be S=\int {\ud^dx\over (2\pi)^d}\phi_0(-k)\mathcal F(r,k)\phi_0(k)\bigg|_{0}^{r_h},\label{d11}\ee
where
\be \mathcal F(r,k)=-{1\over 2}\sqrt{-g}g^{rr}\pi(r,-k)\partial_r\pi(r,k).\label{d12}\ee
Taking the second derivative with respect to the source $\phi_0(k)$ we obtain the Green's function
\be G(k)=-\mathcal F(r,k)|_0^{r_h}-\mathcal F(r,-k)|_0^{r_h}.\label{d13}\ee
The imaginary part of $\mathcal F$
\be Im(\mathcal{F}(r,k))={1\over 4i}\sqrt{-g}g^{rr}\left(\pi^\star(r,k)\partial_r\pi(r,k)-\pi(r,k)\partial_r\pi^\star(r,k)\right)\label{d14}\ee
is radially conserved\footnote{This can be checked by taking a radial derivative and using the equation (\ref{refeq0}).}, \cite{Son}. Therefore, in each term of $G$, the imaginary part at the horizon cancels the imaginary part at the boundary. It follows that $G$ is real, which is unacceptable for a Green's function.

One could guess that discarding the contribution from the horizon would solve this problem. In this case we obtain
\be G(k)=-\mathcal F(r,k)|_0-\mathcal F(r,-k)|_0,\label{d15}\ee
which is still problematic; reality of $\phi$ implies that $\mathcal{F}(r,k)^\star=\mathcal{F}(r,-k)$ and the $G$ defined above is again real.

The prescription proposed in \cite{Son}, which avoids the problems outlined above, is the following
\be G^R(k)=-2\mathcal F(r,k)|_{r\to 0}.\label{ss0}\ee
In \cite{HS} a more general prescription has been formulated in the context of the Schwinger-Keldysh formalism. The two-point (and higher) functions can be calculated by variational derivatives of the on-shell action, contrary to (\ref{ss0}).

\subsection{The Holographic Dictionary\label{dict}}

Using the GKPW formula, (\ref{gkpw0}), every observable in the boundary quantum theory can be translated to a bulk quantity and vice versa. The duality can be summarized in a ``Holographic dictionary". An overview is given in table \ref{HD}.

A source in the boundary theory is a dynamical field in the bulk. This implies that for every operator that can be written down in the QFT there exists a corresponding bulk field. The scaling dimension of each operator is related to the mass of its dual bulk field, as shown in table \ref{SD}.

The simplest example is the map between a scalar operator and a scalar bulk field
\be \phi(x,r)\leftrightarrow \mathcal O(x).\label{hd1}\ee
Consider the expansion in (\ref{iref1})
\be \phi=\phi^{(0)}r^{\Delta_-}+{L^{1-d}\over 2\Delta_+-d} \braket{\mathcal{O}}_s r^{\Delta_+} +\cdots\label{hd2}\ee
under the scaling transformation $x^\mu\to \lambda x^\mu$. Since the left-hand side is a bulk scalar, its scaling dimension is $0$. Then it is clear that $\phi^{(0)}$  and $\mathcal O$ have scaling dimensions $\Delta_-$ and $\Delta_+$ respectively. Since $\Delta_{\pm}$ depend on the mass of the bulk field, the scaling dimension of each operator is directly related to the mass of its dual bulk scalar. The same is true for higher-spin fields (see table \ref{SD}).

A higher-spin operator is dual to a field with the same number of indices. For example a vector operator is dual to a vector field
\be A_\mu(x,r)\leftrightarrow  \mathcal J^a(x).\label{hd3}\ee
If $\mathcal J^a$ is a conserved current ($\partial_a \mathcal J^a=0$), then the field $A_\mu$ is a gauge field. Note that the index $\mu$ runs over $1$ more value than $a$ corresponding to the Holographic coordinate.

A spin-2 operator is dual to a bulk field with two indices
\be g_{\mu\nu}(x,r)\leftrightarrow  T^{ab}(x).\label{hd4}\ee
If $\mathcal T^{ab}$ is conserved ($\partial_a \mathcal T^{ab}=0$), then $g_{\mu\nu}$ is massless. The most important example is a stress-energy tensor dual to the bulk graviton. Since every reasonable translationally invariant theory contains a conserved stress-energy tensor, every Holographic theory must contain gravity. The non-conserved, symmetric, $2$-index operators are dual to massive, symmetric, $2$-index bulk fields.

The above can be generalized for any kind of operator. Fermionic operators are dual to fermionic fields, spin-3 operator are dual to spin-3 fields and so on.

The symmetries on the two sides are also matched. A global space-time symmetry in the field theory corresponds to an isometry in the bulk metric. For example, the symmetry group of a conformal theory in $d$ dimensions is isomorphic to $SO(d-2,d)$, which is precisely the isometry group of AdS$_{d+1}$. The internal symmetry of a boundary field corresponds to a gauged symmetry of the dual bulk field. An example is a $U(1)$ current, dual to a gauged $U(1)$ field.

\begin{table}[H]
	\label{HD}
\begin{center}
	\begin{tabular}{| l | l |}
		\hline
		\multicolumn{2}{|c|}{\textbf{The Holographic Dictionary}} \\
		\hline
		\textbf {Boundary} & \textbf{Bulk}  \\ \hline
		Global space-time symmetry & Local isometry  \\ \hline
		Internal global symmetry & Gauged symmetry\\ \hline
		Energy scale & Holographic coordinate\\ \hline
		Source of the operator & Leading mode of the field at the boundary\\ \hline
		VEV of the operator & Subleading mode of the field at the boundary\\ \hline
		Spin of the operator & Spin of the field\\ \hline
		Scaling dimension of the operator & Mass of the field\\ \hline	
		Temperature & Temperature of the black hole  \\ \hline
		Conserved current $\mathcal J^a$ & Gauge field $A_\mu$  \\ \hline
		Stress-energy tensor $\mathcal T^{ab}$ & Metric $g_{\mu\nu}$  \\
		\hline
	\end{tabular}
\end{center}
\caption{An overview of the Holographic dictionary.}
\end{table}

\begin{table}[H]
			\label{SD}
\begin{center}
	\begin{tabular}{| l  c |}
		 \hline
		\textbf{Spin} & \textbf{Scaling Dimension}  \\ \hline & \\[-1em]
		spin-0 & $(d\pm \sqrt{d^2+4m^2L^2})/2$  \\
		spin-1/2 & $(d+2\abs{m}L)/2$\\
		p-form & $(d\pm \sqrt{(d-2p)^2+4m^2L^2})/2$\\
		3/2 & $(d+2\abs{m}L)/2$\\
		spin-2 massless  & $d$\\
		\hline
	\end{tabular}
\end{center}
\caption{The bulk mass $m$ of a field in AdS$_{d+1}$ of radius $L$ is directly related to the scaling dimension of the corresponding boundary operator. This table was copied from \cite{st}, section 13.5.}
\end{table}

\subsection{Linear Response and Charge Transport Coefficient\label{LRCT}}

Consider a QFT with action $S$ and let $\mathcal{O}$ be one of its operators.
In linear response theory, we add an external source $\phi_0$ coupled to $\mathcal{O}$, by adding the term $\delta S=\int \phi_0\mathcal{O}$ to $S$, and study the response of the system to linear order in $\phi_0$. Define
\be \delta\braket{\mathcal{O}}=\braket{\mathcal{O}}_s-\braket{\mathcal{O}},\label{lr1}\ee
where $\braket{\phantom{O}}_s$ denotes an expectation value in the presence of the source $\phi_0$.
In Fourier space, at the linear level, we have
\be \delta\braket{\mathcal O(k)}=G_R^{\mathcal O\mathcal O}(k)\phi_0(k)\label{gf0}\ee
where $G_R^{\mathcal O\mathcal O}$ is the retarded Green's function
\be G_R^{\mathcal O\mathcal O}(k)=i\int \ud^d x e^{-ikx}\theta(t)\braket{[\mathcal{O}(x),\mathcal{O}(0)]}.\label{lr2}\ee
The Green's functions can be obtained from Holography, as explained in section \ref{2pf}.

The \textit{transport coefficients} are related to Green's functions by Kubo formulas. An example of a transport coefficient is the conductivity $\sigma$, defined by Ohm's law
\be \delta\braket{\mathcal J}=\sigma\mathcal E,\label{Ohm0}\ee where $\mathcal E$ is the external electric field.
 We set the spatial dimensions to $2$, ($x,y$), for simplicity. For a vector source
\be \delta S=\int \mathcal A_i^{(0)}\mathcal J^i\label{lr3}\sp i=x,y\ee equation (\ref{gf0}) becomes
\be \delta\braket{\mathcal J^i(k)}=G^{ij}_R(k) \mathcal A_j^{(0)}(k)\sp i,j=x,y,\label{r0}\ee
where
\be G^{ij}_R(k)=i\int \ud^d x e^{-ikx}\theta(t)\braket{[\mathcal J^i(x),\mathcal J^j(0)]}\sp i,j=x,y.\label{lr4}\ee
In the gauge $\mathcal A_t^{(0)}=0$, the external electric field in Fourier space is
\be \mathcal E_i=i\omega \mathcal A_i^{(0)}\sp i=x,y.\label{lr5}\ee
 Using (\ref{Ohm0}) and (\ref{r0}) we obtain the Kubo formula for the AC conductivity
\be \sigma_{ij}(\omega)={G^{ij}_R(\omega,\vec{k}=0)\over i\omega}\sp i,j=x,y\label{kubo}.\ee
In a rotationally symmetric system we have
\be \sigma_{xx}=\sigma_{yy}\sp \sigma_{yx}=-\sigma_{xy}.\label{lr6}\ee
The DC conductivity can be calculated by the zero-frequency limit
\be \sigma_{ij}^{DC}=\lim\limits_{\omega\to 0}\sigma_{ij}(\omega).\label{lr7}\ee

\subsubsection{Charge Transport in the Drude Model}

The Drude model is a phenomenological model of charge transport based on Newtonian physics. This simple picture describes many metals and other condensed matter systems, \cite{Zaanen:2018edk}.

Consider a particle of mass $m$ and charge $q$ moving in a solid, under the influence of a homogeneous electric field $\vec E$. In the Drude model, the interactions with the solid is contained in a friction term linear in the particle's velocity $\vec v$. Newton's equation reads
\be q\vec E=m {d\vec v\over dt}+{m\over \tau}\vec v,\label{neq0}\ee
where $\tau$ is the \textit{relaxation time}. It can be thought of as the average time between two consecutive collisions of the particle with something in the solid.

If there is a density $n$ of charge carriers, the current propagating in the solid is
\be \vec j=nq\vec v\label{lr8}.\ee
Solving for the velocity and substituting into (\ref{neq0}) we obtain an equation for the current
\be {nq^2\over m}\tau\vec{E}=\tau{d\vec j\over dt}+\vec j.\label{neq00}\ee
If the electric field $\vec E$ is constant in time, the steady-state current ($d\vec j/dt=0$) is obtained from the above
\be \vec j={nq^2\over m}\tau\vec{E},\label{lr9}\ee
hence the DC conductivity is
\be \sigma_{DC}={nq^2\over m}\tau.\label{lr10}\ee

Consider now an alternating electric field $\vec E=\vec E_0 e^{-i\omega t}$. At the linear response level, the current oscillates with the same frequency $\vec j=\vec j_0 e^{-i\omega t}$ and (\ref{neq00}) reads
\be \sigma_{DC}\vec E=(1-i\omega\tau)\vec j,\label{lr11}\ee
hence the AC conductivity is
\be \sigma(\omega)={\sigma_{DC}\over 1-i\omega\tau}.\label{drude}\ee
The real and imaginary parts are
\be Re(\sigma)=\sigma_{DC}{1\over 1+\omega^2\tau^2}\sp  Im(\sigma)=\sigma_{DC}{\omega\over 1+\omega^2\tau^2}.\label{lr13}\ee
At $\omega=0$ there is a peak in the real part, known as the Drude peak, while the imaginary part vanishes.

\newpage

\section{The Holographic Setup and Preliminaries\label{setup}}

In this section we introduce our Holographic model and describe the general form of the background solutions which we will study. In particular, we will consider two-dimensional, rotationally symmetric systems at finite charge density and temperature, in the presence of a magnetic field.

In order to build the holographic model, we first consider the most important operators in the field theory that we want to study. Any field theory contains a stress energy tensor $\mathcal T_{ab}$, therefore the bulk must contain the graviton $g_{\mu\nu}$, which is the minimum ingredient of any Holographic theory. We are interested in a field theory that contains a conserved current $\mathcal J^{a}$, which in our case is the electromagnetic current. By the Holographic dictionary, we need a gauge field in the bulk theory.
We also include a scalar field in the bulk action, dual to a scalar operator $\mathcal O$.
 \be
 \begin{split}
 	\text{Stress-Energy tensor}\; \mathcal T_{ab} &\leftrightarrow g_{\mu\nu}\; \text{Metric}\\
 	\text{Conserved Current}\; \mathcal J_{a} &\leftrightarrow A_{\mu}\; \text{Gauge field}\\
 	\text{Scalar operator}\; \mathcal{O} &\leftrightarrow \phi \; \text{Scalar field}.
 \end{split} \label{aa1}
 \ee

A strongly coupled theory at the large-$N$ limit can be approximated by a two-derivative bulk action.
The two-derivative action containing the above fields is the Einstein-Maxwell-Dilaton (EMD) action:
\be S_{EMD}=M^{2}\int \ud^{4}x \sqrt{-g}\left(R-\frac12(\partial \phi)^2+V(\phi)-{1\over 4}Z(\phi) F^2\right).\label{aa2}\ee
We set the bulk dimensions to $3+1$, since we are interested in $2+1$-dimensional quantum field theories. The bulk fields are dimensionless and $M$ is a characteristic mass scale of the theory.

At the two derivative level we can also add the Peccei-Quinn term, $S_{PQ}$, which is important for systems with intrinsic CP-violation:
\be S_{PQ}=-{1\over 4}M^{2}\int \ud^{4}x\sqrt{-g} W(\phi)F\wedge F\label{PQ}\ee
where
\be F\wedge F={1\over 2\sqrt{-g}}F_{\m\n}\e^{\m\n\r\s}F_{\r\s}.\label{aa3}\ee
Since, by the CPT theorem, this term also violates T-symmetry,  we shall refer to it as T-violating.

We will, therefore, consider the EMD-PQ action in $3+1$ dimensions:
\be \label{action}S=S_{EMD}+S_{PQ}.\ee

The equations of motion are obtained by extremizing (\ref{action}) with respect to the three dynamical fields $g_{\mu\nu},A_\mu,\phi$. The equations corresponding to the metric, gauge field and scalar variations are respectively:
\begin{subequations}
	\begin{align}
	&	R_{\m\n}= -{V(\phi)\over 2}g_{\mu\nu}+{1\over 2}\partial_\mu\phi\partial_\nu\phi+{Z(\phi)\over
		2}\le[F^{\;\rho}_\mu\ F_{\nu\rho}-\frac{g_{\mu\nu}}4 F^2\ri]\label{EE}\\
	&\nabla_\mu\le(Z(\phi)F^{\mu\nu}+{W(\phi)} ^\star F^{\m\n}\ri)= 0\label{GE}\\
	& \square{\phi}+{\partial_\phi}\left(V(\phi)-\frac{Z(\phi)}4F^2-\frac{W(\phi)}{4}F\wedge F\right)= 0,\label{SE}
	\end{align}
\end{subequations}
where
\be {}^\star F^{\mu\nu}={1\over 2\sqrt{-g}}\epsilon^{\mu\nu\rho\sigma}F_{\rho\sigma}.\label{aa4}\ee

We want to calculate the charge transport coefficient of the dual quantum field theory by adding an interaction of the form
\be \int d^3x \mathcal A_a^{(0)}\mathcal J^{a}\label{int}.\ee
The external source $\mathcal A_a^{(0)}$ is the boundary value of the dynamical field $A_\mu$ propagating in the higher-dimensional bulk, with the boundary condition at the AdS boundary:
\be MA_\mu|_{r\to boundary}=\mathcal A_\mu^{(0)}.\label{BC}\ee
We include the mass scale $M$ in the above, since the gauge field on the field theory side has mass dimension $1$ and the bulk fields are dimensionless.

\subsection{Ansatz and Equations of Motion}

We will now motivate our ansatz for the bulk fields and derive the equations of motion.

We are interested in a theory at finite charge density $\mathcal J_t$, hence we need the bulk field $A_t$ to have a finite value at the boundary. Its leading term is proportional to the chemical potential of the boundary theory, which sources the charge density. The charge density appears in coefficient of the subleading mode of $A_t$ near the boundary, as we will see below, because it is the vacuum expectation value of $\mathcal J_t$.

In addition, we turn on a constant magnetic field, $h$, along the holographic coordinate $r$, which is equivalent to turning on a time-independent, homogeneous magnetic field on the two-dimensional boundary
\be A_\m=(A_t(r),0,0,hx).\label{aa5}\ee
A magnetic field $\mathcal B$ has mass dimension $2$, hence it is related to $h$ by
\be \mathcal B=Mh.\label{aa6}\ee

 In a two-dimensional world, a magnetic field field does not introduce any anisotropies, since it is a scalar\footnote{By "scalar" it is meant that it is rotation invariant.}. Therefore we can consider a rotationally symmetric space ansatz.
 We use the following ansatz for the metric and scalar:
 \be ds^2=-D(r)dt^2+B(r)dr^2+C(r)(dx^2+dy^2)\sp \phi=\phi(r).\label{aa7}\ee
 The coordinates $t,x,y$ are identified with the coordinates of the field theory on the boundary, while $r$ is the Holographic coordinate. The fields $D,B,C$ are not completely determined by the equations; there is a residual gauge symmetry related to the freedom of reparametrization of the radial coordinate.

To sum up, the ansatz is
\be ds^2=-D(r)dt^2+B(r)dr^2+C(r)(dx^2+dy^2),\;\; \phi=\phi(r), \;\; A_\m=(A_t(r),0,0,hx).\label{ans}\ee
Substituting (\ref{ans}) into (\ref{EE}), (\ref{SE}), we obtain the independent set of equations
\begin{subequations}\label{eom01}
	\begin{align}
	& BV+h^2Z{B\over 2C^2}+Z{A_t'^2\over 2D}+{B'D'\over 2BD}-{C'D'\over CD}+{D'^2\over 2D^2}-{D''\over D}=0\label{aa8}\\
	& \phi'^2+2{C''\over C}-{C'\over C}\left({C'\over C}+{B'\over B}+{D'\over D}\right)=0\label{aa09}\\
	& -BV+h^2Z{B\over 2C^2}+{ZA_t'^2\over 2D}-{B'C'\over 2BC}+{C'D'\over 2 CD}+{C''\over C}=0\label{aa010}\\
	& \phi''+\log(C\sqrt{D\over B})'\phi'+BV_\phi-W_\phi\sqrt{B\over D}{hA_t'\over C}+Z_\phi\left({A_t'^2\over 2D}-{Bh^2\over 2C^2}\right)=0\label{deq01}
	\end{align}
\end{subequations}
where $V_{\phi}$ indicates a derivative with respect to $\phi$.

Using (\ref{ans}), the gauge field equation (\ref{GE}) can be integrated once to obtain
\be \partial_r A_t(r)=(q-hW(\phi)){\sqrt{D(r)B(r)}\over Z(\phi)C(r)}\label{gfs}\ee
where $q$ is the integration constant, and is related to the boundary charge density
as we will see below.
Since $q$ always appears in the equations shifted by $hW$, it is convenient to define
\be Q(\phi)=q-hW(\phi).\label{newequation}\ee
We can substitute (\ref{gfs}), (\ref{newequation}) into (\ref{eom01}) in order to obtain the following set of independent equations
\begin{subequations}\label{eom1}
	\begin{align}
	&\partial_r A_t(r)=Q{\sqrt{D(r)B(r)}\over Z(\phi)C(r)}\\
	& BV+h^2Z{B\over 2C^2}+{Q^2B\over 2ZC^2}+{B'D'\over 2BD}-{C'D'\over CD}+{D'^2\over 2D^2}-{D''\over D}=0\label{aa19}\\
	& \phi'^2+2{C''\over C}-{C'\over C}\left({C'\over C}+{B'\over B}+{D'\over D}\right)=0\label{aa9}\\
	& -BV+h^2Z{B\over 2C^2}+{Q^2B\over 2ZC^2}-{B'C'\over 2BC}+{C'D'\over 2 CD}+{C''\over C}=0\label{aa10}\\
	& \phi''+\log(C\sqrt{D\over B})'\phi'+BV_\phi-W_\phi{hQB\over C^2Z}+Z_\phi\left({Q^2B\over 2C^2Z^2}-{Bh^2\over 2C^2}\right)=0\label{deq1}
	\end{align}
\end{subequations}

\subsection{Asymptotically AdS Black Hole Backgrounds\label{aads}}

In this subsection we specialize the ansatz (\ref{ans}) to asymptotically-AdS black holes, which are dual to theories at finite temperature. We also study the near-horizon and near-boundary behavior of the bulk fields.

Since the ansatz (\ref{ans}) has a residual gauge freedom, for definiteness we choose
\be S\equiv D=B^{-1}.\label{aa11}\ee
In Holography, a field theory at finite temperature is dual to a black hole (see section \ref{ft0}). This implies that $S(r)$ vanishes at some finite value of $r$. The simplest asymptotically AdS charged black hole solution to the equations of motion (\ref{eom1}) is the AdS-Reissner-Nordstrom black hole (see appendix \ref{rnapp}).
In this coordinate system, the location of the outer horizon $r_h$ is the largest, real, single root of the function $S(r)$ and the AdS boundary is located at $r\to\infty$.

The near-boundary metric is:
\be ds^2\simeq{r^2\over L^2}(-dt^2+dx_idx^i)+{L^2dr^2\over r^2}\sp r\to\infty,\label{aAdS}\ee
hence the asymptotic behavior of the fields is, using (\ref{eom1}),
\be \begin{split}&S(r)=r^2/L^2+\cdots\sp C(r)=r^2/L^2+\cdots\\
	&A_t(r)=\mu-L^2{q-hW_\infty\over Z_\infty }r^{-1}+\cdots\sp \phi(r)={c\over r^{3-\Delta}}+\cdots.\end{split}\label{aAdSf}\ee
 The constant $\Delta$ depends on the mass of the dilaton (which, in turn, depends on the second derivatives of $V,Z,W$). We defined $Z_\infty,W_\infty$ as
 \be Z_\infty=\lim_{r\to\infty}Z(\phi(r))\sp W_\infty=\lim_{r\to\infty}W(\phi(r)).\ee
 If the scalar is constant, $\phi=\phi_0$, then
  \be Z_\infty=Z(\phi_0)\sp W_\infty=W(\phi_0).\ee
 The constant $\mu$ is the chemical potential sourcing the charge density $q$. Because of the gauge invariance, $q$ is not fixed in terms of $\mu$ by the equations. The relationship between them is found by requiring that the $1$-form, $A$, is well-defined at the horizon, which implies $A_t(r_h)=0$, \cite{Koba}.

The temperature and entropy of the field theory are related to the Hawking temperature and entropy of the black hole. To calculate the temperature we consider the near horizon limit of the metric ($r=r_h+\epsilon$) in Euclidean time ($t\to i\tau$)
\be ds^2=S'(r_h)\epsilon d\tau^2+{d\epsilon^2\over S'(r_h)\epsilon}+C(r_h)dx_idx^i.\label{aa12}\ee
Redefining the radial coordinate \be {d\epsilon^2\over S'(r_h)\epsilon}=d\sigma^2
\label{aa13}\ee we obtain
\be ds^2=\sigma^2  {(S'(r_h))^2d\tau^2\over 4}+d\sigma^2+C(r_h)dx_idx^i.\label{aa14}\ee
In order to eliminate the conical singularity we need to identify $\tau\sim \tau+{4\pi\over S'(r_h)}$. The temperature is equal to the inverse period
\be T={S'(r_h)\over 4\pi}\label{temp}.\ee
The entropy density is equal to the area of the horizon
\be s=4\pi C(r_h).\label{ent0}\ee
The near-horizon expansion of the fields is
\be \begin{split} &S(r)\simeq 4\pi T(r-r_h)+\mathcal{O}((r-r_h)^2)\sp C(r)\simeq {s\over 4\pi}+\mathcal{O}((r-r_h)^1)\\
	& A_t(r)\simeq a_0(r-r_h)+\mathcal{O}((r-r_h)^2) \sp \phi(r)\simeq\phi_0+\mathcal{O}((r-r_h)^1).\label{aa15}
\end{split}\ee

  In the following sections we will consider perturbations around a background containing a black hole. The horizon is a singular point of the equations. One set of boundary conditions for the perturbations stems from the requirement that they are regular at the horizon. The near-boundary (\ref{aAdSf}) and near-horizon (\ref{aa15}) expansions will be important in order to set boundary conditions.

\subsection{Charge Density and Current}

We will briefly explain how the background (\ref{ans}) realizes a quantum field theory at finite charge density and describe how an electric current can be created by perturbing the bulk gauge field. We derive a formula for the boundary current $\mathcal J^a$ in terms of the bulk fields, which will be useful in the subsequent sections.

Starting from the action (\ref{action}), we can derive an expression for the boundary current density $\mathcal J^a$ in terms of the bulk fields. Consider an arbitrary fluctuation\footnote{Perturbing $A_\mu$, in general, will backreact on the bulk geometry. We need to turn on additional elements of the metric to the same order as $A_x,A_y$. This will be done in detail in the following sections.} of the components $A_x,A_y$, around the background (\ref{ans}).

 We vary (\ref{action}) with respect to the gauge field and, on-shell, only the boundary term survives
\be \delta S=-M^{2}\int \ud^{3}x  \sqrt{-g} \delta A_{\nu} \left(Z(\phi) F^{r\nu}+ W(\phi) {}^\star F^{r\nu}\right)\label{osai}\bigg|_{r\to\infty},\ee
where we used the fact that the radial coordinate is perpendicular to the boundary and the horizon.
The contribution from the horizon is discarded, as explained in section \ref{2pf}.
Keeping in mind the boundary condition (\ref{BC}), we find the boundary current:
\be \mathcal J^{a}={\delta S\over \delta \mathcal A_a^{(0)}}=-M\sqrt{-g}\left(Z(\phi) F^{r a}+ W(\phi) {}^\star F^{r a}\right)|_{r\to\infty}\sp a=t,x,y\label{curr00}\ee
From the $r$-component of (\ref{GE}) we obtain
\be \partial_a\left(\sqrt{-g}\left(Z(\phi) F^{r a}+ W(\phi) {}^\star F^{r a}\right)\right)+\partial_r(\left(\sqrt{-g}\left(Z(\phi) F^{r r}+ W(\phi) {}^\star F^{r r}\right)\right)=0\sp a=t,x,y.\ee
Note that the last term vanishes by antisymmetry of $F^{\mu\nu}$.
Taking the limit $r\to\infty$ implies that the current, (\ref{curr00}), is conserved
\be \partial_a \mathcal J^a=0.\label{aa16}\ee
Using (\ref{ans}) and (\ref{gfs}) we can confirm that the charge density is
\be \mathcal J_t=M q.\label{aa17}\ee
We define the following ``bulk current"
\be J^a(r)=-\sqrt{-g}\left(Z(\phi) F^{ra}+ W(\phi) {}^\star F^{ra}\right),\label{currents0}\ee
which will prove useful in the following sections. The boundary current is related to the above by
\be \mathcal J^a=M J^a(r\to \infty).\label{aa18}\ee

\section{DC Conductivity in Systems with Translational Symmetry\label{hcond}}

In this section we calculate the DC conductivity in the class of background solutions described in the previous section (\ref{ans}), in the gauge (\ref{aa11}). For clarity we present again the general form of the background
\be ds^2=-S(r)dt^2+{dr^2\over S(r)}+C(r)(dx^2+dy^2),\;\; \phi=\phi(r), \;\; A_\m=(A_t(r),0,0,hx).\label{bg1}\ee

In order to calculate the conductivity, we perturb the boundary theory by a term
\be  S_s=\int \ud^3 x  \left(\mathcal A_x^{(0)}(t) \mathcal J^x+\mathcal A_y^{(0)}(t) \mathcal J^y\right)\label{sourceterm}\ee
and compute the response to linear order. We consider a source that does not depend on $x,y$ and we work in the gauge $\mathcal A_t^{(0)}=0$.
 On the bulk side, this corresponds to perturbing the bulk gauge field
\be \delta A_i(r,t)\sp i=x,y\label{bb1}\ee
such that
\be  \delta A_i|_{r\to\text{boundary}}= M^{-1}\mathcal A_i^{(0)}\sp i=x,y.\label{bb2}\ee
Perturbing $A_x,A_y$ around a background of the form (\ref{ans}), we are forced to turn on additional fields by the bulk equations. A consistent ansatz is the following:
\be  \delta A_i=a_i (r,t)\sp  \delta g_{ti}=C(r)z_i(r,t)\sp i=x,y.\label{pert1}\ee

In Fourier space, at the linear response level, the current is related to the source by \be \mathcal J_i(\omega)=G^R_{ij}(\omega) \mathcal A_j^{(0)}(\omega).\label{bb4}\ee The retarded Green's function $G^R_{ij}$ can be calculated from Holography and the conductivity is obtained by the Kubo formula, (\ref{kubo}),
\be \sigma_{ij}(\omega)={G^R_{ij}(\omega)\over i\omega}.\label{bb5}\ee
We will use this method to compute the AC conductivity in section \ref{ACcond}.

The DC conductivity is obtained in the limit
\be \sigma_{DC}=\lim\limits_{\omega\to 0} \sigma(\omega).\label{bb6}\ee
The standard way of calculating the DC conductivity is to compute the Green's function from Holography and then take the limit shown above. A variety of techniques have been developed to obtain analytic expressions for $\sigma_{DC}$ even when the equations cannot be solved analytically for $\omega\ne 0$ (for example see \cite{hartnoll}, \cite{mg4}). However, as described in \cite{donos}, \cite{blake}, for the purpose of extracting just the DC conductivity we can turn on a constant electric field from the start. In this case we need to turn on additional components of the metric, in order to obtain consistent fluctuation equations.

\subsection{Constant Electric Field Ansatz and Fluctuation Equations}

In this subsection we motivate the perturbation ansatz and derive the linearized fluctuation equations.
Following  \cite{blake}, we turn on a constant electric field on the boundary
\be \mathcal A_i^{(0)}=-\mathcal E_i t\sp i=x,y\label{bb7}\ee
by perturbing the bulk field
\be \delta A_i=-E_i t\sp i=x,y,\label{bb8}\ee
where the bulk and boundary electric fields are related by
\be E_i=\mathcal E_i/M\sp i=x,y.\label{ef}\ee
 By the Holographic dictionary, the currents will appear in the subleading mode of $\delta A_i$ near the boundary. Hence we need an $r$-dependent component in the gauge field
\be \delta A_i(r,t)=-E_it+a_i(r)\label{bb9}.\ee
To obtain consistent equations we also need to turn on $g_{ri},g_{ti}$. We will see this necessity soon by examining the behavior of the equations near the horizon.
A consistent ansatz is the following:
\be  \delta A_i=-E_it+a_i (r)\sp
\delta g_{ti}=C(r)z_i(r)\sp g_{ri}=C(r)g_i(r)\sp i=x,y\label{pert}.\ee

We need to express the current $\vec{\mathcal J}=(\mathcal J_x,\mathcal J_y)$ in terms of the electric field $\vec{\mathcal E}=(\mathcal E_x,\mathcal E_y)$. Then the conductivity can by calculated by Ohm's law:
\be\begin{pmatrix}
	\mathcal J_x\\ \mathcal J_y
\end{pmatrix}=\begin{pmatrix}
	\sigma_{xx}&\sigma_{xy}\\ \sigma_{yx}&\sigma_{yy}
\end{pmatrix} \begin{pmatrix}
	\mathcal E_x\\ \mathcal E_y
\end{pmatrix}.\label{Ohm}\ee
where $\sigma$ is the conductivity matrix. In a rotationally symmetric system we have
\be \sigma_{xx}=\sigma_{yy}\sp \sigma_{xy}=-\sigma_{yx}.\label{bb10}\ee
In terms of the bulk fields (\ref{currents0}), (\ref{ef}), equation (\ref{Ohm}) reads
\be J_i|_{r\to\infty}=\sigma_{ij} E_j\sp i,j=x,y.\label{Ohm2}\ee

From (\ref{pert}) it is clear that the Maxwell tensor $F_{\mu\nu}$ depends only on the radial coordinate, $r$. A quick glance at (\ref{GE}) and (\ref{currents0}) shows that the bulk current is radially conserved.	
Linearizing in the perturbation amplitudes we find
 \be J_i=-ZS a_i'-Q z_i-\epsilon^{ij}hZSg_j+\epsilon^{ij}E_jW\sp \partial_r J_i=0\sp i,j=x,y. \label{curr1}\ee
From the Einstein equation (\ref{EE}) we obtain:
\begin{subequations}
	\begin{align}
	&(Cz_i)''-\left(C''+{h^2Z\over CS}\right)z_i+{Q\over C}a_i'-h{Z\over S C}\epsilon^{ij}E_j+\epsilon^{ij}{hQ\over C}g_j=0
	\label{EE2}\\
	&{hQ}z_i+hSZa_i'+\epsilon^{ij}{Q}E_j+\epsilon^{ij}h^2SZg_j=0\sp i,j=x,y
	\label{EE1}.
	\end{align}
\end{subequations}
We used the totally antisymmetric symbol $\epsilon^{ij}$ with $\epsilon^{xy}=1$, in order to write the equations in a more compact form.

An important observation is that the bulk current (\ref{curr1}) is radially conserved, hence it can be evaluated at any value of $r$. In particular, its boundary value (which is related to the current of the dual theory) is equal to its horizon value. The latter is straightforward to calculate by demanding that the fields are regular at the horizon. This will be important in section \ref{condax}, in which the equations are not as simple as in the present case.

\subsection{Boundary Conditions}

In this subsection we study the consistency of the fluctuation equations. In particular we examine the near-horizon and near-boundary behavior of the fields. The insights from this section will be helpful when we tackle the case of broken translational symmetry in section \ref{condax}, in which more fields participate in the fluctuations equations.

 \subsubsection{Near the Boundary}

 First we check the structure of the equations near the boundary $r\to\infty$, keeping in mind the asymptotic behaviors of the background fields (\ref{aAdSf}). In order to calculate the charge transport coefficient, we need the transport current. This means we are interested in the current that is sourced purely by the electric field and not by other fields, such as the metric. Hence we require that $z_i,g_i$ fall off sufficiently fast as $r\to\infty$, in order not to have additional sources that contribute to the boundary current, (\ref{curr1}).
 From (\ref{curr1}) we can see that $g_i$ does not contribute if
 \be g_i\simeq \mathcal{O}(r^{-3})\label{lbg}.\ee
 From (\ref{EE2}) we find that $z_i$ has two independent solutions near the boundary
 \be z_i=c_i+b_i r^{-3}+\cdots. \label{lbz}\ee

 The leading mode is a heat source, \cite{Donos:2014cya}, for the boundary theory and we must require that $c_i=0$, so that there is no additional contribution to $J_i$ from thermoelectric effects.

From (\ref{curr1}) we find that $a_i$ has two independent solutions that behave as $r^0$ and $r^{-1}$. We can gauge away the former and the latter is proportional to the current, offset by the PQ term (\ref{PQ})
 \be a_i=L^2Z_{\infty}^{-1}(J_i-W_\infty \epsilon^{ij}E_j)r^{-1}+\cdots.\label{lba}\ee
 In addition, from (\ref{pert}), the gauge field contributes a constant electric field at the boundary, which is the relevant deformation sourcing the current.

 The final check is the consistency of the constraint equations (\ref{EE1}). It is straightforward to confirm that the leading behaviors (\ref{lbg}), (\ref{lbz}) and (\ref{lba}) are consistent with (\ref{EE1}).

 \subsubsection{Near the Horizon}

 Near the horizon, $r=r_h$, we require that the fields are regular.
 In order to impose the regularity condition we switch to Eddington-Finkelstein coordinates $(u,r)$, where
 \be du=dt+{dr\over S},\label{bc1}\ee
 or, near the horizon,
 \be u\simeq t+{1\over 4\pi T}\log (r-r_h)+\mathcal{O}((r-r_h)^0),\label{bc2}\ee
 where we used that $4\pi T=S'(r_h)$, (\ref{temp}).
 From (\ref{pert}), we have
 \be \delta A_i\simeq-E_iu+{E_i\over 4\pi T}\log(r-r_h)+a_i(r)+\mathcal{O}((r-r_h)^0).\label{bc3}\ee
 Regularity of $\delta A_i$ at the horizon implies
 \be a_i(r)\simeq-{E_i\over 4\pi T}\log(r-r_h)+\mathcal{O}((r-r_h)),\label{bc4}\ee
 so that the logarithms cancel out. The above condition can be written in the following form, which is more convenient for our calculations
\be a_i'(r)\simeq -{E_i\over S(r)}+\mathcal{O}((r-r_h)^0).\label{gfnh1}\ee
For the metric we require
\be z_i(r)\simeq z_i(r_h)+\mathcal{O}((r-r_h)^1).\label{mnh1}\ee
To impose the regularity condition on $g_i$, consider the following equation, which is a linear combination of (\ref{EE1}) and (\ref{EE2})
\be CSQ(Cz_i)''+(h^2Z^2+Q^2)(Sa_i'+hS\epsilon^{ij}g_j)=0.\label{bc5}\ee
If $g_i=0$, then the near-horizon limit of the above equation implies that $E_i=0$. This justifies the choice to turn on the $g_{ri}$ components of the metric in (\ref{pert}). Solving for $g_i$ and taking the near-horizon limit, keeping in mind (\ref{gfnh1}), (\ref{mnh1}), we obtain
\be
g_i\simeq -\epsilon^{ij}{E_j\over hS}+\mathcal{O}((r-r_h)^1)\sp i,j=x,y.
\label{ghor}\ee

Substituting these leading behaviors into (\ref{EE1}) and (\ref{curr1}) we obtain the following horizon values for $z_i$
\
\be z_i(r_h)=-\epsilon^{ij}{E_j\over h}.\label{bc6}\ee
We can use this information to write (\ref{ghor}) in a simpler form
\be g_i\simeq{z_i\over S}+\mathcal{O}((r-r_h)^1).\label{gnh1}\ee
Finally, the near-horizon behavior of (\ref{curr1}) implies
\be J_i=\epsilon^{ij}{q\over h}\sp i=x,y,\label{r1}\ee
which is in agreement with the solution (\ref{soln1}).

\subsection{The Conductivity\label{hconds}}
Solving (\ref{EE1}) for $z_i$ and substituting into (\ref{curr1}) we obtain
\be J_i=\epsilon^{ij}E_j\left({Q\over h}+W\right)=\epsilon^{ij}E_j{q\over h}.\label{soln1}\ee
Note that the contributions from the PQ term (\ref{PQ}) exactly cancel out.
Comparing with (\ref{Ohm2}), we can read:
\boxedeq{hcond1}{\begin{split}
		&\sigma_{xx}=\;\;\,\sigma_{yy}=0\\
		&\sigma_{xy}=-\sigma_{yx}={q\over h}
\end{split}}

The expression (\ref{hcond1}) has been previously derived in \cite{hartnoll}. As explained there, the DC conductivity in a Lorentz invariant theory with a magnetic field is constrained to the form (\ref{hcond1}). The argument is simple: suppose we have a system with charge density $q$ and magnetic field $h$. In a reference frame boosted by a small-magnitude velocity $-\vec v=-(v_x,v_y)$, there is a current $J_i=q v_i$ and an electric field
\be E_i=-h\epsilon^{ij}v_j=-{h\over q} \epsilon_{ij}J_j\sp i,j=x,y.\label{bc7}\ee
Inverting the above and using Ohm's law, we obtain precisely (\ref{hcond1}).

For $h=0$, going back to the fluctuation equations, we find $E_i=0$ from (\ref{EE1}). Taking the limit $r\to\infty$ in (\ref{curr1}) we find $J_i=0$. This indicates that our approach is not valid in this case. On grounds of Lorentz invariance, we would expect that the resistivity vanishes, since we can get arbitrarily large currents from boosting a system at finite charge density. Indeed in sections \ref{condax} and \ref{chnum} we will see that this is the case.

To conclude, the Peccei-Quinn term (\ref{PQ}) does not contribute to the conductivity at all in this case. This is expected, since it does not break any continuous symmetry.

\section{EMD with Momentum Relaxation\label{EMDax}}

In section \ref{hcond} we calculated the DC conductivity in a translationally invariant system. In the presence of a magnetic field, it is constrained to the form (\ref{hcond1}), as explained in \cite{hartnoll}. In the absence of a magnetic field, the DC conductivity diverges. This happens because translational symmetry implies momentum conservation, therefore there is no current dissipation. In order to calculate finite DC conductivity we must break the translational symmetry.

 In order to break the translational symmetry, the bulk fields must depend on the coordinates $x,y$. Then, in general, we will  obtain a system of partial differential equations (PDEs) which is hard to solve. There are several ways to introduce momentum dissipation without turning the equations into PDEs.  For example in \cite{mg4}, \cite{zhou} it is done in the context of massive gravity. A simpler way to achieve this is by adding two scalar fields $\chi_x,\chi_y$
 to the action \cite{donos}, \cite{blake}. These fields should have no potential so that their action is invariant under a constant shift
 \be \chi_i\to \chi_i+c_i\sp i=x,y\sp c_i=\text{constant}.\ee
Such scalars are usually called ``axions".
Therefore we add the following axion terms to the action (\ref{action})
\be S_{ax}=-{1\over 2}M^{2}\int \ud^{4}x \sqrt{-g}~\Psi(\phi)\left((\partial \chi_x)^2+(\partial \chi_y)^2)\right).\label{cc1} \ee
We take the coupling $\Psi$ of the axions to depend on the dilaton. A solution of the form
\be
\chi_x=kx\sp \chi_y=ky
 \ee
 breaks the translation symmetry of the boundary theory in a very simple way, while keeping the rotational symmetry intact. The constant $k$ is the \textit{momentum relaxation scale}. For $k=0$ we recover the translational symmetry.

The theory is now governed by the action:
\be \label{actionax} S=M^{2}\int \ud^{4}x \sqrt{-g}\left(R-\frac12(\partial \phi)^2-{1\over 2}\Psi(\phi)\sum_{i=x,y}(\partial \chi_i)^2+V(\phi)-{1\over 4}Z(\phi) F^2-{1\over 4} W(\phi)F\wedge F\right).\ee
Varying (\ref{actionax}) with respect to the metric, the gauge field, the dilaton and the axions we obtain respectively:
\begin{subequations}\label{eom001}
	\begin{align}
	&R_{\m\n}= -{V(\phi)\over 2}g_{\mu\nu}+{1\over 2}\partial_\mu\phi\partial_\nu\phi+{\Psi(\phi)\over 2}\left(\partial_\mu\chi_x\partial_\nu\chi_x+\partial_\mu\chi_y\partial_\nu\chi_y\right)+{Z(\phi)\over
		2}\le[F^{\;\rho}_\mu\ F_{\nu\rho}-\frac{g_{\mu\nu}}4 F^2\ri]\label{EEQ}\\
	&\nabla_\mu\le(Z(\phi)F^{\mu\nu}+{W(\phi)} ^\star F^{\m\n}\ri)= 0\label{GFEQ}\\
	& \square{\phi}+{\partial_\phi}\left(V(\phi)-\frac{Z(\phi)}4F^2-\frac{W(\phi)}{4}F\wedge F-{1\over 2}\Psi(\phi)((\partial \chi_x)^2+(\partial \chi_y)^2)\right)= 0	 \label{DEQ}\\
	&\nabla_\mu \left(\Psi(\phi)\nabla^\mu \chi_i\right)=0,\;\;\; i=x,y\label{AEQ}
	\end{align}
\end{subequations}

We consider the following class of solutions as the background:
\be ds^2=-D(r)dt^2+B(r)dr^2+C(r)dx_idx^i,\;\; \phi=\phi(r),\;\; \chi_i=kx_i, \;\; A_\m=(A_t(r),0,0,hx)\label{ans2}\ee
where $i=x,y$ above. The axion parameter $k$ is constant (the momentum relaxation scale).

Using the ansatz (\ref{ans2}), the equations of motion, (\ref{eom001}), read
\begin{subequations}\label{eom21}
	\begin{align}
		& BV+h^2Z{B\over 2C^2}+Z{A_t'^2\over 2D}+{B'D'\over 2BD}-{C'D'\over CD}+{D'^2\over 2D^2}-{D''\over D}=0\label{EEQ11}\\
		& \phi'^2+2{C''\over C}-{C'\over C}\left({C'\over C}+{B'\over B}+{D'\over D}\right)=0\label{EEQ21}\\
		& k^2{B\Psi\over C}-BV+h^2Z{B\over 2C^2}+{ZA_t'^2\over 2D}-{B'C'\over 2BC}+{C'D'\over 2 CD}+{C''\over C}=0\label{EEQ31}\\
		& A_t'={Q\sqrt{BD}\over CZ}\label{GFEQ11}\\
		& \phi''+\log(C\sqrt{D\over B})'\phi'+BV_\phi+W_\phi{\sqrt{B\over D}}{hA_t'\over C}+Z_\phi\left({A_t'^2\over 2D}-h^2{B\over 2C^2}\right)-\Psi_\phi {k^2B\over C}=0\label{DEQ11}
	\end{align}
\end{subequations}
where $Q=q-hW$ and $V_\phi$ indicates a derivative with respect to $\phi$. The axion equations are trivially satisfied. We can substitute $A_t'$ from (\ref{GFEQ11}) into the rest of the equations to obtain
\begin{subequations}\label{eom2}
	\begin{align}
	& BV+h^2Z{B\over 2C^2}+{Q^2B\over 2C^2Z}+{B'D'\over 2BD}-{C'D'\over CD}+{D'^2\over 2D^2}-{D''\over D}=0\label{EEQ1}\\
	& \phi'^2+2{C''\over C}-{C'\over C}\left({C'\over C}+{B'\over B}+{D'\over D}\right)=0\label{EEQ2}\\
	& k^2{B\Psi\over C}-BV+h^2Z{B\over 2C^2}+{Q^2B\over 2C^2Z}-{B'C'\over 2BC}+{C'D'\over 2 CD}+{C''\over C}=0\label{EEQ3}\\
	& A_t'={Q\sqrt{BD}\over CZ}\label{GFEQ1}\\
	& \phi''+\log(C\sqrt{D\over B})'\phi'+BV_\phi+W_\phi{hBQ\over ZC^2}+Z_\phi\left({Q^2B\over 2C^2Z^2}-h^2{B\over 2C^2}\right)-\Psi_\phi {k^2B\over C}=0\label{DEQ1}
	\end{align}
\end{subequations}

\section{DC Conductivity in Systems with Broken Translational Symmetry\label{condax}}

In this section we calculate the DC conductivity in the background \ref{ans2}, using the same techniques as in section \ref{hcond}. We expect the DC conductivity to be finite, since the translational symmetry is broken by the background axion fields, as described in section \ref{EMDax}.

We use the gauge $D=B^{-1}\equiv S$ (see section \ref{aads}). For clarity we present the general form of the background
\be ds^2=-S(r)dt^2+{dr^2\over S(r)}+C(r)(dx^2+dy^2),\;\; \phi=\phi(r), \;\; A_\m=(A_t(r),0,0,hx)\sp \chi_i=kx_i.\label{bg2}\ee

As in section \ref{hcond}, we perturb the system by a small, constant electric field $(E_x,E_y)$ and calculate the currents in the $x-y$ plane. This time we also perturb the axions. A consistent ansatz is the following:
\be \begin{split}& \delta a_i(r,t)=-E_it+a_i (r)\sp \delta \chi_i(r,t)=k_i(r)\\
	& \delta g_{ti}(r,t)=C(r)z_i(r)\sp \delta g_{ri}(r,t)=C(r)g_i(r)\sp i=x,y\label{pert2}\end{split}\ee
 The expression for the bulk current is the same as in (\ref{curr1})
\be J_i=-ZS a_i'-Q z_i-\epsilon^{ij}hZSg_j+\epsilon^{ij}E_jW\sp i=x,y. \ee
The independent set of equations is:
\begin{subequations}
	\begin{align}
	&J_i=-ZS a_i'-Q z_i-\epsilon^{ij}hZSg_j+\epsilon^{ij}E_jW\sp~~~~ J_i'=0\label{curr2}\\
	&CS(Cz_i)''-\left(C''CS+h^2Z+k^2C{\Psi}\right) z_i+{QS} a_i'-\epsilon^{ij}h{Z}E_j+\epsilon^{ij}h{QS} g_j=0\label{Eti}\\
	&\left(k^2C\Psi +h^2{Z}\right)Sg_i-\epsilon^{ij}{hQ}z_j-\epsilon^{ij}{hSZ}a_j'-kC\Psi Sk_i'+{E_iQ}=0,\label{Eri}\\
	&\left(C\Psi S(k_i'-kg_i)\right)'=0\label{Ax}
	\end{align}
\end{subequations}
where $i,j$ take the values $x,y$ and $\epsilon^{ij}$ is totally antisymmetric with $\epsilon^{xy}=1$. The axion fluctuation equations are implied by the equations above.

We must find the dependence of $J_x,J_y$ on the electric fields $E_x,E_y$ in order to compute the conductivity. As we will see in the following subsection, the conductivity can be calculated in terms of horizon data.

\subsection{Boundary Conditions}

 We briefly check the leading behavior of the equations near the boundary $r\to\infty$. As in section \ref{hcond} we require that $z_i,g_i$ fall off at least as $r^{-3}$, in order not to have additional contributions to the current (\ref{curr2}). This is consistent with equation (\ref{Eti}). Using the above with (\ref{curr2}) we find
\be a_i=L^2Z_{\infty}^{-1}(J_i-W_\infty \epsilon^{ij}E_j)r^{-1}+\cdots.\label{lba2}\ee
Finally (\ref{Eri}) implies that near the boundary
\be k_i'\simeq kg_i+\mathcal{O}(r^{-4}),\label{601}\ee
which is in agreement with (\ref{Ax}).

The regularity condition on $\delta A_i$ is obtained by switching to Eddington-Finklestein coordinates, as in section \ref{hcond}. For the scalars $k_i$ we require that they do not diverge at horizon. Adding the equations (\ref{Eti}), (\ref{Eri}) and taking the near-horizon limit, we find that $Sg_i-z_i$ needs to vanish at the horizon, which is consistent with (\ref{Ax}). This is the same condition as (\ref{gnh1}) in section \ref{hcond}.
To sum up, the regularity conditions can be written as
\begin{subequations}\label{reg}
	\begin{align} & a_i'(r)\simeq-{E_i/ S(r)}+\mathcal{O}\left((r-r_h)^0\right)\label{reg1} \\
	& g_i(r)\simeq z_i(r)/S(r)+\mathcal{O}\left((r-r_h)^1\right)\label{reg2}\\
	& k_i(r)\simeq \mathcal{O}\left((r-r_h)^0\right)\label{reg3}.\end{align}
\end{subequations}

\subsection{The Conductivity}

Substituting (\ref{reg}) into (\ref{Eri}) we obtain the following equations for the values of $z_i$ at the horizon
\be(C_hk^2\Psi_h+h^2Z_h)z_i(r_h)-\epsilon^{ij}hQ_hz_j(r_h)=-Q_hE_i-\epsilon^{ij}hZ_hE_j\label{systemeq}\sp i,j=x,y\ee
and we can also simplify the expression (\ref{curr2}) for the current
\be J_i=-Q_hz_i-\epsilon^{ij}hZ_hz_j+Z_hE_i+\epsilon^{ij}E_jW_h \label{curreq}\ee
where all the fields above are calculated at the horizon $r=r_h$ and we defined
\be Q_h=Q(\phi(r_h))\sp Z_h=Z(\phi(r_h))\sp W_h=W(\phi(r_h))\sp \Psi_h=\Psi(\phi(r_h))\sp C_h=C(r_h)\label{602}.\ee
Define \be J=\begin{pmatrix}
J_x\\ J_y
\end{pmatrix}\sp z=\begin{pmatrix}
z_x(r_h)\\ z_y(r_h)
\end{pmatrix}\sp
E=\begin{pmatrix}
E_x\\ E_y
\end{pmatrix}\label{603}\ee
and
\be H=\begin{pmatrix}
Z_h & W_h\\ -W_h & Z_h
\end{pmatrix},
N=\begin{pmatrix}
Q_h & hZ_h\\ -hZ_h & Q_h
\end{pmatrix},
L=\begin{pmatrix}
C_hk^2\Psi_h+h^2Z_h & -hQ_h\\ hQ_h & C_hk^2\Psi_h+h^2Z_h
\end{pmatrix},\label{604}\ee
then we can write (\ref{curreq}) and (\ref{systemeq}) respectively as
\be J=HE-Nz\label{605}\ee
\be Lz=-NE\label{606}.\ee
Eliminating $z$, we obtain (note that $\det L\ne 0$, unless $h=k=0$)
\be J=(H+NL^{-1}N)E.\label{607}\ee
Comparing with (\ref{Ohm2}), we identify the conductivity matrix
\be \sigma=H+NL^{-1}N.\label{608}\ee
Each of the matrices $H,L,N$ is invariant under the rotation group. They each have the form
\be
M^{ij}=a\delta^{ij}+b\epsilon^{ij}
\ee
in terms of the invariant tensors of SO(2).
These two matrices belong in the rotation group $O(2)$ and, since $O(2)$ is Abelian, they commute with every element of $O(2)$, hence they remain invariant under rotations.
Now let $R\in O(2)$ be arbitrary.
 We have
\be R\sigma R^T=R(H+NL^{-1}N)R^T=RHR^T+RNR^T(RLR^T)^{-1}RNR^T=H+NL^{-1}N=\sigma,\label{609}\ee
where we used the group relation $R^TR=1$. It follows that the conductivity matrix is rotationally invariant, as expected.

Performing the matrix multiplications we find
\boxedeq{cond}{
	\begin{split}
		&\sigma_{xx}=\;\;\,\sigma_{yy}=k^2\Psi_h C_h {Q_h^2+h^2Z_h^2+k^2\Psi_h C_hZ_h\over h^2Q_h^2+(k^2C_h\Psi_h+h^2Z_h)^2}\\
		&\sigma_{xy}=-\sigma_{yx}=W_h+hQ_h{Q_h^2+h^2Z_h^2+2k^2C_hZ_h\Psi_h\over h^2Q_h^2+(k^2C_h\Psi_h+h^2Z_h)^2}
	\end{split}
}
We remind that $Q=q-hW$. The formulas above have been derived in \cite{blake} for $W=0$.  When the momentum dissipation is weak (small $k/q,k/h$) the leading behavior of the conductivity is
\begin{subequations}\label{610}
	\begin{align}
		&\sigma_{xx}={k^2\over h^2}\Psi_h{C_h}+\cdots\label{611}\\
	&\sigma_{xy}={q\over h}-{C_h^2\Psi_h^2Q_hk^4\over h^3(Q_h^2+h^2Z_h^2)}+\cdots.\label{612}
	\end{align}
\end{subequations}
Note that setting $k=0$ (which restores the translational symmetry in the $x-y$ plane) we obtain $\sigma_{xx}=0$ and $\sigma_{xy}=q/h$, as in (\ref{hcond1}). Setting first $h=0$ and then $k=0$ we find that the resistivity is zero, as was expected from the discussion in section \ref{hconds}.

The following subsections are dedicated to studying special cases and implication of (\ref{cond}).

\subsection{The DC Conductivity at Zero Magnetic Field\label{dccond0}}

Setting $h=0$ in (\ref{cond}) we obtain:
\be \sigma_{xx}=Z_h+{q^2\over k^2\Psi_h C_h} \label{dccond}.\ee
\be \sigma_{xy}=W_h.\label{613}\ee
The formula for $\sigma_{xx}$ was previously derived in \cite{donos}. The first term has been named ``charge conjugation symmetric", $\sigma_{ccs}=Z_h$, \cite{blake}, because it appears even in the absence of (net) charge density ($q=0$). The proposed physical explanation is that charged particle-antiparticle pairs are created in the presence of the electric field, \cite{Karch:2007pd}. This term does not appear in the thermal conductivity, \cite{Donos:2014cya}. This result supports the pair-creation interpretation, as the pairs carry charge but no momentum.

There is also a constant contribution to the Hall conductivity, equal to the horizon value of the coupling $W$, in the absence of the magnetic field. This term is analogous to $\sigma_{ccs}=Z_h$ that appears in the longitudinal conductivity, and is a consequence of T-violation.

The second term in the longitudinal conductivity,
\be\sigma_{cd}={q^2\over k^2\Psi_h C_h},\label{614}\ee is the current dissipation term and appears because of the axions (\ref{cc1}). As we will see in sections \ref{chnum} and \ref{axnum}, for $k=0$ the DC conductivity diverges. For small values of $k$ there is a sharp Drude peak near $\omega=0$. Charge and heat is transported by collective excitations, \cite{DG}, \cite{Kim:2014bza}. As $k$ increases the peak becomes shorter and wider, as is expected from strong momentum dissipation. In this regime the transport becomes "incoherent" (no collective excitations), \cite{DG}, \cite{Kim:2014bza}, and there are deviations from the Drude model.

\subsection{The Hall Angle}

An important observable is the Hall angle $\theta_H$, which is given by the formula
\be \tan\theta_H={\sigma_{xy}\over\sigma_{xx}}={W_h(h^2Q_h^2+(k^2C_h\Psi_h+h^2Z_h)^2)\over k^2\Psi_h C_h(Q_h^2+k^2\Psi_h C_hZ_h+h^2Z_h^2)}+{hQ_h(Q_h^2+h^2Z_h^2+2k^2C_hZ_h\Psi_h )\over k^2\Psi_h C_h(Q_h^2+h^2Z_h^2+k^2\Psi_h C_hZ_h)}.\label{hangle}\ee
Since the ratio
\be \xi={Q_h^2+h^2Z_h^2+2k^2\Psi_hC_hZ_h \over Q_h^2+h^2Z_h^2+k^2\Psi_h C_hZ_h}\label{615}\ee
is a number between $1$ and $2$, we can write (\ref{hangle}) in the form
\be \tan\theta_H=\xi{hq\over k^2\Psi_hC_h}+W_h{k^2\Psi_hC_h\over Q_h^2+k^2\Psi_hC_hZ_h+h^2Z_h^2}.\label{hangle2}\ee

The first term has already appeared in \cite{blake}, while the second term stems from the T-odd part of the action (\ref{PQ}).
Note that the latter is non-zero even in the absence of a magnetic field ($h=0$) or charge density ($q=0$). This is a consequence of intrinsic T-violation. The contribution from this term becomes more apparent for small values of $q/k$ and $h/k$. For $h=0$, the Hall angle is:
\be
	\tan\theta_H=W_h{k^2\Psi_h C_h\over q^2+k^2\Psi_h C_hZ_h}\sp h=0.\label{616}\ee
At zero charge density ($q=0$) we have:
\be \tan\theta_H=W_h{k^2\Psi_hC_h\over (W_h^2+Z_h^2)h^2+k^2\Psi_hC_hZ_h}\sp q=0.\label{617}\ee
The expansion for large magnetic field $h$ is
\be\tan\theta_H\simeq{qh\over k^2C_h\Phi_h }+{qZ_h\over W_h^2+Z_h^2}h^{-1}+\mathcal{O}(h^{-2}).\label{618}\ee

In conclusion, there is a non-trivial contribution to the Hall angle, stemming from the PQ term (\ref{PQ}). This contribution becomes more apparent when the charge density and magnetic field are small compared to the momentum relaxation $k$ (small $q/k$ and $h/k$).

\subsection{Special Values for the Magnetic Field}

\begin{itemize}
	\item For $h=q/W_h$ the background gauge field vanishes ($A_t'=0$) and we have:
	\begin{subequations}\label{619}
		\begin{align}
		&\sigma_{xx}=Z_h{k^2W_h^2C_h\Psi_h \over k^2W_h^2C_h\Psi_h +q^2Z_h}\label{620}\\
		&\sigma_{xy}=W_h.\label{wqh}
		\end{align}
	\end{subequations}
    The Hall conductivity has reverted to its ``neutral" state, with the only contribution coming from pair-creation.
	For $k\to 0$ we obtain the same expression as in (\ref{hcond1}), since $\sigma_{xx}=0$ and $W_h=q/h$, but now $q,h$ are constrained.
	
	\item Consider now the longitudinal conductivity (\ref{cond}), set $W=0$ for simplicity and rewrite it as
	\be \sigma_{xx}= Z_h+{C_h\Psi_h k^2(q^2-h^2Z_h^2)-h^2Z_h(q^2+h^2Z_h^2)\over h^2q^2+(C_h\Psi_h k^2+h^2Z_h)^2}.\label{621}\ee
	The first term $Z_h$ is the pair-creation term, which is always there even at zero charge density and magnetic field. The second term vanishes when the magnetic field, $h$, assumes specific values in terms of the charge density and relaxation parameter
	\be h_c^2={1\over 2Z_h^2}\left(\sqrt{(q^2+C_hk^2\Psi_hZ_h)^2+4C_h\Psi_hZ_hk^2q^2}-(q^2+C_hk^2\Psi_hZ_h)\right).\label{622}\ee
	Note that the above quantity is always positive, so there are always two critical values $h=\pm h_c$, symmetric around $h=0$, at which only pair-creation contributes to the longitudinal conductivity.
	
	Consider now the case $W_h\ne 0$:
	\be \sigma_{xx}= Z_h+{C_h\Psi_h k^2((q-hW_h)^2-h^2Z_h^2)-h^2Z_h((q-hW_h)^2+h^2Z_h^2)\over h^2(q-hW_h)^2+(C_h\Psi_h k^2+h^2Z_h)^2}\label{623}.\ee
	For $0< W_h\leq Z_h$ there are exactly $2$ real values of $h$ at which the second term vanishes, however they are no longer symmetric around $0$.
	For values $W_h>Z_h$ there are either $2$ or $4$ real roots, depending on the values of $k,q$.
	 The expressions are not presented, as they are significantly more convoluted.
	
\end{itemize}

In conclusion, there are critical values of the magnetic field $h$ at which the pair-creation terms in either the longitudinal or Hall conductivity dominate. There is no value of $h$ that makes the longitudinal conductivity vanish, since the numerator of $\sigma_{xx}$ in (\ref{cond}) is always positive (unless $k=0$).

\section{Optical Conductivity\label{ACcond}}

In this section the goal is to calculate the full $\omega$-dependence of the AC conductivity in various backgrounds. In \ref{NBH} we calculate the AC conductivity in the finite-temperature neutral AdS-Schwarzschild black hole, which turns out to be frequency independent, \cite{Herzog:2007ij}. In this case, it is straightforward to find an analytic solution to the fluctuation equations, however, in the following subsections the solutions are found numerically or perturbatively in certain limits.

In \ref{chnum} we consider the AdS-RN charged black hole. We find that the conductivity diverges as $\omega\to 0$. This is expected because there is no momentum dissipation. There is also a constant contribution to the Hall conductivity stemming from the PQ term \ref{PQ}.

In \ref{axnum} we include axions in the background solution to break the translational invariance, as described in section \ref{EMDax}. We find that the DC conductivity is now finite and there is a Drude peak at $\omega=0$ for small values of momentum relaxation $k$. As $k$ increases, there are deviations from the Drude model, \cite{DG}.

In \ref{dynum} we consider the dyonic AdS-RN black hole in the presence of axions. There are peaks in the conductivity, related to the cyclotron poles in the complex $\omega$ plane \cite{Hartnoll:2007ip}. Damping is present even at $k=0$ because of the collisions of particles and anti-particles, which are executing cyclotron orbits in opposite directions \cite{Hartnoll:2007ih}. As $k$ increases, the damping becomes stronger. There is a constant contribution to the Hall conductivity stemming from the PQ term, (\ref{PQ}). The charge density appears shifted by a value proportional to the magnetic field and PQ coupling.

In order to calculate the AC conductivity, we perturb the background (\ref{ans2}) with a time-dependent electric field. The perturbations in $A_x,A_y$ force us to turn on additional fields. More details will be given in each of the following subsections.

Suppose that we perturb the gauge field as follows
\be \delta A_i(r,t)\sp i=x,y.\label{pertax}\ee
The Maxwell sector of the action (\ref{actionax}) is:
\be S_{M}=M^{2}\int \ud^{4}x \sqrt{-g}\left(-{1\over 4}Z(\phi) F^2-{1\over 4} W(\phi)F\wedge F\right)=M^2\int \ud^{4}x  \partial_\mu (K^{\mu\nu} A_\mu).\label{701}\ee
where \be K^{\mu\nu}=-{1\over 2}\sqrt{-g}Z(\phi)F^{\mu\nu}-{1\over 4}W(\phi)\epsilon^{\mu\nu\rho\sigma}F_{\rho\sigma}\label{702}\ee
and we also used that $\partial_{\mu}K^{\mu\nu}=0$, (\ref{GFEQ}), on-shell.
Only the boundary term remains:
\be S_{on-shell}=M^{2}\int\ud^{3}x (K^{r\nu} A_\nu)|_{r\to\text{boundary}},\label{703}\ee
where we used the fact that the radial direction is perpendicular to the boundary and horizon, and discarded the contribution from the horizon according to the prescription, \cite{Son}, (also see section \ref{2pf}).

Using (\ref{pertax}), the background solution (\ref{ans2}), and keeping the terms that are quadratic in the gauge field fluctuations we obtain
\be S_{AA}={1\over 2}V_2M^{2}\int \ud t \left(-ZS\delta A'_i \delta A_i+W\epsilon^{ij}\delta\dot{A}_i\delta A_j\right)|_{r\to\text{boundary}}\label{osa0}\ee
where we defined the $2$-dimensional volume
\be V_2=\int \ud x\ud y.\label{705}\ee
We Fourier transform the temporal part of the fields as follows
\be \delta A_i(r,t)=M\int {\ud\omega\over 2\pi}e^{-i\omega t}a_i(r,\omega)\label{706}\ee
and substitute into (\ref{osa0}) to obtain
\be S_{AA}=-{1\over 2}V_2\int {\ud\omega\over 2\pi} \left(ZS\delta^{ij}a'_i(\omega)a_j(-\omega)+i\omega W\epsilon^{ij}a_i(\omega)a_j(-\omega)\right)|_{r\to\text{boundary}}.\label{onshell}\ee
We used the mass scale $M$ in the definition above so that the dimensions of the transformed bulk fields agree with the field theory side.

In order to calculate the retarded Green's function we need to solve the fluctuation equations with the in-going boundary condition at the horizon, \cite{Son}. Then, using the prescription from \cite{Son}, we obtain the current-current correlator from (\ref{onshell}).

\subsection{Neutral AdS-Schwarzschild Black Hole\label{NBH}}

Consider the AdS-Schwarzschild background solution to the theory governed by (\ref{actionax})
\be ds^2=L^2r^{-2}\left(-f(r)dt^2+{dr^2\over f(r)}+d\vec{x}^2\right)\sp A_t(r)=0\sp\phi(r)=\phi_0\sp \chi_i=0\label{sol1}\ee
where \be  L^2=6/V(\phi_0)\sp \partial_\phi V(\phi_0)=0\sp Z(\phi_0)=Z_0\sp W(\phi_0)=W_0\label{707}\ee
and the blackening factor is
\be f(r)=1-{r^3\over r_h^3}.\label{708} \ee
This solution describes a translationally invariant system at zero charge density and finite temperature:
\be T={3\over 4\pi r_h}.\label{709}\ee

\subsubsection{Fluctuation Equations}

To calculate the conductivity in this background we need only perturb the gauge field:
\be  \delta A_i(r,t)=M\int {\ud\omega\over 2\pi}  a_i(r,\omega)e^{-i\omega t}.\label{710}\ee
The fluctuation equations are decoupled and read
\be \left(fa_i'\right)'+{\omega^2\over f}a_i=0\label{gfx1}\ee
By redefinition of the radial coordinate
\be dr/f(r)=d\rho,\label{redef}\ee
equation (\ref{gfx1}) becomes a simple harmonic oscillator
\be  \ddot{a}_i+\omega^2 a_i=0\label{711}\ee
where the dots denote derivatives with respect to $\rho$.
The general solution is
\be a_i(\rho)=C_1^i e^{i\omega \rho}+C_2^i e^{-i\omega\rho}.\label{712}\ee
Integrating (\ref{redef}) we find
\be \rho(r)={r_h\over 6}\left(2\sqrt{3}\text{atan}\left(2r+r_h\over \sqrt{3}r_h\right)+\log{r_h^3-r^3\over (r_h-r)^3}\right)-{\pi r_h\over 6\sqrt{3}},\label{713}\ee
where the integration constant was chosen so that $\rho(0)=0$.

The ingoing boundary condition at the horizon $r=r_h$ is satisfied by setting $C_2=0$. Additionally, $C_1^i=a_i^{(0)}$ is the value of $a_i$ at the boundary $\rho=0$. The full solution is
\be a_i(r)=a_i^{(0)}e^{-{\pi\over 6\sqrt{3}}}\left({r_h^3-r^3\over (r_h-r)^3 }\right)^{i\omega r_h/6}\left({i\sqrt{3}+1+2r/r_h\over i\sqrt{3}-1-2r/r_h}\right)^{-{\omega r_h\over 2\sqrt{3}}}.\label{714}\ee
The expansion of $a_i$ near the boundary ($r=0$) is
\be a_i(r)=a_i^{(0)}(1+i\omega  r-\omega^2 r^2+\mathcal{O}(r^3)).\label{715}\ee
Substituting this expansion into (\ref{onshell}) we obtain
\be S_{AA}=-{1\over 2}V_2\int {\ud \omega\over 2\pi} a_i^{(0)}(\omega)\left(i\omega Z_0\delta^{ij}+i\omega W_0 \epsilon^{ij}\right)a_j^{(0)}(-\omega)\label{716}\ee
from which we can read the retarded Green's function, \cite{Son},
\be G^R_{ij}(\omega)=i\omega (Z_0\delta^{ij}+ W_0 \epsilon^{ij}).\label{717}\ee
The conductivity is given by the Kubo formula (\ref{kubo})
\be \sigma_{ij}(\omega)={G^R_{ij}(\omega)\over i\omega}=Z_0\delta^{ij}+W_0 \epsilon^{ij},\ee
therefore
\boxedeq{ACneutral}{\begin{split}
& \sigma_{xx}=~~\sigma_{yy}=Z_0\\
& \sigma_{xy}=-\sigma_{yx}=W_0
\end{split}}
This background has zero charge density, but the conductivity is finite. This is because of pair-creation, as explained in section \ref{dccond0}.
There is no dependence on the frequency (this was first discovered in \cite{Herzog:2007ij}), which is expected from dimensional analysis. The only dimensionful parameter of the field theory is the temperature $T$. The conductivity, being dimensionless, may only depend on the ratio $\omega/T$. Since the field theory is scale invariant, all temperatures are equivalent (except for $T=0$), therefore the conductivity cannot depend on $\omega/T$.
Turning on a magnetic field or chemical potential introduces additional dimensionful scales and the frequency dependence of $\sigma$ becomes non-trivial.

\subsection{Charged AdS-RN Black Hole\label{chnum}}

We now place the system at finite charge density. The relevant solution to (\ref{actionax}) is the charged AdS-RN black hole. This background describes a translationally invariant system at finite charge density, $q$, and finite temperature, $T$.
\be ds^2=L^2r^{-2}\left(-f(r)dt^2+{dr^2\over f(r)}+d\vec{x}^2\right)\sp A_t(r)=\mu+{q\over Z_0}r\sp\phi(r)=\phi_0\sp \chi_i=0\label{720}\ee
\be  L^2=6/V\sp \partial_\phi V=\partial_\phi Z=0\sp Z(\phi_0)=Z_0\sp W(\phi_0)=W_0\label{721}\ee
\be f(r)=1+{q^2\over 4 Z_0L^2} r^4-mr^3.\label{722}\ee
For $q=0$, the we obtain the Schwarzschild black hole of \ref{NBH}. For $q\ne 0$ the polynomial, (\ref{722}), has exactly two real roots $r_1,r_2$, as long as the black hole mass is big enough
\be m^4>{4q^6\over 27Z_0^3L^6}.\label{bhmc}\ee
We factor the polynomial in the following way
\be f(r)=\left(1-{r\over r_1}\right)\left(1-{r\over r_2}\right)\left(1+{r_1+r_2\over r_1r_2}r+{r_1^2+r_1r_2+r_2^2\over r_1^2r_2^2}r^2\right),\label{723}\ee
where
\be {q^2\over 4 Z_0L^2}={r_1^2+r_1r_2+r_2^2\over r_1^3r_2^3}\label{ineq0}\ee
\be m={r_1^3+r_1^2r_2+r_1r_2^2+r_2^3\over r_1^3r_2^3}\label{724},\ee
and the third factor in (\ref{723}) has no positive real roots.
Define
\be r_h=\min (r_1,r_2)\sp r_\star=\max(r_1,r_2).\label{725}\ee
Then, from (\ref{ineq0}) we find the condition
\be {q^2r_h^4\over 4Z_0L^2}={r_h\over r_\star}\left({r_h^2\over r_\star^2}+{r_h\over r_\star}+1\right)\leq 3,\label{ineq}\ee
which will be useful later.

Since we are interested in the region from the boundary $r=0$ up to the outer horizon $r=r_h$ we can drop $r_\star$ in favor of $q$ and write $f$ in the more convenient form
\be f(r)=\left(1-{r\over r_h}\right)\left(1+{r\over r_h}+{r^2\over r_h^2}-{q^2r_h\over 4 Z_0L^2}r^3\right).\label{726} \ee
The temperature is calculated at the outer horizon
\be T={1\over 4\pi r_h}\left(3-{q^2r_h^4\over 4 Z_0L^2}\right),\label{tempcbh}\ee
and is nonnegative for all values of $q$, according to (\ref{ineq}). Note that the equality in (\ref{ineq}) holds in the extremal case, in which $r_1=r_2$ and the temperature vanishes.
The black hole mass in terms of the outer horizon $r_h$ is
\be m=r_h^{-3}\left(1+{q^2r_h^4\over 4Z_0L^2}\right)\label{bhm}.\ee
Substituting (\ref{bhm}) into (\ref{bhmc}),  it is straightforward to check that the inequality is satisfied for all values of $q$, except for the extremal value, at which (\ref{bhmc}) becomes an equality.

\subsubsection{Fluctuation Equations}

The linearized equations in this background do not mix the $x,y$ components of the gauge field, however we have to turn on the $g_{tx},g_{ty}$ components of the metric
\be \delta A_i=M\int {\ud\omega\over 2\pi} a_i (r,\omega)e^{-i\omega t}\sp \delta g_{ti}=M{L^2\over r^2}\int {\ud\omega\over 2\pi}  z_i(r,\omega)e^{-i\omega t}.\label{727}\ee
From the gauge field equations we obtain
\be \left(Z_0fa_i'+qz_i\right)'+\omega^2{Z_0\over f}a_i=0\label{gfx2}\ee
and from the Einstein equations
\be  L^2{z_i'}+ r^2{q}a_i=0\label{Erx2}\ee
\be \left({r^{-2}}L^2z_i'+qa_i\right)'=0.\label{728}\ee
The latter is obviously implied by (\ref{Erx2}).
Eliminating $z_i$ from (\ref{gfx2}), (\ref{Erx2}) we obtain the decoupled equations for the gauge field fluctuations
\be Z_0(fa_i')'+\left(\omega^2{Z_0\over f}-{q^2r^2\over L^2}\right)a_i=0\label{gfx3}.\ee
We rescale the radial coordinate
\be \rho=r/r_h\label{729}\ee
so that the horizon is at $\rho=1$. We also define the dimensionless parameters:
\be\tilde q=qr_h^2/(\sqrt{Z_0}L)\sp \tilde\omega=\omega r_h\label{dpar}.\ee
Note that (\ref{ineq}) implies
\be \tilde q^2\leq 12,\label{730}\ee
where the equality holds in the extremal case $T=0$.

The equation (\ref{gfx3}) now becomes
\be (fa_i')'+\left({\tilde\omega^2\over f}-\tilde q^2\r^2\right)a_i=0\label{gfx4}\ee
where primes now denote derivatives with respect to $\rho$ and
\be f(\rho)=\left(1-{\rho}\right)\left(1+\rho+\rho^2-{1\over 4}\tilde q^2\rho^3\right)\label{chargednum}.\ee
We choose the following as the independent parameters
\be \tilde q,\tilde\omega, r_h.\label{731}\ee
According to (\ref{dpar}) we are measuring $q,\omega$ in units of $r_h$.
This choice is convenient for calculations, but for the plots in \ref{nmchnum}, we will drop $r_h$ in favor of the temperature $T$, which has a direct physical meaning for the field theory.

Since the equations for $a_x,a_y$ are identical, from now on we will focus on the $a_x$ component.

\subsubsection{Hydrodynamic Limit (small $\tilde\omega$)\label{elhydr}}

We derive the formula for the AC conductivity at low frequencies, $\tilde \omega$.

We start from (\ref{gfx4}), dropping the index for convenience
\be (fa')'+\left({\tilde\omega^2\over f}-\tilde q^2\r^2\right)a=0.\label{732}\ee
We transform the field as follows
\be a(\r)=g(\r)Y(\r)\label{733}\ee
to obtain
\be Y''+Y'\left({f'\over f}+2{g'\over g}\right)+Y\left({\tilde\omega^2\over f^2}-{q^2r^2\over f}+{f'g'\over fg}+{g''\over g}\right)=0.\label{gfx5}\ee
We require that $g$ satisfies
\be {g''}+{f'\over f}g'-{q^2r^2\over f}g=0\label{734}\ee
so that the coefficient of $Y$ is negligible at the limit $\tilde\omega\to 0$. It is straightforward to check that
\be g(\r)=1-{4\tilde q^2\over 3(4+\tilde q^2)}\rho\label{735}\ee
satisfies the above equation. Now (\ref{gfx5}) becomes
\be Y''+Y'\left({f'\over f}+2{g'\over g}\right)+Y{\tilde\omega^2\over f^2}=0.\label{736}\ee
The equation is singular at the horizon. Using the ansatz $Y(\rho)=f(\rho)^n$ near $\rho=1$ we obtain $n_{\pm}=\pm i\omega/f'(1)$. The mode $n_+$ corresponds to the ingoing solution at the horizon, and is the correct behavior if we are interested in the retarded Green's function.
We use
\be Y(\rho)=f(\rho)^{i\tilde\omega/f'(1)}X(\rho)\label{737}\ee
to remove the leading behavior at the horizon. The equation for $X$ is
\be X''+X'\left({f'\over f}+2{g'\over g}+{2i\tilde\omega f'\over f'(1)f}\right)+X\left(\frac{i \tilde\omega  f''(r)}{f'(1) f}+\frac{2 i \tilde\omega  f' g'}{f'(1) f g}-\frac{\tilde\omega ^2
	f'^2}{f'(1)^2 f^2}+\frac{\tilde\omega ^2}{f^2}\right)=0.\ee
Now $X$ needs to be regular at $\rho=1$.
In order to solve the equation perturbatively, we expand $X$ for small $\tilde\omega$ as
\be X(\rho)=X_0(\rho)+\tilde\omega X_1(\rho)+\mathcal{O}(\tilde\omega^2).\label{738}\ee
We obtain
\begin{subequations}\label{739}
	\begin{align}
	&X_0''+X_0'\left({f'\over f}+2{g'\over g}\right)=0\label{eqX0}\\
	&X_1''+X_1'\left({f'\over f}+2{g'\over g}\right)=-{2if'\over f'(1)f}X_0'-{i\over f'(1)}\left(2{f'g'\over fg}+{f''\over f}\right)X_0.\label{eqX1}
	\end{align}
\end{subequations}
We will solve the equations with the following boundary conditions at the horizon $X_0(1)=C, X_1(1)=0$, in addition to the regularity requirement.
Equation (\ref{eqX0}) is straightforward to integrate
\be X_0'= {c_1\over fg^2}.\label{740}\ee
Since the right-hand side has a simple pole at $\rho=1$, $X_0$ diverges logarithmically at the horizon ($\rho=1$), unless we set $c_1=0$. Therefore $X_0$ is constant.

Since $X_0$ is constant, (\ref{eqX1}) is simplified and we find
\be X_1'={c_3\over fg^2}-{iX_0\over f'(1)}{f'\over f}.\label{741}\ee
Both terms have poles at $\rho=1$. We must choose $c_3$ so that they cancel out, and $X_1$ is finite at the horizon. Taking the near-horizon limit of the right-hand side we find
\be c_3={iX_0g(1)^2}\label{742},\ee
hence
\be X_1(\rho)=iX_0\int_1^\rho \ud\rho'{1\over f(\rho')}\left({(g(1))^2\over (g(\rho'))^2}-{f'(\rho')\over f'(1)}\right)\equiv iX_0H(\rho).\label{743}\ee

The solution up to first order in $\tilde\omega$ is
\be a(\r)=X_0g(\rho)\left(1+i\tilde\omega H(\rho)+i\tilde\omega {\log f(\rho)\over f'(1)}\right).\label{744}\ee
The integration constant $X_0$ can be written in terms of the boundary value of $a$, $a_0=a(0)$:
\be X_0={a_0\over 1+i\tilde\omega H(0)}.\label{745}\ee

Substituting into (\ref{onshell}), we obtain the Green's function and using the Kubo formula (\ref{kubo}), we obtain the conductivity
\be \sigma(\tilde\omega)=Z_0\left(\left({12-\tilde q^2\over 12+3\tilde q^2}\right)^2+i{4\tilde q^2\over 3\tilde\omega(4+\tilde q^2)}\right).\label{746}\ee
This formula is valid for $0<\tilde\omega<<1$ and arbitrary $\tilde q$.

We find that the imaginary part has a pole at $\tilde\omega=0$. The Kramers-Kronig relations imply that there is an additional delta-function in the real part
\boxedeq{hydro}{ \sigma(\tilde\omega)=Z_0\left(\left({12-\tilde q^2\over 12+3\tilde q^2}\right)^2+\delta(\tilde\omega){4\tilde q^2\pi\over 3(4+\tilde q^2)}+i{4\tilde q^2\over 3\tilde\omega(4+\tilde q^2)}\right)}
The DC conductivity is divergent, as is expected from a translationally invariant theory.

\subsubsection{Numerical Solution\label{nmchnum}}

To obtain the full frequency-dependence we solve the equation numerically with the dimensionless parameters $\tilde\omega$, $\tilde q$, using the method described in \cite{Hartnoll:2009sz}.

From equation (\ref{gfx4}) we find that
$a_x\sim f(\rho)^{n_\pm}$ with $n_\pm={\pm i\tilde\omega/f'(1)}$ near the horizon. The mode $n_+$ corresponds to the ingoing solution at the horizon. To solve the equation numerically we need to remove the leading behavior near the horizon. As in the previous section \ref{elhydr},
we substitute
\be a_x(\rho)=f(\rho)^{i\tilde\omega/f'(1)}\tilde{a}_x(\rho)\label{747}\ee
into (\ref{gfx4}) to obtain a differential equation for $\tilde{a}_x$
\be f\tilde a_x''+f' \tilde a_x'\left(1+{2i\tilde\omega\over f'(1)}\right)+\left({i\tilde\omega f''\over f'(1)}-\tilde q^2\rho^2+{\tilde\omega^2\over f}\left(1-{f'^2\over f'(1)^2}\right)\right)\tilde a_x=0,\label{dif1}\ee
where
\be f'(1)=-3+{\tilde q^2\over 4}.\label{748}\ee
Now $\tilde{a}_x$ is regular at the horizon
\be \tilde{a}_x(\rho)=a_0+a_1(1-\rho)+a_2(1-\rho)^2+\cdots.\label{749}\ee
The choice of the mode $n_+$ means that we set one boundary condition. The equation is homogeneous and linear, so by rescaling we can set $a_0=1$, which is the second boundary condition. Then the solution is completely specified. Since the equation is singular at $\rho=1$, we need to set the boundary conditions a small distance, $\epsilon$, from the horizon. To that end we find the Taylor expansion of $\tilde{a}_x$ up to first order in $(1-\rho)$, by solving equation (\ref{dif1})
\be \tilde{a}_x=1+\left(24i{\tilde{q}^2-4\over (\tilde{q}^2-12)^2}-{4\tilde{q}^2\over \tilde{q}^2+8i\tilde{\omega}-12}\right)(1-\rho)+\mathcal{O}((1-\rho)^2)\label{750}\ee
and set the boundary conditions at $\rho=1-\epsilon$, where $\epsilon$ is a small number ($\epsilon=0.01$ was used for the graphs below), instead of $\rho=1$.

The conductivity is again obtained from the on-shell action (\ref{onshell}).
From the equations (\ref{gfx2}), (\ref{Erx2}) we obtain the near-boundary behavior ($\rho\to 0$)
\be a_x=a_x^{(0)}+a_x^{(1)}\rho-{1\over 2}\tilde{\omega}^2 a_x^{(0)}\rho^2-{1\over 6}\tilde{\omega}^2a_x^{(1)}\rho^3+\cdots\label{751}\ee
\be z_x=z_x^{(0)}-{1\over 3}\tilde{q}a_x^{(0)}\rho^3+\cdots.\label{752}\ee
Since $z_x$ appears in the equations only through its derivative, we can gauge away the constant term $z_x^{(0)}$, so that it does not contribute to the current $a_x^{(1)}$.
The relevant term from (\ref{onshell}) becomes
\be S_{AA}= -{1\over 2}V_2\int {\ud \omega\over 2\pi} \left(Z_0\delta^{ij}a_i^{(1)}(\omega)a_j^{(0)}(-\omega)+i\omega W_0 \epsilon^{ij}a_i^{(0)}(\omega)a_j^{(0)}(-\omega)\right).\label{753}\ee
Since there is no mixing of $a_x,a_y$ in the equations, $a_x^{(1)}$ does not depend on $a_y^{(0)}$ and $a_y^{(1)}$ does not depend on $a_x^{(0)}$. Since we can choose a gauge with $z_x^{(0)}=0$, there is no contribution to $a_x^{(1)}$ from the metric. Keeping the above in mind, we obtain the Green's function
\be G^R_{ij}=i\omega W_0\epsilon_{ij}+\delta_{ij}Z_0{a_i^{(1)}\over a_i^{(0)}}.\label{754}\ee
The off-diagonal conductivity is the same as in the neutral case
\be \sigma_{xy}=W_0\label{755}\ee
and, since $a_x^{(1)}=a_x'(0)$, the longitudinal conductivity can be calculated from the numerical solution as follows
\be \sigma_{xx}(\tilde\omega)=Z_0{\tilde{a}_x'\over i\tilde{\omega}\tilde{a}_x}\bigg|_{\rho=\varepsilon}\label{756}\ee
where $\varepsilon$ is a small number ($\varepsilon=10^{-3}$ was used for the graphs below).

\begin{figure}[H]
	\centering
	\includegraphics[width=0.49\textwidth]{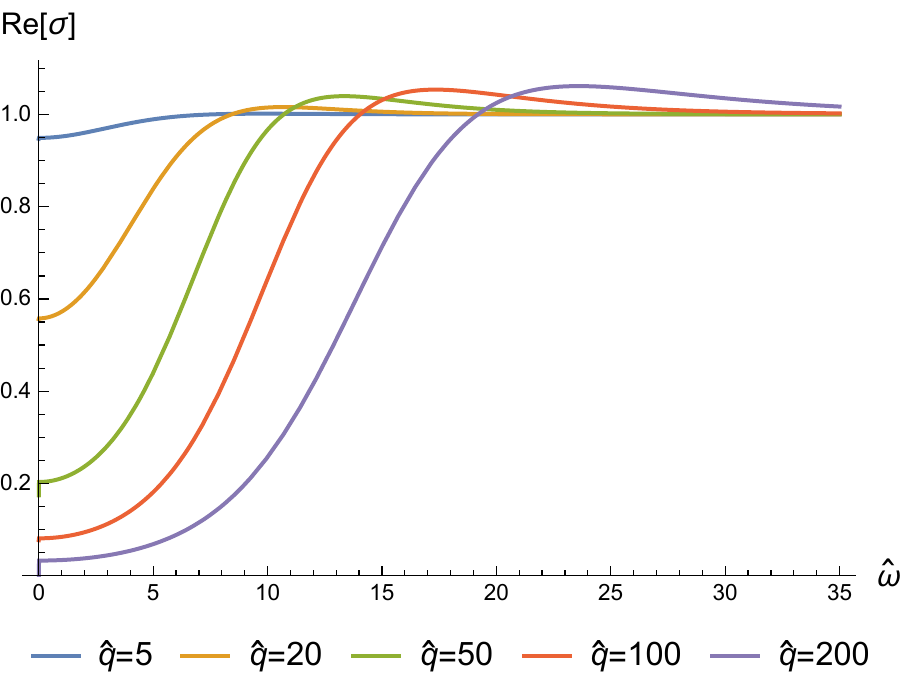}
	\includegraphics[width=0.49\textwidth]{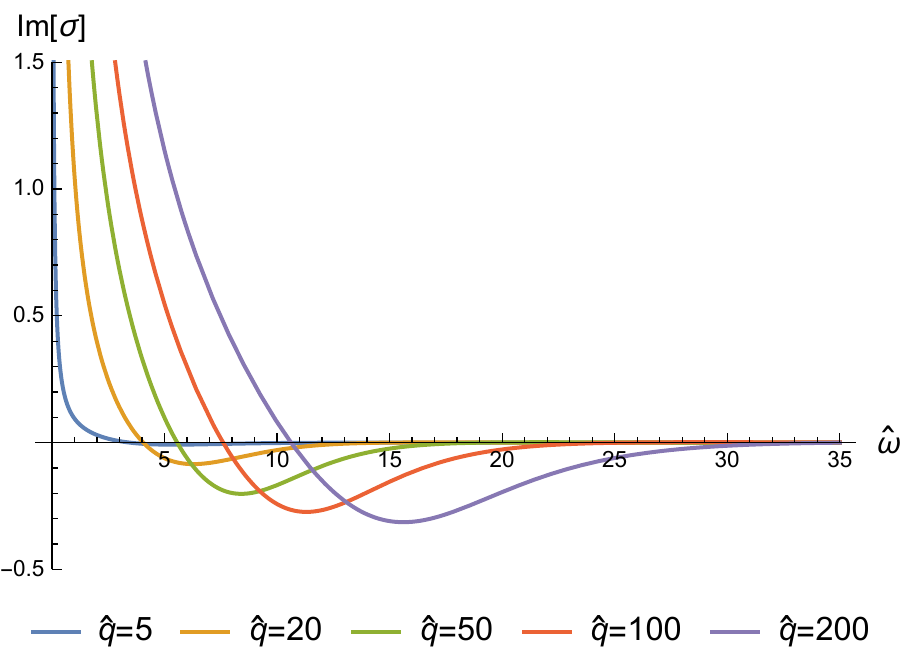}
	\caption{\small The real and imaginary parts of the AC conductivity are plotted against the dimensionless frequency $\hat{\omega}=\omega/T$, for fixed values of the dimensionless charge density $\hat{q}=(L\sqrt{Z_0})^{-1}q/T^2$. The vertical axes have units of $Z_0$. The imaginary part has a pole at $\hat\omega=0$. The corresponding delta-function in the real part is not drawn.}
	\label{fig1}
\end{figure}

In (\ref{dpar}) we defined our units in terms of the horizon location $r_h$ for convenience. It is more natural to measure the physical quantities of the field theory in terms of the temperature. For the plots we will use the following units
\be \hat q={\tilde q\over r_h^2 T^2}={q\over T^2\sqrt{Z_0}L}\sp \hat \omega={\tilde \omega\over r_h T}={\omega\over T}.\label{757}\ee
Expressing the observables in the new units involves solving the polynomial (\ref{tempcbh}) numerically. In figure \ref{fig1} we plot the conductivity against $\omega/T$, while keeping the ratio $q/T^2$ fixed.

As the frequency increases, the longitudinal conductivity approaches the value $\text{Re}(\sigma)=Z_0$ and $\text{Im}(\sigma)=0$, as in (\ref{ACneutral}), for all values of $\hat q$. Note that the imaginary part of the conductivity has a pole at $\hat\omega=0$, as in section \ref{elhydr}. In the real part there is a delta-function at $\hat\omega=0$ (see \ref{elhydr}), which is not drawn.

The Hall conductivity remains constant
\be \sigma_{xy}=W_0.\ee
It is clear that this constant is contributed by the PQ term (\ref{PQ}).
Since there is no magnetic field, the fluctuation equations (\ref{gfx4}) do not mix the $x,y$ components of the gauge field, hence there is no additional contribution. This term can be thought of as analogous to the constant contribution in the longitudinal conductivity (\ref{ACneutral}) in the absence of charge density.

\subsection{Charged AdS-RN Black Hole with Axions\label{axnum}}

We now include two axionic fields, in order to break the translational symmetry. This background is characterized by a finite charge density $q$, finite temperature $T$ and momentum relaxation scale $k$.
The relevant solution to (\ref{actionax}) is the following:
\be ds^2=L^2r^{-2}\left(-f(r)dt^2+{dr^2\over f(r)}+d\vec{x}^2\right)\sp A_t(r)=\mu+{q\over Z_0}r\sp\phi(r)=\phi_0\sp \chi_i=kx_i\label{sol2}\ee
where \be  L^2=6/V\sp \partial_\phi V=\partial_\phi\Psi=\partial_\phi Z=0\sp Z(\phi_0)=Z_0\label{758}\ee
and
\be f(r)=1+{q^2\over 4 Z_0L^2} r^4-{1\over 2}k^2\Psi_0r^2-mr^3.\label{759}\ee

For $k=0$, we obtain the charged black hole of section \ref{chnum}. For $q\ne 0$ the polynomial, (\ref{759}), has exactly two positive, distinct real roots $r_1,r_2$, as long as the black hole mass is big enough
\be 108m^2>k^2\Psi_0\left(k^4\Psi_0^2-36{q^2\over Z_0L^2}\right)+\left(k^4\Psi_0^2+12{q^2\over Z_0L^2}\right)^{3/2} .\label{bhmc2}\ee
We factor the $f$ in the following way
\be f(r)=\left(1-{r\over r_1}\right)\left(1-{r\over r_2}\right)\left(1+{r_1+r_2\over r_1r_2}r+{r_1^2+r_1r_2+r_2^2\over r_1^2r_2^2}r^2\right),\label{760}\ee
where
\be {1\over 2}k^2\Psi_0 +{q^2\over 4 Z_0L^2}r_1r_2={r_1^2+r_1r_2+r_2^2\over r_1^2r_2^2}\label{ineq1}\ee
\be m={r_1^3+r_1^2r_2+r_1r_2^2+r_2^3\over r_1^3r_2^3}.\label{761}\ee
Define
\be r_h=\min (r_1,r_2)\sp r_\star=\max(r_1,r_2).\label{762}\ee
Then, from (\ref{ineq1}) we find the condition
\be {1\over 2}k^2\Psi_0r_h^2+{q^2r_h^4\over 4Z_0L^2}\leq{1\over 2}k^2\Psi_0r_h^2+{q^2r_h^3r_\star\over 4Z_0L^2}={r_h^2\over r_\star^2}+{r_h\over r_\star}+1\leq 3.\label{ineq11}\ee
We drop $r_\star$ in favor of $q,k$ and write $f$ in the more convenient form
\be f(r)=\left(1-{r\over r_h}\right)\left(1+{r\over r_h}+{r^2\over r_h^2}-{1\over 2}k^2\Psi_0r^2-{1\over 4}{q^2r_h\over Z_0 L^2}r^3\right).\label{763} \ee
The temperature is
\be T={1\over 4\pi r_h}\left(3-{1\over 2}k^2\Psi_0r_h^2-{1\over 4}{q^2r_h^4\over Z_0 L^2}\right)\label{tempax},\ee
which is nonnegative for all values of $k,q$, according to (\ref{ineq11}).
The equality in (\ref{ineq11}) holds in the extremal case, in which $r_1=r_2$ and the temperature vanishes.

The black hole mass in terms of the outer horizon $r_h$ is
\be m=r_h^{-3}\left(1-{1\over 2}k^2\Psi_0r_h^2+{q^2r_h^4\over 4Z_0L^2}\right)\label{bhm1}.\ee

\subsubsection{Fluctuation Equations}

The linearized equations in this background do not mix the $x,y$ components of the gauge field, however we need to perturb the $g_{tx},g_{ty}$ components of the metric and the axions
\be  \delta A_i=M\int {\ud\omega\over 2\pi} a_i (r)e^{-i\omega t}\sp \delta \chi_i=M\int {\ud\omega\over 2\pi} k_i(r)e^{-i\omega t}\sp
\delta g_{ti}=M{L^2\over r^2}\int {\ud\omega\over 2\pi} z_i(r)e^{-i\omega t}.\label{764}\ee
We focus on the $x$ components of the fluctuation equations. The equations for the $y$ components are identical. The independent fluctuation equations are
\be Z_0\left(fa_x'\right)'+qz_x'+{Z_0\omega^2\over f}a_x=0\label{765}\ee
\be r^2f{(f r^{-2}k_x')'}+{\omega^2}k_x-i\omega {k}z_x=0\label{766}\ee
\be	i\omega L^2z_x'-kL^2\Psi_0fk_x'+i\omega q r^2 a_x=0\label{767}\ee
Defining the dimensionless parameters
\be \tilde\omega=\omega r_h\sp \tilde k=kr_h\sqrt{\Psi_0}\sp \tilde q={qr_h^2\over L\sqrt{Z_0}} \label{redefax}\ee
and rescaling the radial coordinate and the fields
\be \rho=r/r_h\sp z_x\to {\sqrt{Z_0}r_h\over L} z_x\sp k_x\to {r_h\over L}\sqrt{Z_0\over \Psi_0} k_x\label{768}\ee
we obtain
\begin{subequations}\label{axfluc}
	\begin{align}
	&f\left(fa_x'\right)'+\tilde{q}fz_x'+\tilde{\omega}^2a_x=0\label{axbh}\\
	&\rho^2f{(f \rho^{-2}k_x')'}+{\tilde{\omega}^2}k_x-i\tilde{\omega} {\tilde{k}}z_x=0\label{axk}\\
	&	i\tilde{\omega }z_x'-\tilde{k}fk_x'+i\tilde{q}\tilde{\omega} \rho^2 a_x=0\label{axcons}
	\end{align}
\end{subequations}
where primes now denote derivatives with respect to $\rho$, and
\be f(\rho)=(1-\rho)\left(1+\rho+\rho^2-{1\over 2}\tilde{k}^2\rho^2-{1\over 4}\tilde{q}^2\rho^3\right).\label{769}\ee

\subsubsection{Hydrodynamic Limit (small $\tilde\omega,\tilde k$)}
We adapt the method from \cite{Ge:2010yc} to calculate the conductivity analytically at small frequencies, $\tilde\omega$, and weak momentum dissipation, $\tilde k$.
Starting from the system (\ref{axfluc})
we
define the following functions
\be \phi_\pm={z_x'\over \r^2}+qa_x+{C_\pm\over \r}a_x,\label{phipm}\ee
where
\be C_\pm={6\tilde k^2-3\tilde q^2-12\over 8\tilde q}\pm {\sqrt{64\tilde k^2\tilde q^2+(12-6\tilde k^2+3\tilde q^2)^2}\over 8\tilde q}.\label{cpm}\ee
We obtain a decoupled system for $\phi_\pm$
\be (\r^2f\phi_\pm')'+\left({\r^2\omega^2\over f}+\lambda_\pm \r\right)\phi_\pm=0,\label{770}\ee
where
\be \lambda_+={C_+f'+\r(C_-+\tilde q\r)(\tilde k^2-C_+\tilde q\r)\over C_+-C_-}\label{771}\ee
\be \lambda_-={-C_-f'-\r(C_++\tilde q\r)(\tilde k^2-C_-\tilde q\r)\over C_+-C_-}.\label{772}\ee
To first non-trivial order in $\tilde k$ we have
\be \lambda_+=\tilde k^2\left({\rho(-12-3\tilde q^2+4\rho\tilde q^2)\over 12+3\tilde q^2}\right)+\mathcal{O}(\tilde k^4)\label{773}\ee
\be \lambda_-=-{3\over 4}\rho^2(4+\tilde q^2)+\tilde k^2\left({\rho(-24-6\tilde q^2+(36+\tilde q^2)\rho)\over 24+6\tilde q^2}\right)+\mathcal{O}(\tilde k^4)\label{774}.\ee
We also write $f$ as follows
\be f(\r)=f_0(\r)+\tilde k^2 f_1(\r),\label{775}\ee
where
\be f_0(\r)=(1-\r)\left(1+\r+\r^2-{1\over 4}\tilde q^2\r^3\right)\sp f_1(\r)=-{1\over 2}\r^2(1-\r).\label{776}\ee

We start from the equation for $\phi_+$. Using
\be \phi_+=\psi f^{i\omega\over f'(1)}\label{777}\ee
removes the leading behavior at the horizon. Now $\psi$ must be regular at the horizon. We expand for small $\tilde\omega,\tilde k$ as follows
\be \psi=\psi_0+\tilde\omega \psi_1+\tilde k^2\psi_2+\mathcal{O}(\tilde\omega^2,\tilde k^4,\tilde\omega\tilde k^2).\label{778}\ee
The equation for $\psi_0$ is
\be r^2f_0\psi_0'=c_0\label{779}\ee
for which regularity at the horizon implies $c_0=0$, hence $\psi_0$ is constant.
Using this fact we obtain the following equations
\begin{subequations}\label{780}
	\begin{align}
	&(\r^2f_0\psi_1')'+{i\psi_0\over f_0'(1)}(\r^2f_0')'=0\label{p1}\\
	&(\r^2f_0\psi_2')'+\rho \psi_0 B_1=0\sp B_1={\rho(-12-3\tilde q^2+4\rho\tilde q^2)\over 12+3\tilde q^2}\label{p2}.
	\end{align}
\end{subequations}
From (\ref{p1}) we find
\be \psi_1=i\psi_0\int_1^\rho {1\over f_0}\left({1\over\r^2}-{f_0'\over f_0'(1)}\right)\equiv i\psi_0 P_1(\rho).\label{781}\ee
From (\ref{p2}) we find
\be \psi_2= -\psi_0 P_2(\rho)\sp P_2(\rho)={4(\r-1)\over (12+3\tilde q^2)\r}.\label{782}\ee

Now we find a perturbative solution for $\phi_-$. We first use the transformation
\be \phi_-=gY\label{783}\ee
with
\be g={1\over\r}-{4\tilde q^2\over 3(4+\tilde q^2)}\label{784}\ee
so that the coefficient of $Y$ vanishes at the limit $\tilde\omega\to 0\sp\tilde k\to 0$.
We use
\be Y=X f^{i\omega\over f'(1)}\label{785}\ee
to remove the leading behavior at the horizon. We require that $X$ is regular at the horizon and expand it as follows
\be X=X_0+\tilde\omega X_1+\tilde k^2X_2+\mathcal{O}(\tilde\omega^2,\tilde k^4,\tilde\omega\tilde k^2).\label{786}\ee
For $X_0$ we find
\be \r^2f_0g^2X_0'=c\label{787}\ee
which, by regularity at the horizon, implies that $X_0$ is constant. Using this fact we obtain the following equations
\begin{subequations}\label{788}
	\begin{align}
	&(\r^2f_0g^2X_1')'+{iX_0\over f_0'(1)}(\r^2f_0'g^2)'=0\label{x1}\\
	&(\r^2f_0g^2X_2')'+\rho X_0 B_2=0\sp B_2={2\rho\tilde q^2(36+\tilde q^2)(\tilde q^2(4\r-3)-12)\over 27(4+\tilde q^2)^3}\label{x2}.
	\end{align}
\end{subequations}
From (\ref{x1}) we obtain
\be X_1=iX_0\int_0^\rho \left({g(1)^2\over \r^2f_0g^2}-{f_0'\over f_0f_0'(1)}\right)\equiv iX_0 Q_1(\rho).\label{789}\ee
From (\ref{x2}) we find
\be X_2= -X_0 Q_2(\rho)\sp Q_2(\rho)={8\tilde q^2(36+\tilde q^2)\over 3(4+\tilde q^2)(-12+\tilde q^2(-3+4\r))^2}.\label{790}\ee

We now need to fix the integration constants $X_0,\psi_0$ in terms of the boundary values
\be a_x^{(0)}=a_x(0)\sp k_x^{(0)}=k_x(0)\sp z_x^{(0)}=z_x(0).\label{791}\ee
The system (\ref{axfluc}) implies the equation
\be f\r^2\left(\rho^{-2}z_x'+\tilde q a_x\right)'+{\tilde k^2 z_x}+{i\tilde k\tilde \omega k_x}=0.\label{zpp}\ee
Using (\ref{phipm}) and (\ref{zpp}) we find
\be f\left(\r^2\phi_\pm'-C_\pm a_x'\r+C_\pm a_x\right)=\tilde k^2z_x+i\tilde k\tilde\omega k_x.\label{792}\ee
Near the boundary we obtain
\be \lim\limits_{\r\to 0}(\r^2\phi_\pm')=-C_\pm a_x^{(0)}+\tilde k^2 z_x^{(0)}+i\tilde k\tilde\omega k_x^{(0)}.\label{nbeq}\ee
The expansion of $\phi_\pm$ near the boundary is\footnote{there are no logarithms in the expansion; one can check from the solution that $\phi_\pm'$ do not contain any $1/\rho$ terms.}
\be \phi_\pm=-{W_\pm\over \r}+D_\pm+\cdots\label{793}\ee
where
\begin{subequations}\label{WD}
	\begin{align}
	&W_+=\psi_0(i\tilde\omega- {4\tilde k^2\over 12+3\tilde q^2})+\mathcal{O}(\tilde\omega^2,\tilde\omega\tilde k^2,\tilde k^4)\label{794}\\
	&W_-=-X_0+\mathcal{O}(\tilde\omega^2,\tilde\omega\tilde k^2,\tilde k^4)\label{795}\\
	& D_+=\psi_0+\mathcal{O}(\tilde\omega,\tilde k^2)\label{796}\\
	&D_-=X_0\left(-{4\tilde q^2\over 12+3\tilde q^2}+i\tilde\omega {(\tilde q^2-12)^2\over (12+3\tilde q^2)^2}-\tilde k^2 {8\tilde q^2(36+\tilde q^2)\over (12+3\tilde q^2)^3}\right)+\mathcal{O}(\tilde\omega^2,\tilde\omega\tilde k^2,\tilde k^4).\label{797}
	\end{align}
\end{subequations}
Then (\ref{nbeq}) implies
\be W_\pm=-C_\pm a_x^{(0)}+\tilde k^2 z_x^{(0)}+i\tilde k\tilde \omega k_x^{(0)}+\mathcal{O}(\tilde\omega^2,\tilde\omega\tilde k^2,\tilde k^4),\label{798}\ee
therefore
\be \psi_0={-C_+ a_x^{(0)}+\tilde k^2 z_x^{(0)}+i\tilde k\tilde \omega k_x^{(0)}\over i\tilde\omega-\tilde k^2 {4\over 12+3\tilde q^2}}+\mathcal{O}(\tilde\omega^2,\tilde\omega\tilde k^2,\tilde k^4)\label{799}\ee
\be X_0={C_- a_x^{(0)}-\tilde k^2 z_x^{(0)}-i\tilde k\tilde \omega k_x^{(0)} }+\mathcal{O}(\tilde\omega^2,\tilde\omega\tilde k^2,\tilde k^4).\label{7100}\ee
From (\ref{phipm}) we can solve for $a_x$:
\be a_x=\r{\phi_--\phi_+\over C_--C_+}\label{7101}\ee
which implies
\be a_x'(0)={D_--D_+\over C_--C_+},\label{7102}\ee
where $D_\pm$ are given in (\ref{WD}). Using also (\ref{cpm}), we obtain the terms relevant to the electric conductivity
\be \delta a_x'(0)/\delta a_x^{(0)}=i\omega {(12-\tilde q^2)^2\over (12+3\tilde q^2)^2}-{4i\omega\tilde q^2\over (12+3\tilde q^2)i\omega -4\tilde k^2}+\mathcal{O}(\tilde\omega^2,\tilde\omega\tilde k^2,\tilde k^4)\label{7103}\ee
hence
\boxedeq{hydrcax}{\sigma(\tilde\omega)= {Z_0{\tilde q^2\over\tilde k^2}\over 1 -{(12+3\tilde q^2)\over 4\tilde k^2}i\tilde\omega}+Z_0{(12-\tilde q^2)^2\over (12+3\tilde q^2)^2}.}
The pole has moved away from the real axis; it is now located at
\be \omega_0=-i{4\tilde k^2\over 12+3\tilde q^2}.\label{7104}\ee
The expression (\ref{hydrcax}) is identical to the Drude conductivity (\ref{drude}) offset by a constant, due to pair-creation. If $\tilde k<<\tilde q$ (coherent regime \cite{DG},\cite{Kim:2014bza}), the Drude term dominates
\be \sigma(\tilde\omega)={\sigma_{DC}\over 1-i\omega\tau},\label{7105}\ee
where
\be \sigma_{DC}=Z_0{\tilde q^2\over \tilde k^2}+\mathcal{O}(\tilde k^0)\sp \tau=r_h{12+3\tilde q^2\over 4\tilde k^2}+\mathcal{O}(\tilde k^0).\label{7106}\ee
In this limit, the first term in (\ref{dccond}) is negligible, and the DC conductivity agrees with the expression above. The relaxation time $\tau$ is inversely proportional to $\tilde k^2$, which makes sense; as $\tilde k\to 0$, the momentum dissipation becomes weak, hence the particles collide less often with impurities in the material. For $\tilde q>\tilde k$ (incoherent regime \cite{DG},\cite{Kim:2014bza}) there is a deviation from the standard Drude peak. For $\tilde k>>\tilde q$ the constant term in (\ref{hydrcax}) dominates and the curve is approximately flat at small $\tilde\omega$.

\subsubsection{Numerical Solution}

We now find the full $\tilde\omega,\tilde k$ dependence of the conductivity numerically.
We have a system of $3$ linear coupled equations (\ref{axfluc}). Counting the derivatives we see that we need $5$ boundary conditions to specify a solution.  From (\ref{axfluc}) we find that the leading behaviors at the horizon $\rho=1$, such that the in-going boundary conditions are satisfied, are
\be a_x(\rho)=f(\rho)^{i\tilde{\omega}/f'(1)}(a_0+a_1(1-\r)+\cdots)\label{7107}\ee
\be z_x(\rho)=f(\rho)^{1+i\tilde{\omega}/f'(1)}(z_0+z_1(1-\r)+\cdots)\label{7108}\ee
\be k_x(\rho)=f(\rho)^{i\tilde{\omega}/f'(1)}(k_0+k_1(1-\r)+\cdots)\label{7109}.\ee
This is equivalent to setting $3$ boundary conditions. In addition, we find the relation
\be z_0=\frac{4 (\tilde k k_0-\tilde qa_0 )}{2 \tilde k^2+\tilde q^2+4 i \tilde\omega -12}.\label{7110}\ee
The constants $k_0,a_0$ are free and all the subleading terms are fixed in terms of them.
Therefore to build a solution from the horizon we need $2$ boundary conditions, in addition to the regularity condition.

Expanding the fields near the boundary, $\r\to 0$, and substituting into the fluctuation equations, (\ref{axfluc}), we find
\be a_x(\rho)=a^{(0)}+a^{(1)}\rho+\cdots\label{7111}\ee
\be z_x(\rho)=z^{(0)}-\frac{i\tilde k (\tilde\omega k^{(0)}  -i\tilde k z^{(0)} )}{2} \r^2+z^{(3)}\r^3+\cdots\label{7112}\ee
\be k_x(\rho)=k^{(0)}+\frac{\tilde\omega  (\tilde\omega k^{(0)}  -i \tilde k z^{(0)})}{2} \r^2+k^{(3)}\r^3+\cdots\label{7113}\ee
with the constraint coming from (\ref{axcons})
\be k^{(3)}={i\tilde\omega\over 3\tilde k}qa^{(0)}+{i\tilde\omega\over\tilde k}z^{(3)}\label{axconst}.\ee
The coefficients $a^{(0)},z^{(0)},k^{(0)}$ and $a^{(1)},z^{(3)},k^{(3)}$ are the sources and responses respectively. Due to the constraint (\ref{axconst}) there are $5$ independent parameters at the boundary.

Setting the in-going boundary conditions at the horizon, the arbitrary constants above are fixed. Since the equations are linear, each response is a linear combination of the sources. For example, the current $a_x^{(1)}$ has the form
\be a_x^{(1)}=\lambda_1 a_x^{(0)}+\lambda_2 z_x^{(0)}+\lambda_3 k_x^{(0)}.\label{7114}\ee
In order to compute the conductivity, we need the coefficient of the first term in the above expression.

To solve the equations numerically it is much more convenient to set all the boundary conditions at the horizon. Then the sources $a^{(0)},z^{(0)},k^{(0)}$ as well as the responses $a^{(1)},z^{(3)},k^{(3)}$ are completely determined.
The numerical method described in \cite{Kim:2014bza} allows us to extract the correlators from numerical solutions built from the horizon, by exploiting the linearity of the equations. The basic idea is that one solves the equations numerically with different sets of boundary conditions at the horizon, in order to obtain $3$ linearly independent solutions. Then, given an arbitrary set of sources $a^{(0)},z^{(0)},k^{(0)}$, one can find a linear combination of solutions with the given boundary values. The response of the system is the corresponding linear combinations of the responses. For systems with constraints, such as (\ref{axfluc}), solutions along the gauge orbit are used to obtain an invertible system of equations.

Let
\be J^a=\begin{pmatrix}
	a_x^{(0)}\\ z_x^{(0)}\\ k_x^{(0)}
\label{7115}\end{pmatrix}\ee
be an arbitrary set of sources and let
\be \Phi^a(\r)=\begin{pmatrix}
	a_x(\r)\\ z_x(\r)\\ k_x(\r)
\end{pmatrix}\label{7116}\ee
be the solution  with the ingoing boundary conditions at the horizon that satisfies $\Phi^a(0)=J^a$. Since there are only two free parameters at the horizon, we can solve the system of equations (\ref{axbh}) for only two sets of linearly independent boundary conditions. Let
\be \begin{split} a_0=1\sp k_0=~~1\\ a_0=1\sp k_0=-1 \end{split}\label{axBC}\ee
be the two sets
and let $\Phi_{\;1}^a(r),\Phi_{\;2}^a(r)$ be the corresponding solutions. The system (\ref{axbh}) also has a constant solution
\be \Phi_{\;3}^a(\r)=\begin{pmatrix}
	0\\
	1\\
	{i\tilde k\over\tilde\omega}
\end{pmatrix}.\label{7117}\ee
This solution is related to the residual gauge freedom in equations (\ref{axbh}); if $\Phi^a$ is a solution, then $\Phi^a+\lambda\Phi_{\;3}^a$ is also a solution, where $\lambda$ is an arbitrary constant. Therefore $\Phi_{\;3}^a$ is a vector along the gauge orbit and is not a true degree of freedom. However it can be used to obtain a basis $\{\Phi_{\;1}^a(\r),\Phi_{\;2}^a(\r),\Phi_{\;3}^a(\r)\}$ for the space of solutions, and thus, the full solution $\Phi^a(\r)$ can be written as a unique linear combination
\be \Phi^a(\r)=\Phi^a_{\;i}(\r)c^i.\label{7118}\ee
The coefficients $c^i$ can be determined by the boundary values
\be \Phi^a(0)=J^a=\Phi^a_{\;i}(0)c^i\label{7119}\ee
hence
\be c^i=(\Phi(0)^{-1})^i_{\;a} J^a\label{7120}\ee
and the full solution can be written in terms of the numeric solutions
\be \Phi^a(\r)=\Phi^a_{\;i}(\r)(\Phi(0)^{-1})^i_{\;b} J^b.\label{7121}\ee
From the above we can calculate (for the index $a=1$)
\be  a_x'(0)=\partial_\r \Phi^1_{\;i}(\r)(\Phi(0)^{-1})^i_{\;b} J^b\bigg|_{\r= 0}\label{7122}\ee
and thus
\be {\delta a_x'(0)\over \delta a_x^{(0)}}=\partial_\r \Phi^1_{\;i}(r)(\Phi(0)^{-1})^i_{\;1}\bigg|_{\r= 0},\label{7123}\ee
therefore, from the on-shell action (\ref{onshell}) we find that
the longitudinal conductivity is
\be \sigma_{xx}=\sigma_{yy}=Z_0{1\over i\omega}f(\rho)\partial_\r \Phi^1_{\;i}(\r)(\Phi(0)^{-1})^i_{\;1})\bigg|_{\r= 0}.\label{condf}\ee
Since there is no mixing of the $x,y$ components of the fields in the equations, the off-diagonal part of the conductivity is the same as in the neutral case
\be \sigma_{xy}=-\sigma_{yx}= W_0.\label{7124}\ee

All that is left now is to find the two numerical solutions $\Phi^1(\r),\Phi^2(\r)$. As in previous sections, we use
\be a_x(\rho)=f(\rho)^{i\tilde{\omega}/f'(1)}\tilde{a}_x(\rho)\label{7125}\ee
\be k_x(\rho)=f(\rho)^{i\tilde{\omega}/f'(1)}\tilde{k}_x(\rho)\label{7126}\ee
\be z_x(\rho)=f(\rho)^{1+i\tilde{\omega}/f'(1)}\tilde{z}_x(\rho)\label{7127}\ee
to remove the leading behavior near the horizon and obtain a system of equations for $\tilde{k}_x,\tilde{a}_x,\tilde{z}_x$. We set the boundary conditions a small distance from the horizon $\rho=1-\epsilon$ and solve the above equations numerically for various values of the dimensionless parameters $\tilde{\omega},\tilde{k},\tilde{q}$.

\begin{figure}[H]
	\centering
	\includegraphics[width=0.49\textwidth]{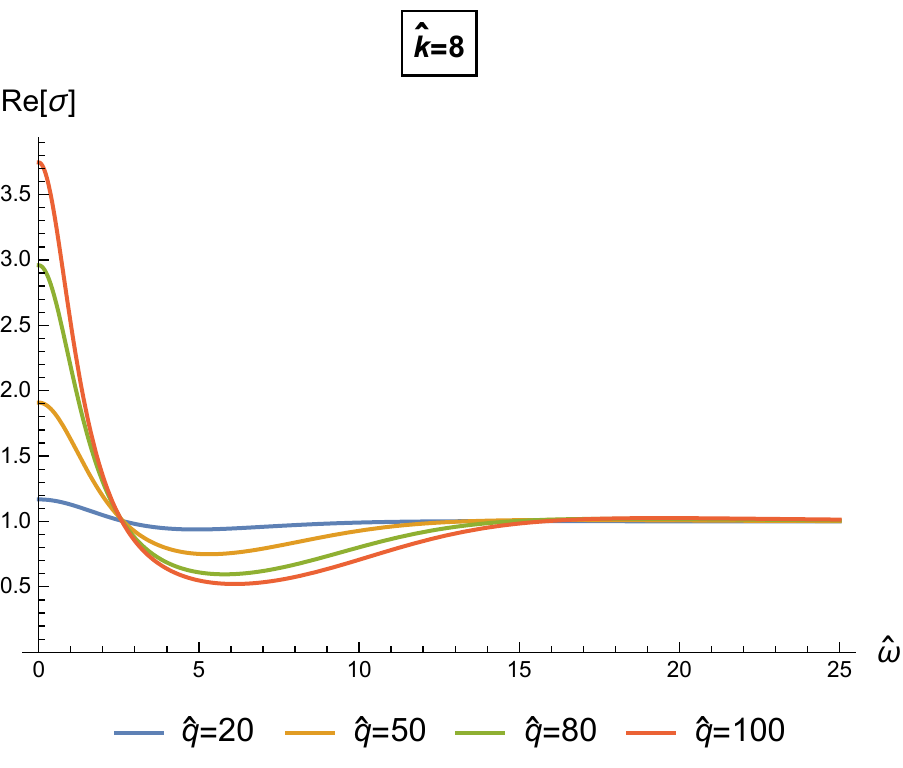}
	\includegraphics[width=0.49\textwidth]{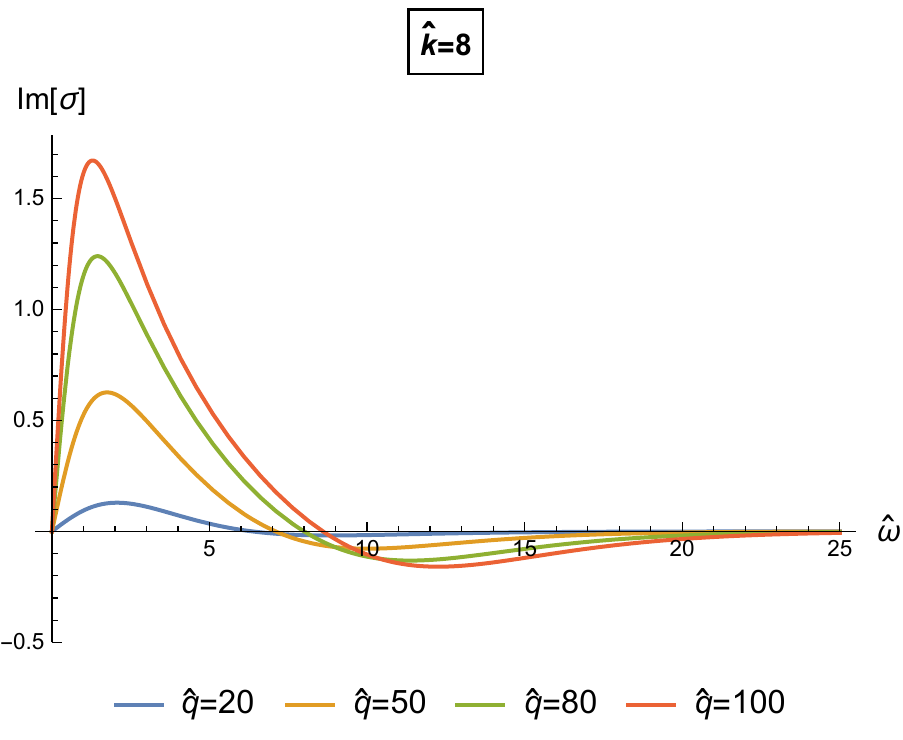}
	\caption{\small The real and imaginary parts of the AC conductivity are plotted against the dimensionless frequency $\hat{\omega}=\omega/T$ for various values of the dimensionless charge density $\hat{q}=(L\sqrt{Z_0})^{-1}q/T^2$, while the dimensionless momentum relaxation parameter is fixed at $\hat k=\sqrt{\Psi_0}k/T=8$. The vertical axes have units of $Z_0$. The pole that appeared in the translationally symmetric background of section \ref{chnum} has moved away from the real axis and the delta-function in the real part is now spread out into a peak around $\hat\omega=0$.}\label{fig2}
\end{figure}

For the plots we use the following dimensionless units
\be \hat\omega={\omega\over T}\sp  \hat k=\sqrt{\Psi_0}{k\over T}\sp \hat q={q\over T^2L\sqrt{Z_0}}. \label{redefax2}\ee
We keep $\hat k,\hat q$ fixed and plot the conductivity against $\hat\omega$. In figure \ref{fig2} we show the dependence of the conductivity on $\hat q$. The DC conductivity no longer diverges; the pole at $\hat \omega=0$ has moved away from the real $\hat\omega$ axis.

\begin{figure}[H]
	\centering
	\includegraphics[width=0.49\textwidth]{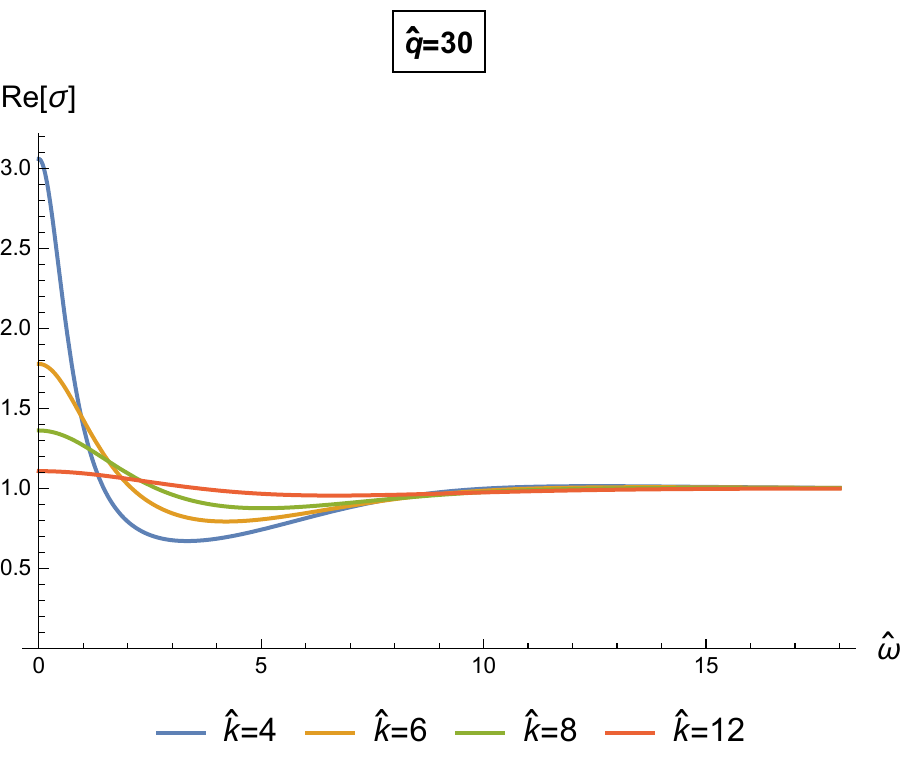}
	\includegraphics[width=0.49\textwidth]{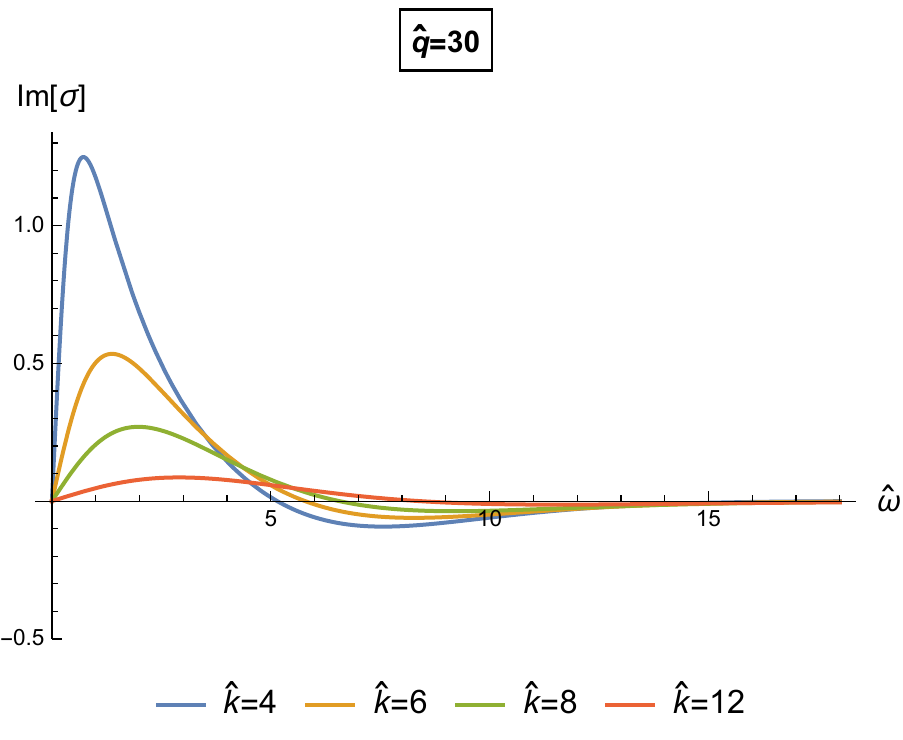}
	\caption{\small The real and imaginary parts of the AC conductivity are plotted against the dimensionless frequency $\hat{\omega}=\omega/T$ for various values of the dimensionless momentum relaxation parameter $\hat k=\sqrt{\Psi_0}k/T$, while the dimensionless charge density is fixed $\hat{q}=(L\sqrt{Z_0})^{-1}q/T^2=30$. The vertical axes have units of $Z_0$. As $\hat k$ increases, the peak in the real part becomes shorter and wider.}\label{fig3}
\end{figure}

In figure \ref{fig3} we show the dependence of the conductivity on the momentum relaxation parameter $\hat k$. As $\hat k$ increases, the pole moves further away from the real axis, causing the curves to become flatter.

\subsection{Dyonic AdS-RN Black Hole with Axions\label{dynum}}

In this final case we also turn on a magnetic field.
This background is characterized by a finite charge density $q$, magnetic field $h$, finite temperature $T$ and momentum relaxation scale $k$.
The AdS-RN dyonic black hole solution to (\ref{actionax}), modified by the axions, is
\be ds^2=L^2r^{-2}\left(-f(r)dt^2+{dr^2\over f(r)}+d\vec{x}^2\right)\sp A_t(r)=\mu+{Q_0\over Z_0}r\sp\phi(r)=\phi_0\sp \chi_i=kx_i\label{n1}\ee
with \be \begin{split} Q_0=q-hW_0\sp W(\phi_0)=W_0\sp Z(\phi_0)=Z_0\sp \Psi(\phi_0)=\Psi_0\\
	L^2=6/V(\phi_0)\sp \partial_\phi V=\partial_\phi\Psi=0\sp 2hQZ_0\partial_\phi W=(Q^2-h^2Z_0^2)\partial_\phi Z\end{split}\label{n2}\ee
and blackening factor
\be f(r)=1+{Q_0^2+h^2Z_0^2\over 4 Z_0L^2} r^4-{1\over 2}k^2\Psi_0r^2-mr^3.\label{n3}\ee
This solution for $W=0$ has been studied in \cite{Kim:2015wba}.
The polynomial (\ref{n3}) has the same form as in the previous section \ref{axnum} and the root analysis is identical.
We write it in the more convenient form
\be f(r)=\left(1-{r\over r_h}\right)\left(1+{r\over r_h}+{r^2\over r_h^2}-{1\over 2}k^2\Psi_0r^2-{1\over 4}{Q_0^2+h^2Z_0^2\over Z_0 L^2}r_hr^3\right), \label{n4}\ee
where $r_h$ is the location of the outer horizon.
The temperature is
\be T={1\over 4\pi r_h}\left(3-{1\over 2}\Psi_0k^2r_h^2-{Q_0^2+h^2Z_0^2\over 4L^2Z_0}r_h^4\right)\label{tempmagax},\ee
and
the black hole mass in terms of the outer horizon $r_h$ is
\be m=r_h^{-3}\left(1-{1\over 2}k^2\Psi_0r_h^2+{Q_0^2+h^2Z_0^2\over 4Z_0L^2}r_h^4\right)\label{bhm2}.\ee

\subsubsection{Fluctuation Equations}
In order to calculate the electric AC conductivity, we use the same perturbation ansatz as in section \ref{axnum}
\be  \delta A_i=M\int {\ud\omega\over 2\pi} a_i (r)e^{-i\omega t}\sp \delta \chi_i=M\int {\ud\omega\over 2\pi} k_i(r)e^{-i\omega t}\sp
\delta g_{ti}=M{L^2\over r^2}\int {\ud\omega\over 2\pi} z_i(r)e^{-i\omega t}.\label{n5}\ee
This time the equations mix the $x$ and $y$ components of the gauge field.
We have $6$ independent fluctuation equations:
\begin{subequations}\label{n6}
	\begin{align}
	& f\left(Z_0f a_i'+Q_0 z_i\right)'+\omega^2{Z_0} a_i+i\omega{hZ_0}\epsilon^{ij} z_j=0\label{n7}\\
	& kL^2r^{-2}\Psi_0 fk_i'-i\omega L^2r^{-2}{ { z}_i}'-i\omega{Q_0} a_i+{hQ_0} \epsilon^{ij}z_j+hfZ_0 \epsilon^{ij}a_j'=0\label{n8}\\
	& r^2\left(r^{-2}\Psi_0 f k_i'\right)'-i\omega k{\Psi_0\over f}{z}_i+\omega^2{\Psi_0\over f} k_i=0\sp i,j=x,y.\label{n9}
	\end{align}
\end{subequations}
We define the dimensionless parameters
\be \tilde\omega=\omega r_h\sp \tilde k=kr_h\sqrt{\Psi_0}\sp \tilde q={qr_h^2\over L\sqrt{Z_0}}\sp \tilde h={h\sqrt{Z_0}r_h^2\over L}\sp \tilde Q=\tilde q-\lambda\tilde h\sp \lambda={W_0\over Z_0} \label{n10}\ee
and rescale the radial coordinate and the fields
\be \rho=r/r_h\sp z_x\to {\sqrt{Z_0}r_h\over L} z_x\sp k_x\to {r_h\over L}\sqrt{Z_0\over \Psi_0} k_x\label{n11}\ee
to obtain
\begin{subequations}\label{magfluc}
	\begin{align}
	&f\left(f a_i'+\tilde{Q} z_i\right)'+\tilde\omega^2 a_i+i\tilde\omega{\tilde h}\epsilon^{ij} z_j=0\label{magaxbh}\\
	&\tilde k\r^{-2}fk_i'-i\tilde\omega\r^{-2}{ { z}_i}'-i\tilde\omega\tilde{Q} a_i+{\tilde h\tilde Q} \epsilon^{ij}z_j+\tilde hf\epsilon^{ij}a_j'=0\label{magaxc}\\
	& f\r^2\left(\r^{-2} f k_i'\right)'-i\tilde\omega \tilde k{z}_i+\tilde\omega^2 k_i=0\sp i,j=x,y\label{magaxk}
	\end{align}
\end{subequations}
where primes now denote derivatives with respect to $\rho$ and
\be f(\r)=\left(1-\r\right)\left(1+\r+\r^2-{1\over 2}\tilde k^2\r^2-{1\over 4}(\tilde Q^2+\tilde h^2)\r^3\right).\label{n12}\ee

\subsubsection{Numerical Solution}

The system (\ref{magfluc}) contains $4$ second order and $2$ first order equations, hence it needs $10$ boundary conditions.
The ingoing condition on the $6$ fields at the horizon $\rho\to 1$
\begin{subequations}
	\begin{align}
	& a_i(\rho)=f(\rho)^{i\tilde{\omega}/f'(1)}(a_0^i+a_1^i(1-\r)+\cdots)\label{nn1}\\
	& z_i(\rho)=f(\rho)^{1+i\tilde{\omega}/f'(1)}(z_0^i+z_1^i(1-\r)+\cdots)\label{nn2}\\
	& k_i(\rho)=f(\rho)^{i\tilde{\omega}/f'(1)}(k_0^i+k_1^i(1-\r)+\cdots)\sp i=x,y\label{nn3}
	\end{align}
\end{subequations}
is equivalent to $6$ boundary conditions. Therefore only $4$ of the leading terms $a_0^i,z_0^i,k_0^i$ can be independent. Indeed from (\ref{magfluc}) we find
\be k_0^i={1 \over\tilde k}\left({\tilde Q a_0^i-\tilde h\epsilon^{ij}a_0^j+(f'(1)+i\tilde\omega)z_0^i}\right)\sp i,j=x,y\label{n13}\ee
and all the subleading terms are fixed in terms of $a_0^i,z_0^i$.

From (\ref{magfluc}) we find the following behavior near the boundary $\r\to 0$
\be a_i(\rho)=a^{(0)}_i+a^{(1)}_i\rho+\cdots\label{n14}\ee
\be z_i(\rho)=z^{(0)}_i-\frac{i\tilde k (\tilde\omega k^{(0)}_i  -i\tilde k z^{(0)}_i )}{2} \r^2+z^{(3)}_i\r^3+\cdots\label{n15}\ee
\be k_i(\rho)=k^{(0)}_i+\frac{\tilde\omega  (\tilde\omega k^{(0)}_i  -i \tilde k z^{(0)}_i)}{2} \r^2+k^{(3)}_i\r^3+\cdots\label{n16}\ee
where $a^{(0)}_i,z^{(0)}_i,k^{(0)}_i$ are the $6$ sources and $a^{(1)}_i,z^{(3)}_i,k^{(3)}_i$ are the $6$ responses. Since we need only $10$ boundary conditions, not all of them can be independent. Indeed, from (\ref{magfluc}), we also find the two relations
\be k_i^{(3)}={i\tilde\omega\over\tilde k}\left({1\over 3}\tilde Q a_i^{(0)}+z_i^{(3)}\right)-{h\over 3k}\epsilon^{ij}\left(a_j^{(1)}+\tilde Q z_j^{(0)}\right)\sp i,j=x,y.\label{n17}\ee

Similar to section \ref{axnum}, let
\be J^a=\begin{pmatrix}
	a_x^{(0)}\\ a_y^{(0)}\\z_x^{(0)}\\ z_y^{(0)}\\k_x^{(0)}\\k_y^{(0)}
\end{pmatrix}\sp \Phi^a(\r)=\begin{pmatrix}
	a_x(\r)\\ a_y(\r)\\z_x(\r)\\z_y(\r)\\ k_x(\r)\\ k_y(\r)
\end{pmatrix}\label{n18}\ee
be an arbitrary set of sources and
the solution with ingoing boundary conditions at the horizon which satisfies $\Phi^a(0)=J^a$. We can solve the system of equations (\ref{magfluc}) for four sets of boundary conditions at the horizon. We cannot obtain more than $4$ linearly independent solutions, since the six values $a_0^i,k_0^i,z_0^i$ have two constraints. Let
\be\begin{matrix} &a_0^x=1,& a_0^y=~~1,& k_0^x=~~1,& k_0^y=~~1\\
	&a_0^x=1,& a_0^y=-1,& k_0^x=~~1,& k_0^y=~~1\\
	&a_0^x=1,& a_0^y=~~1,& k_0^x=-1,& k_0^y=~~1\\
	&a_0^x=1,& a_0^y=~~1,& k_0^x=~~1,& k_0^y=-1\label{magaxBC}\end{matrix}\ee
be the four sets of boundary conditions
and let $\Phi_{\;1}^a(r),\dots, \Phi_{\;4}^a(r)$ be the corresponding solutions. In addition, there is a constant solution to the system (\ref{magaxbh})
\be a_i=-\epsilon^{ij}{ih\over \omega}z_j\sp k_i={ik\over \omega }z_i\label{n19}\ee
which spans a two-dimensional vector space and can provide the two extra linearly independent vectors that we need
\be \Phi_{\;5}^a(\r)=\begin{pmatrix}
	-{i\tilde h\over\tilde\omega}\\{i\tilde h\over\tilde \omega}\\
	1\\1\\
	{i\tilde k\over\tilde\omega}\\{i\tilde k\over\tilde\omega}
\end{pmatrix}\sp \Phi_{\;6}^a(\r)=\begin{pmatrix}
	{i\tilde h\over\tilde\omega}\\{i\tilde h\over\tilde \omega}\\
	1\\-1\\
	{i\tilde k\over\tilde\omega}\\-{i\tilde k\over\tilde\omega}
\end{pmatrix}.\label{n20}\ee
These are vectors along the gauge orbit and not actual degrees of freedom, \cite{Kim:2015wba}.
Now the vectors $\Phi_{\;1}^a(\r),\dots,\Phi_{\;6}^a(\r)$ form a basis for the space of solutions, and the full solution $\Phi^a(\r)$ can be written as a unique linear combination
\be \Phi^a(\r)=\Phi^a_{\;i}(\r)c^i.\label{n21}\ee
The coefficients $c^i$ can be determined by the boundary values
\be \Phi^a(0)=J^a=\Phi^a_{\;i}(0)c^i\label{n22}\ee
hence
\be c^i=(\Phi(0)^{-1})^i_{\;a} J^a\label{n23}\ee
and the full solution can be written in terms of the numeric solutions
\be \Phi^a(\r)=\Phi^a_{\;i}(\r)(\Phi(0)^{-1})^i_{\;b} J^b.\label{n24}\ee
From the above we can calculate (for the indices $i=1,2$)
\be  a_i'(0)=\partial_\r \Phi^i_{\;l}(\r)(\Phi(0)^{-1})^l_{\;b} J^b\bigg|_{\r= 0}.\label{n25}\ee
Set
\be \mathcal{B}^{ij}(\r)=\partial_\r \Phi^i_{\;l}(\r)(\Phi(0)^{-1})^{lj},\label{Bmat}\ee
then the part of the on-shell action (\ref{onshell}) that is quadratic in the gauge field fluctuations is
\be  S_{AA}=-M^2V_2{1\over 2}\int {\ud \tilde\omega\over 2\pi} a_i(\tilde\omega)\left(i\tilde\omega W_0\epsilon^{ij}+Z_0f(\rho)\mathcal{B}^{ji}(\r)\right)a_j(-\tilde\omega)\bigg|_{r\to 0}\label{n26} \ee
from which we can obtain the electric conductivity
\be \sigma_{i j}=(\epsilon_{ij}W_0+{Z_0\over i\omega}f(\rho)\mathcal{B}_{ji}(\r))\bigg|_{\r= 0}\label{n27}.\ee
The off-diagonal part of the conductivity is
\be \sigma_{xy}=-\sigma_{yx}=Z_0(\lambda+{1\over i\omega}f(\rho)\mathcal{B}_{21}(\r))\bigg|_{\r= 0}\label{condxy}\ee
and the longitudinal part is
\be \sigma_{xx}=\sigma_{yy}=Z_0{1\over i\omega}f(\rho)\mathcal{B}_{11}(\r))\bigg|_{\r= 0}\label{condxx}\ee

All that is left now is to find the numerical solutions $\Phi^1(\r),\dots,\Phi^4(\r)$. We use
\be a_i(\rho)=f(\rho)^{i\tilde{\omega}/f'(1)}\tilde{a}_i(\rho)\label{n28}\ee
\be k_i(\rho)=f(\rho)^{i\tilde{\omega}/f'(1)}\tilde{k}_i(\rho)\label{n29}\ee
\be z_i(\rho)=f(\rho)^{1+i\tilde{\omega}/f'(1)}\tilde{z}_i(\rho)\label{n30}\ee
to remove the leading behavior near the horizon. We find four sets of solutions with boundary conditions given in (\ref{magaxBC}), calculate the matrix $\mathcal{B}$ from (\ref{Bmat}) and from (\ref{condxx}), (\ref{condxy}) we find the conductivities.

For the plots we use the following dimensionless units
\be \hat\omega={\omega\over T}\sp  \hat k=\sqrt{\Psi_0}{k\over T}\sp \hat q={q\over T^2L\sqrt{Z_0}}\sp \hat h={h\sqrt{Z_0}\over T^2L}\sp \hat Q=\hat q-\lambda\hat h\sp \lambda={W_0\over Z_0}. \label{redefax3}\ee
The set of independent dimensionless parameters is chosen to be $\hat\omega,\hat k,\hat h,\hat Q,\lambda$. For all the plots below we set $\lambda=0.2$. Changing $\lambda$ offsets $Re(\sigma_{xy})$ by the value of $\lambda$, as well as changing the charge density $\hat q$ according to (\ref{redefax3}).

\begin{figure}[H]
	\centering
	\includegraphics[width=0.49\textwidth]{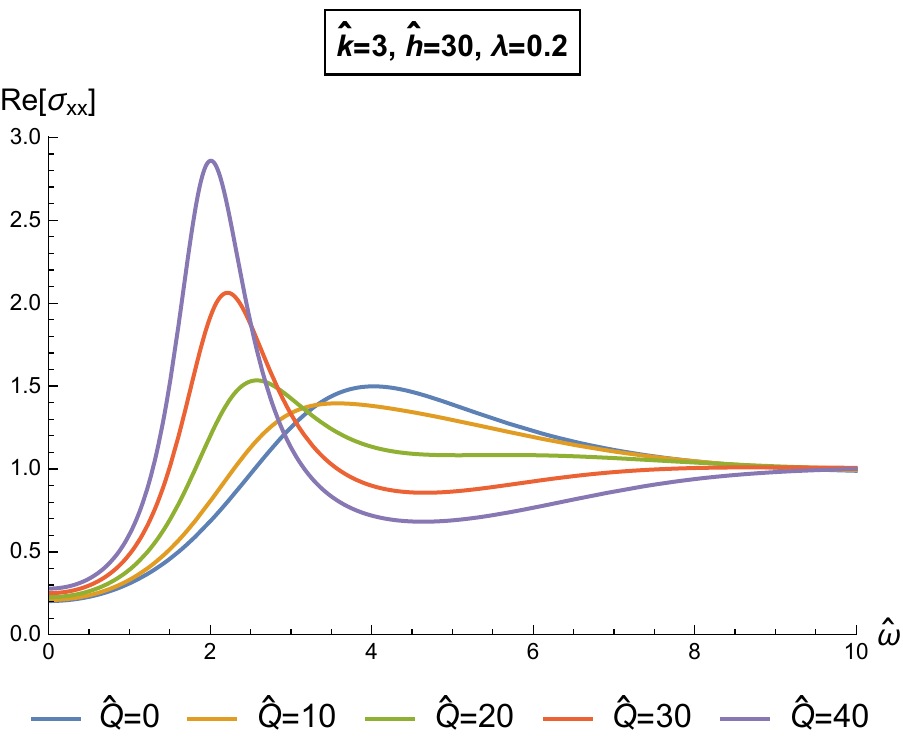}
	\includegraphics[width=0.49\textwidth]{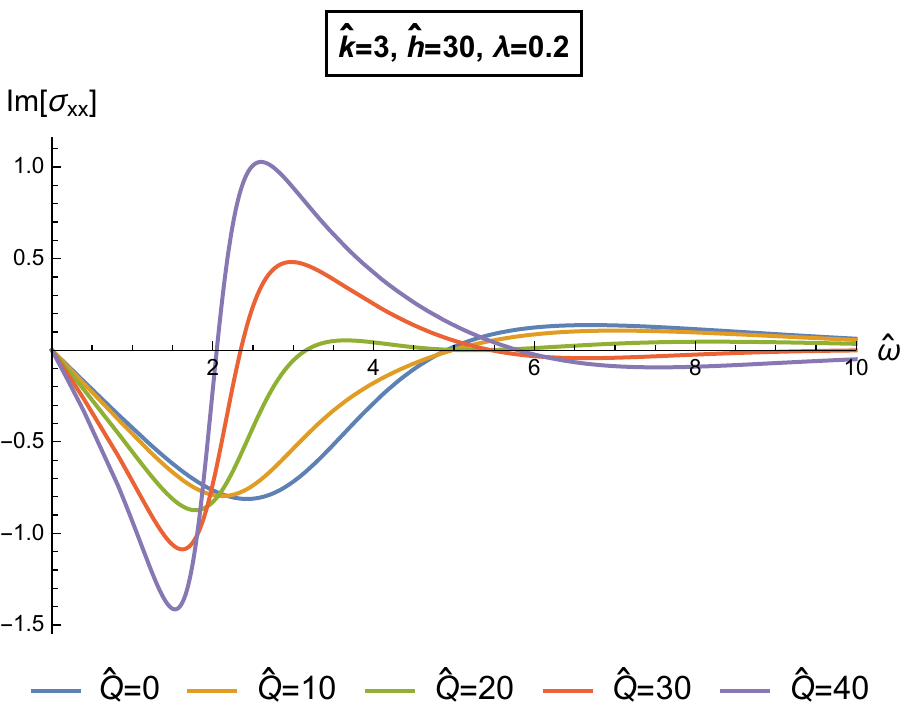}
	\includegraphics[width=0.49\textwidth]{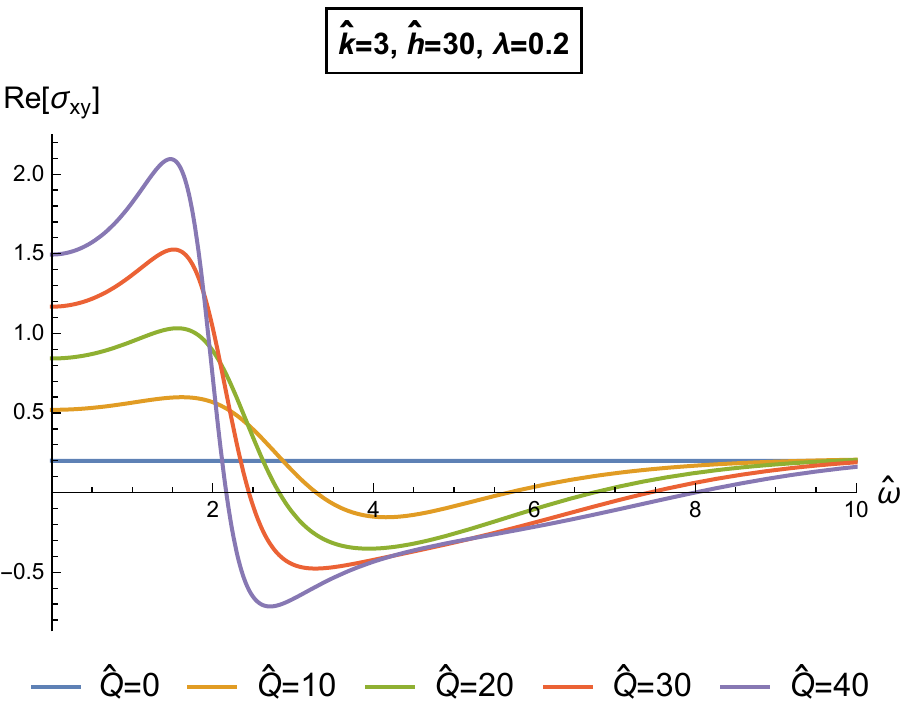}
	\includegraphics[width=0.49\textwidth]{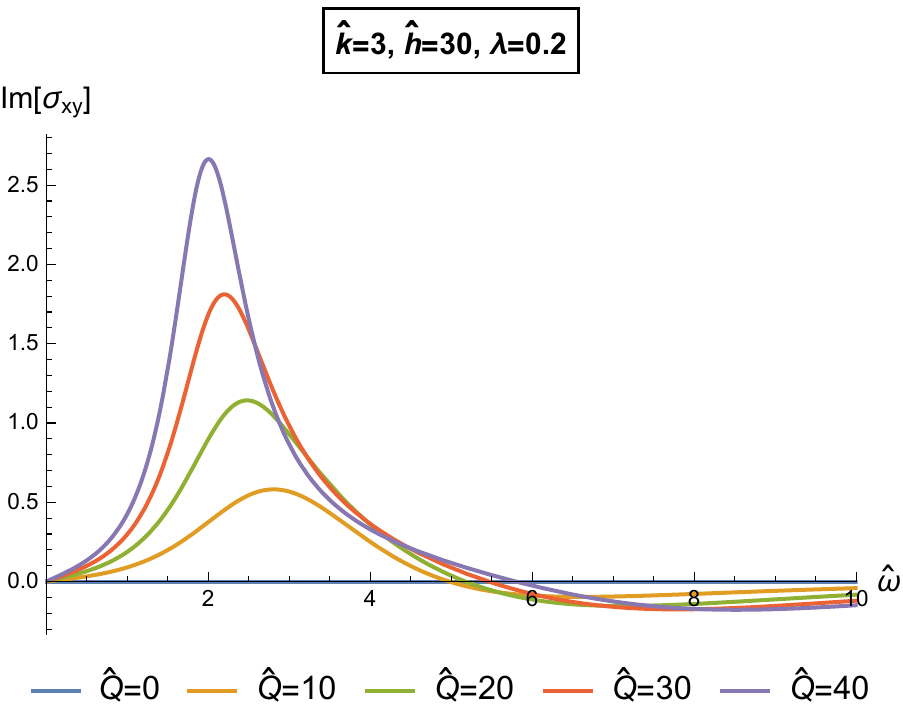}
	\caption{\small  The real and imaginary parts of the longitudinal (top) and Hall (bottom) conductivity are plotted against the dimensionless frequency $\hat{\omega}=\omega/T$ for various values of the dimensionless, shifted charge density $\hat{Q}=\hat q-\lambda\hat h$, while the dimensionless magnetic field  and momentum relaxation parameter are fixed at $\hat h=h\sqrt{Z_0}/ (T^2L)=30$, $\hat k=\sqrt{\Psi_0}k/T=3$. The vertical axes have units of $Z_0$. The peak in the longitudinal conductivity corresponds to the cyclotron pole in the complex $\hat\omega$ plane, \cite{Hartnoll:2007ip}. As $\hat Q$ gets smaller, the peak becomes shorter and eventually disappears. The Hall conductivity has a constant contribution equal to $\lambda$, stemming from the PQ term (\ref{PQ}). Finally, note that at $\hat Q=0$ there is finite charge density and magnetic field, related by $\hat q=\lambda\hat h$.\label{mag1}}
\end{figure}

\begin{figure}[H]
	\centering
	\includegraphics[width=0.49\textwidth]{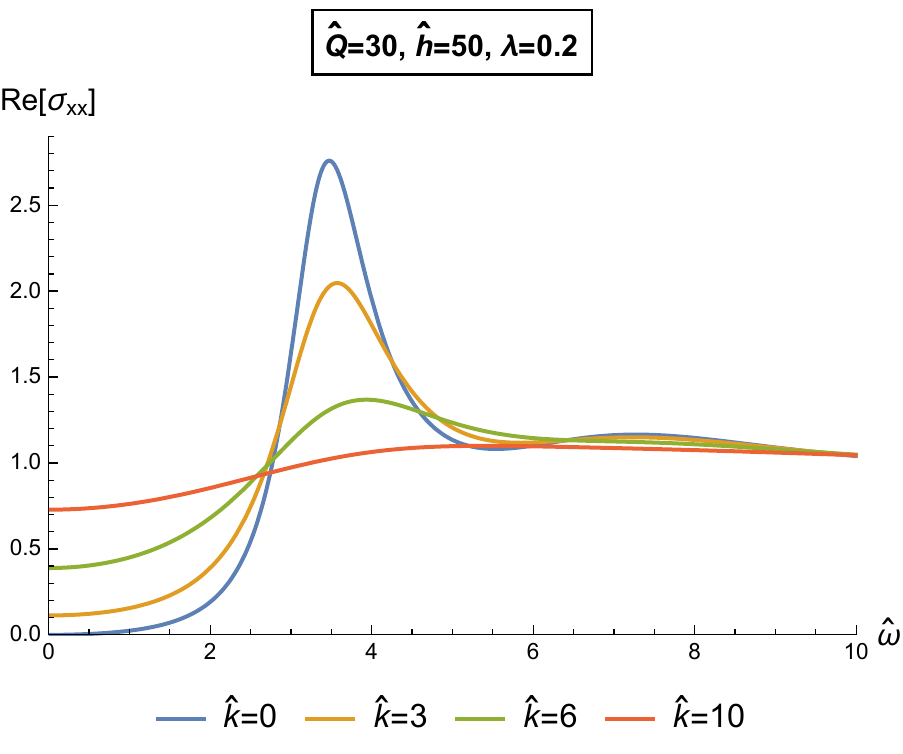}
	\includegraphics[width=0.49\textwidth]{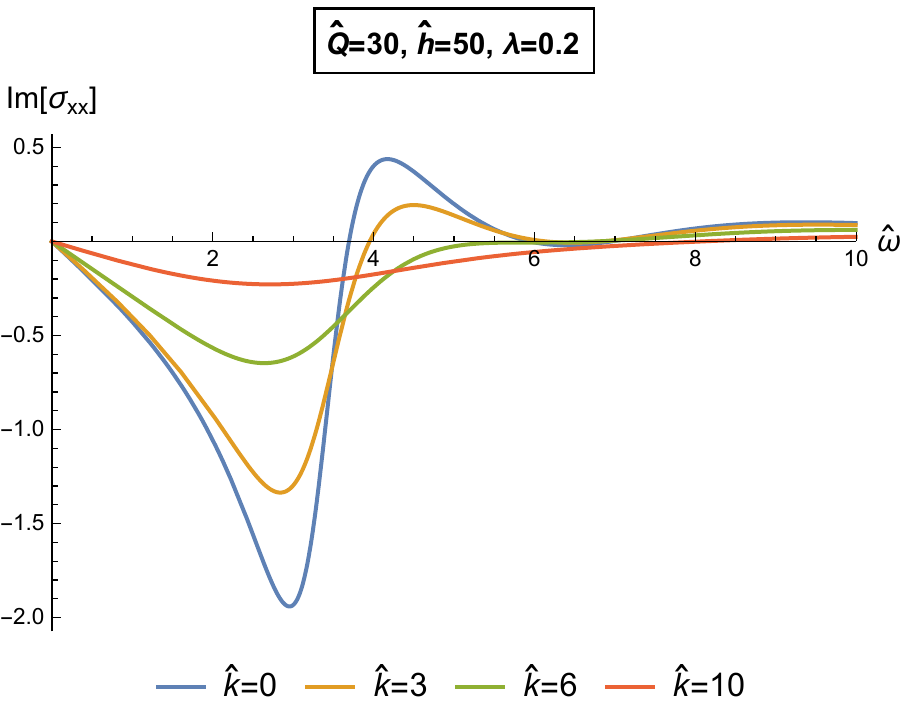}
	\includegraphics[width=0.49\textwidth]{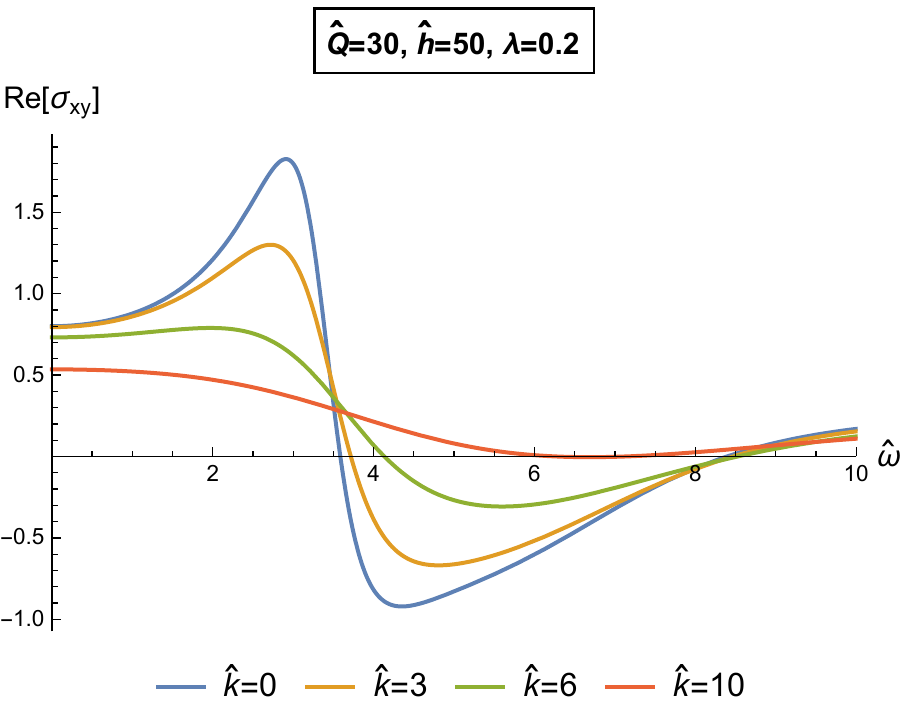}
	\includegraphics[width=0.49\textwidth]{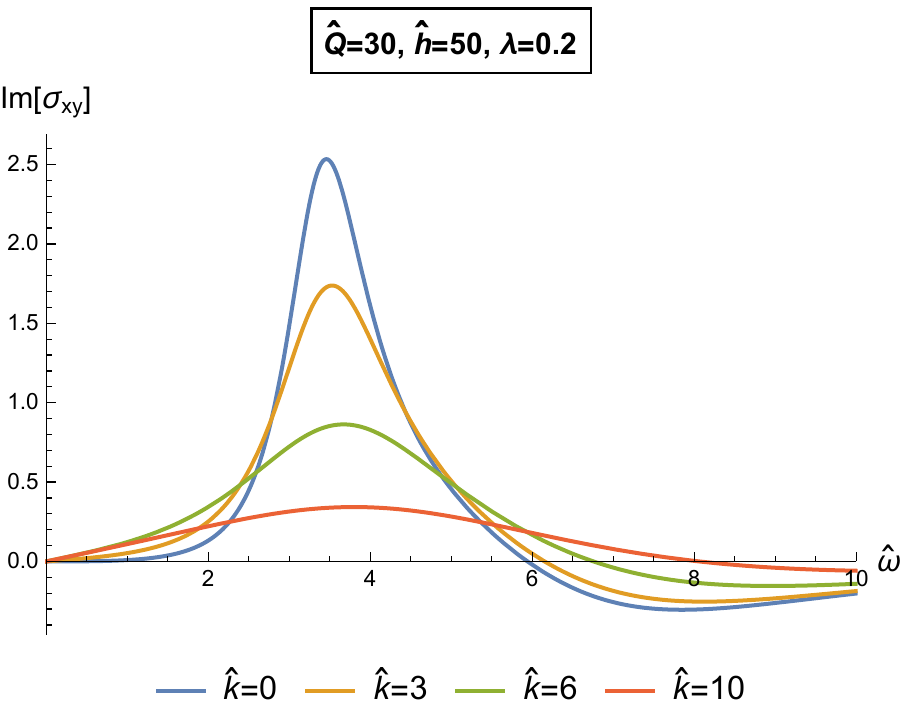}
	\caption{\small  The real and imaginary parts of the longitudinal (top) and Hall (bottom) conductivity are plotted against the dimensionless frequency $\hat{\omega}=\omega/T$ for various values of the dimensionless momentum relaxation parameter $\hat k=\sqrt{\Psi_0}k/T$, while the dimensionless, shifted charge density and magnetic field are fixed at $\hat{Q}=\hat q-\lambda \hat h=30$, $\hat h=h\sqrt{Z_0}/ (T^2L)=50$. The vertical axes have units of $Z_0$. The peak becomes shorter as $\hat k$ increases and eventually disappears, indicating that the pole moves further away from the real axis, as expected from the effect of strong momentum dissipation. Note that for $\hat k=0$ the DC conductivity vanishes, in agreement with (\ref{hcond1}). The Hall conductivity has a constant contribution equal to $\lambda$, stemming from the PQ term (\ref{PQ}).}\label{mag2}
\end{figure}

\begin{figure}[H]
	\centering
	\includegraphics[width=0.49\textwidth]{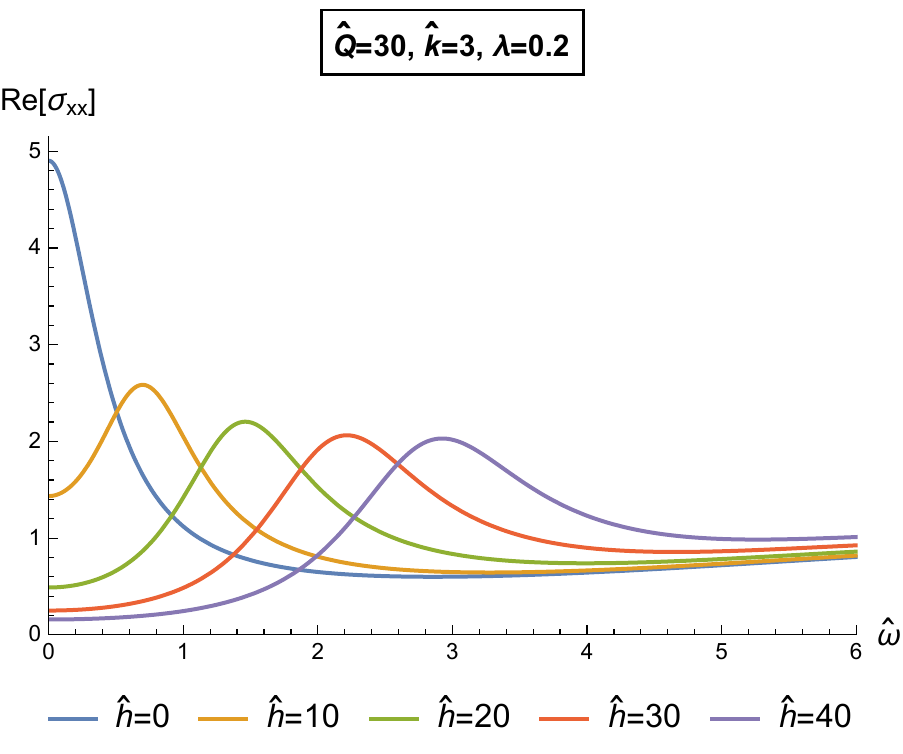}
	\includegraphics[width=0.49\textwidth]{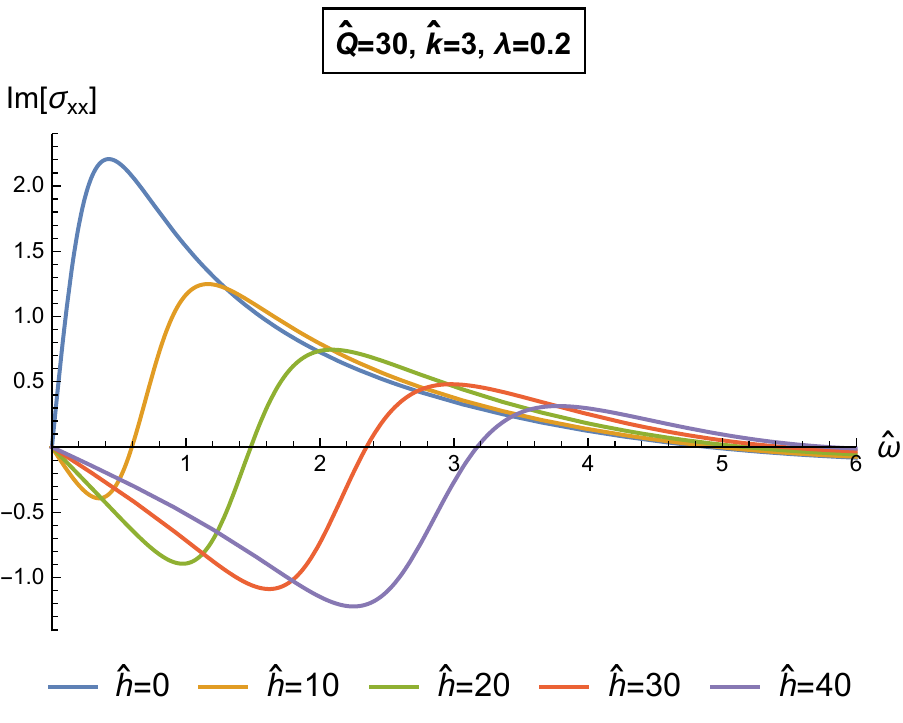}
	\includegraphics[width=0.49\textwidth]{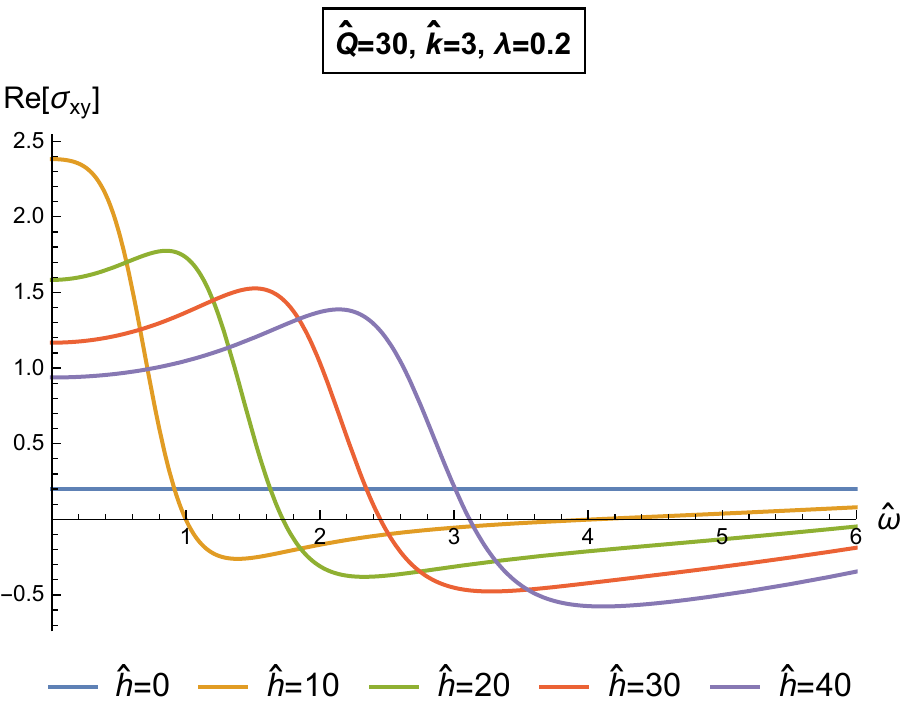}
	\includegraphics[width=0.49\textwidth]{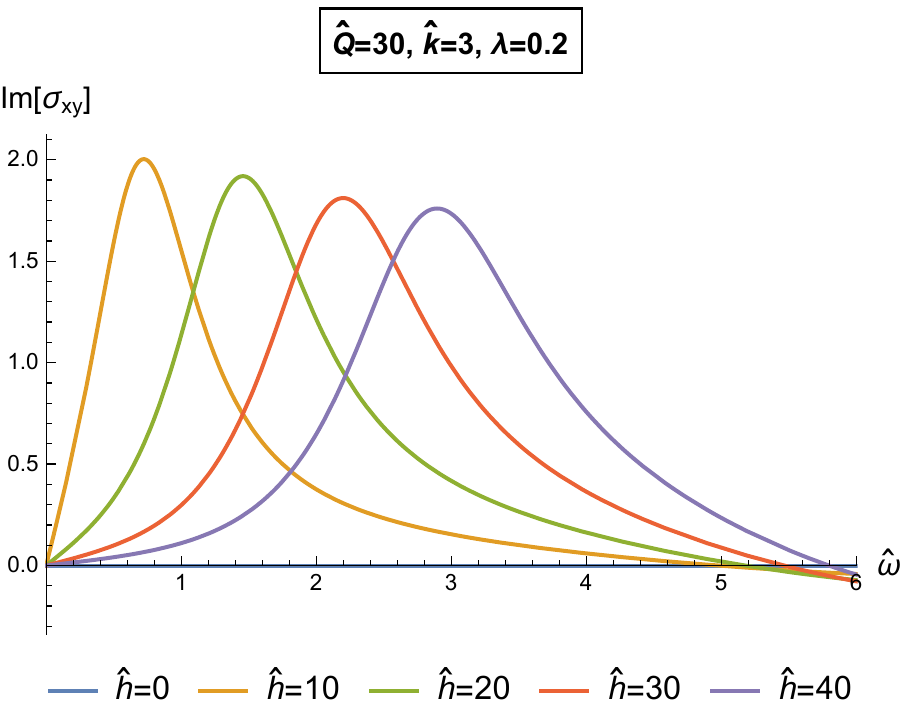}
	\caption{\small  The real and imaginary parts of the longitudinal (top) and Hall (bottom) conductivity are plotted against the dimensionless frequency $\hat{\omega}=\omega/T$ for various values of the dimensionless magnetic field $\hat h=h\sqrt{Z_0}/ (T^2L)$, while the dimensionless momentum relaxation parameter and shifted charge density are fixed at $\hat k=\sqrt{\Psi_0}k/T=3$, $\hat{Q}=\hat q-\lambda \hat h=30$. The vertical axes have units of $Z_0$. The peak moves to higher frequencies as $\hat h$ increases. Moreover the dependence seems to be linear, as in the cyclotron formula, (\ref{ccf}), while the width seems to remain approximately constant. A precise comparison is given in figure \ref{mag4}. The Hall conductivity has a constant contribution equal to $\lambda$, stemming from the PQ term (\ref{PQ}).}\label{mag3}
\end{figure}

We plot the conductivity against $\hat\omega$, keeping $\hat k,\hat q,\hat h,\lambda$ fixed. In figure \ref{mag1} we show the dependence of the conductivity on $\hat Q$. Note that for $\hat Q=0$, the charge density is finite, $\hat q=\lambda\hat h$. There is a peak in the real part of the longitudinal conductivity. Precisely at that frequency the real part of the Hall conductivity is equal to $\lambda$. The formula for the peak's location was found in \cite{Hartnoll:2007ip} for small frequencies, small magnetic fields and in the absence of momentum relaxation ($k=0$). It is
\be \omega_c={B\rho\over \varepsilon+\mathcal P},\label{ccf}\ee
where $B,\rho$ are the magnetic field and charge density, and $\epsilon,\mathcal P$ are the energy and pressure density. This formula has been identified with the relativistic cyclotron frequency, \cite{Hartnoll:2007ip}, \cite{Hartnoll:2007ih}. We will translate the above formula to our units in the next subsection.  As $\hat Q$ decreases, the peak becomes flatter and wider and eventually disappears. The location of the peak also decreases slightly as $\hat Q$ increases. Finally, note that the Hall conductivity has a constant offset of $\lambda$, because of the T-odd term, (\ref{PQ}).

In figure \ref{mag2} we show the dependence of the conductivity on the momentum relaxation parameter $\hat k$. For $\hat k=0$, the longitudinal DC conductivity is zero, as found in (\ref{hcond1}). As $\hat k$ increases, the cyclotron peak becomes flatter and eventually disappears.

\subsubsection{Comparison with the Hydrodynamic Formulas\label{chf}}

The analytic formula for the electric conductivity in the small $\tilde\omega, \tilde h$ limit has been derived in \cite{Hartnoll:2007ip}, in the absence of momentum relaxation ($\tilde k=0$). We quote the formulas here, written in the conventions of \cite{Hartnoll:2007ip}, which we will later match to ours.
\begin{subequations}
	\begin{align}
	&\sigma_{xx}=\sigma_Q {\omega(\omega+i\gamma+i\omega_c^2/\gamma)\over (\omega+i\gamma)^2-\omega_c^2}\sp
	\sigma_{xy}=-{\rho\over B}{-2i\gamma\omega+\gamma^2+\omega_c^2\over (\omega+i\gamma)^2-\omega_c^2}\\
	&\omega_c={B\rho\over \varepsilon+\mathcal P}\sp \gamma=\sigma_Q{B^2\over \varepsilon+\mathcal P}\sp \sigma_Q={1\over g^2}{(sT)^2\over (\varepsilon+\mathcal P)^2}.
	\end{align}
\end{subequations}
In the formulas above $B,\rho$ are the magnetic field and charge density, $\varepsilon,\mathcal P$ are the energy and pressure density and $s,T$ are the entropy density and temperature. The constant $1/g^2$ is the electromagnetic coupling. Note that the pole in the complex $\omega$ plane has negative imaginary part, which, keeping in mind the $e^{-i\omega t}$ time-dependence, leads to damping. The frequency $\omega_c$ is the relativistic cyclotron frequency and $\gamma$ is the damping frequency.

\begin{figure}[H]
	\centering
	\includegraphics[width=0.49\textwidth]{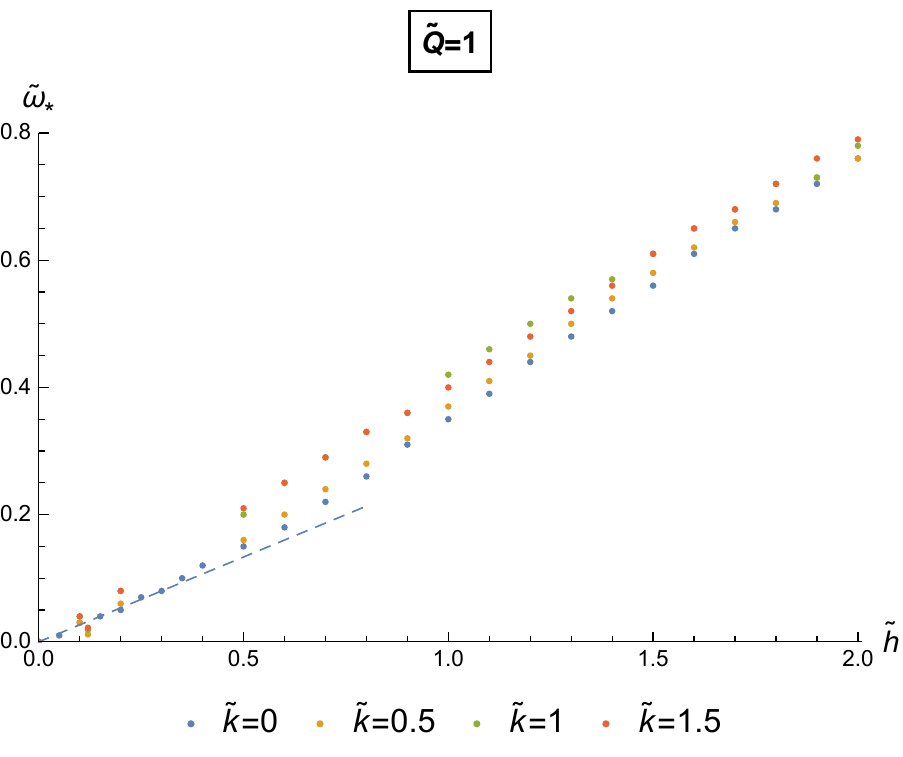}
	\includegraphics[width=0.49\textwidth]{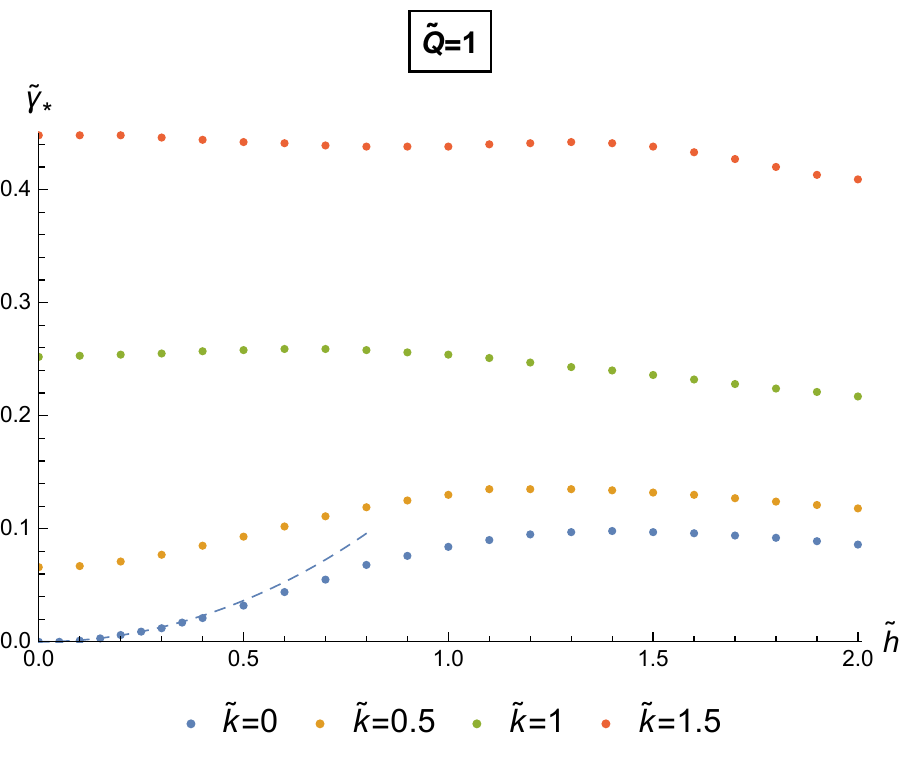}
	\caption{\small  Numerical values (dots) for the pole closest to the origin, $\tilde\omega=\tilde\omega_\star-i\tilde\gamma_\star$, plotted against the dimensionless magnetic field $\tilde h$ for various values of the dimensionless momentum relaxation $\tilde k$, while $\tilde Q=1$ is kept fixed. The dashed blue line corresponds to the analytic formula (\ref{anl}) for $\tilde k=0$ and is in agreement with the numerical data for small $\tilde h$ and $\tilde k=0$ (blue dots). Increasing the momentum relaxation rate $\tilde k$ makes the damping stronger, as expected.}\label{mag4}
\end{figure}

The cyclotron pole arises from the motion of particles and anti-particles, which are circulating in opposite directions.
Damping is present even in the translationally symmetric case ($k=0$) and can be thought of as arising from collisions between particles and anti-particles which are orbiting in opposite directions, \cite{Hartnoll:2007ih}. For $k\ne 0$ we expect stronger damping, as the particles collide not only between themselves, but also with impurities in the material. In figure \ref{mag4} we plot the location of the pole closest to the origin, calculated numerically, as a function of the magnetic field $\tilde h$ for various values of momentum relaxation $\tilde k$.

In order to use the formulas in \cite{Hartnoll:2007ip} for $k=0$ we need to carefully match their parameters with ours. To leading order in $\tilde h$ they read:
\begin{subequations}
	\begin{align}
	 &\sigma_{xx}=Z_0\sigma_Q{\tilde\omega(\tilde\omega+i\tilde\gamma+i\tilde\omega_c^2/\tilde\gamma)\over (\tilde\omega+i\tilde\gamma)^2-\tilde\omega_c^2}\sp \sigma_{xy}=W_0+Z_0{\tilde Q\over\tilde h}{2i\tilde\gamma\tilde\omega-\tilde\omega_c^2-\tilde\gamma^2\over (\tilde\omega+i\tilde\gamma)^2-\tilde\omega_c^2}\\
	&\tilde\omega_c={4\tilde h\tilde Q\over 12+3\tilde Q^2} \sp \tilde\gamma={\sigma_Q}{4\tilde h^2\over 12+3\tilde Q^2}\sp \sigma_Q=\left({12-\tilde Q^2\over 12+3\tilde Q^2}\right)^2.\label{anl}
	\end{align}
\end{subequations}
The above formulas are valid for small $\tilde h,\tilde\omega$ and $\tilde k =0$. We compare (\ref{anl}) with our numerical solution for $\tilde k=0$, in figure \ref{mag4}. For small values of $\tilde h$ they are in agreement. We also plot the full $\tilde h$ dependence of the location of the pole closest to the origin for various values of momentum relaxation $\tilde k$. We find that the damping increases significantly with $\tilde k$, as is expected.

To conclude, there are two kinds of contributions of the PQ term (\ref{PQ}) to the electric conductivity. The first is a constant offset in the Hall conductivity, analogous the the pair-creation term in the longitudinal conductivity, (\ref{ACneutral}). The second is that the charge density, $q$, appears in every formula shifted as $Q=q-hW_0$. This causes a system at finite density, $q$, and magnetic field, $h$, to behave effectively as a system at zero density when $q=hW_0$, since this happens at zero chemical potential, (\ref{chemp}). This simple behavior can be seen directly from the fluctuation equations (\ref{magfluc}), in which the PQ term does not participate explicitly, but only through $Q$. Equations (\ref{GFEP}) show that the coupling $W$ appears explicitly in the fluctuation equations if it has radial dependence (through the dilaton $\phi(r)$). We would expect non-trivial contributions from the PQ term in such cases.

\newpage

\section{Quantum Critical Solutions\label{HVsol}}

The goal of this section is to find and classify the IR-asymptotic hyperscaling violating solutions of (\ref{actionax}), describing holographic systems with a $U(1)$ symmetry and intrinsic T-violation. Most, but not all, of these solution have been studied in the past. Citations are given in the relevant subsections.

A large class quantum critical points in condensed matter physics are characterized by two scaling exponents $\theta,z$. The Lifshitz exponent $z$ signifies a deviation from Lorentz symmetry (for $z\ne 1$) and the hyperscaling violation exponent $\theta$ indicates deviation from scale invariance (for $\theta\ne 0$). Geometries with the same scaling symmetries have been realized in Holography, \cite{eht}, \cite{GK1}, \cite{GK2}, \cite{Kachru:2008yh}, \cite{Taylor:2015glc},
\be ds^2=r^\theta(-r^{-2z}dt^2+L^2r^{-2}dr^2+r^{-2}dx_idx^i)\label{q1}\ee
\be r\to\lambda r\sp x_i\to\lambda x_i\sp t\to \lambda^{z}t\sp ds^2\to \lambda^{\theta}ds^2.\label{q2}\ee
 In particular, such geometries can be obtained from the action (\ref{actionax}) by letting the scalar run to $\pm\infty$ in the IR.
Such solutions can be thought of as the IR limit of a geometry with AdS boundary. In the field theory language, such solutions are IR saddle points.

We assume that $\phi\to\pm\infty$ in the IR and that the scalar potential $V$ and the couplings run exponentially with the dilaton:
\be V(\phi)=V_0e^{\delta\phi}\sp Z(\phi)=Z_0e^{\beta\phi}\sp \Psi(\phi)=\Psi_0e^{\gamma\phi}\sp W(\phi)=W_0 e^{\eta\phi}.\label{q3}\ee
Exponential couplings arise naturally from string theory, \cite{st}. In each of the functions above we considered only the dominant exponential as $\phi\to\pm\infty$. The parameters $\delta,\beta,\gamma,\eta$ are, in general, not arbitrary. For our purposes, however, we will assume that they are independent, continuous parameters. Upon finding a solution, one can check if it can be embedded in the higher-dimensional string theory, as in \cite{GK1}, \cite{Li:2016rcv}.
The ansatz that we use is the following
\be ds^2=r^\theta(-r^{-2z}dt^2+L^2r^{-2}dr^2+r^{-2}dx_idx^i)\label{q4}\ee
\be A=(A_t(r),0,0,hx)\sp \phi=\phi(r)\sp \chi_i=kx_i\label{q5}\ee
where $r$ above is dimensionless.

The solution for the scalar is obtained from (\ref{EEQ2})
\be \phi(r)=\alpha\log r\sp \alpha^2=(2+\theta-2z)(\theta-2)\label{dilsol0}\ee
The gauge field equation (\ref{GFEQ1}) becomes
\be A_t'={qL\over Z_0}r^{1-z-\beta\alpha}-{hLW_0\over Z_0}r^{1-z+(\eta-\beta)\alpha}\label{gfhv}.\ee
We define the exponents $\zeta,\xi$ as follows
\be \zeta=2-\beta\alpha\sp \xi=2+(\eta-\beta)\alpha\label{q6}\ee
so that (\ref{gfhv}) becomes
\be A_t={qL\over Z_0(\zeta-z)} r^{\zeta-z}-{hLW_0\over Z_0(\xi-z)}r^{\xi-z},\label{gfs1}\ee
where we set the integration constant to $0$.

From the Einstein (\ref{EEQ1}),(\ref{EEQ3}) and scalar (\ref{DEQ1}) equations, using (\ref{gfhv}), we obtain the following system of independent equations
\be L^2V_0r^{\alpha\delta+\theta}=\underbrace{{1\over 2}L^2k^2\Psi_0r^{2+\alpha\gamma}}_\text{axion}+(2+z-\theta)(1+z-\theta)\label{ireq1}\ee
\be \begin{split} V_0r^{\alpha\delta+\theta}+\underbrace{{1\over 2}{h^2Z_0}r^{4+\alpha\beta-\theta}}_\text{magnetic}+{\left(\underbrace{q}_\text{electric}-\underbrace{hW_0r^{\alpha\eta}}_{PQ}\right)^2}{r^{4-\alpha\beta-\theta}\over 2Z_0}={1\over L^2}(2 + z - \theta) (2 z - \theta).\label{ireq2}\end{split}\ee

We will find the asymptotic solutions in the IR (which is located either at $r\to 0$ or $r\to\infty$) by matching the exponents and coefficients on both sides.
The scalar potential has to be relevant in the IR, therefore the LHS of (\ref{ireq1}) always participates in the leading order solution. In addition, the terms stemming from the Einstein tensor (the constant terms on the RHS above) also have to participate in the leading order solution. If these terms are subleading the solution is problematic \cite{GK2}.

\subsection{Classification of Solutions}

We consider the following cases for the classification of the solutions:
\begin{itemize}
	\item \textbf{The axion kinetic term participates in (\ref{ireq1}) to leading order.} In this case we call the solution axionic. This term is leading when $\alpha\gamma=-2$, otherwise $r^{2+\alpha\gamma}$ must asymptote to $0$ in the IR.
	
	If the axion is leading in the IR, from (\ref{ireq1}) we obtain \be \alpha\delta=-\theta\sp \alpha\gamma=-2\label{q7}\ee
	\be L^2={2(1 + z - \theta) (2 + z - \theta)\over 2V_0-k^2\Psi_0}\label{ref1}.\ee
	This, along with the scalar solution (\ref{dilsol0}), fixes the parameters $\alpha$, $\theta$ and $z$:
	\be \alpha=-2/\gamma\sp \theta=2{\delta/\gamma}\sp z={\delta^2-\gamma^2-1\over \gamma\delta-\gamma^2}.\label{param0}\ee
	Now (\ref{ireq2}) becomes
	\be {h^2Z_0}r^{4+\alpha\beta-\theta}+{(q-hW_0r^{\alpha\eta})^2\over Z_0}r^{4-\alpha\beta-\theta}={2(z-1)V_0-k^2\Psi_0(2z-\theta)\over 1+z-\theta}\label{hvs1}\ee
	
	From (\ref{hvs1}) we can see that there is a special value of the dissipation parameter
	\be k^2={V_0\over\Psi_0}{2(z-1)\over 2z-\theta}={V_0\over\Psi_0}{1+\g\d-\d^2\over 1+\g^2-\g\d}\label{q8}\ee
	at which the right-hand side vanishes. Since the left-hand side of (\ref{hvs1}) is a sum of squares, every term must vanish, hence $q=0,\; h=0$. Therefore we obtain a solution without charge density and magnetic field. This solution has been found in \cite{Gouteraux:2014hca} and is presented in \ref{HVaxneutral}.
	
	From (\ref{ref1}), there seems to be another special value of the dissipation parameter
	\be k^2={2V_0\over \Psi_0}\label{q9}\ee
	which implies $\theta=1+z$. However, in this case, the location of the IR is not well-defined, since the exponents of $g_{tt}$ and $g_{xx}$ have opposite signs, (\ref{IRwd}).
	
	\item \textbf{The PQ term (\ref{PQ}) participates in (\ref{ireq2}) to leading order.} In this case we call the solution intrinsically T-violating. The PQ term can be leading when $\eta=0$ or $r^{\alpha\eta}\to\infty$ in the IR. If $r^{\alpha\eta}\to 0$ the term does not participate in the leading order solution. Note that this term may lead in the gauge field (\ref{gfs1}) even if it is not leading in (\ref{ireq2}). This happens in solutions \ref{HVmag} and \ref{HVaxmag}.
	
	In the case $\eta=0$, the coupling $W=W_0$ is constant and the solutions are connected with the ones in which the PQ term is subleading, by $q\to q-hW_0$. In this class of solutions, there is a value of the magnetic field $h=q/W_0$ which makes the $t$-component of the gauge field vanish. Since these solutions exhibit no other difference, they will not be presented explicitly.
	
	\item \textbf{The electric term participates in (\ref{ireq2}) to leading order.} The term coming from the charge density $q$ is leading when when $\alpha\beta=4-\theta$. We call these solutions electric.
	
	\item \textbf{The magnetic term participates in (\ref{ireq2}) to leading order.} The term stemming from the magnetic field $h$ is leading when when $\alpha\beta=\theta-4$. In this case we call the solution magnetic. If both electric and magnetic terms participate (which happens only for $\beta=0$) we call the solution dyonic.
\end{itemize}

To sum up:
\begin{itemize}
	\item\textbf{Axionic}: $\alpha\gamma=-2$
	\item\textbf{Electric}: $\alpha\beta=4-\theta$
	\item\textbf{Magnetic}: $\alpha\beta=\theta-4$
	\item\textbf{T-violating}: $\alpha(2\eta-\beta)=\theta-4$.
\end{itemize}

\subsection{Location of the IR and constraints on the parameters\label{HVIR}}

The constraints on the parameters $\theta,z$ stem from the following requirements:
\begin{itemize}
	\item \textbf{Null Energy Condition}\\
	The NEC is satisfied when
	\be (2-2z+\theta)(\theta-2)\ge 0\label{nec1}\ee
	\be (z-1)(2+z-\theta)\ge 0.\label{nec2}\ee
	\item \textbf{The location of the IR is well-defined}\\
	The location of the IR is unambiguous when the exponents of $g_{tt}$ and $g_{xx}$ have the same sign
	\be (\theta-2z)(\theta-2)>0.\label{IRwd}\ee
	In addition we require that the finite temperature mode $r^{2+z-\theta}$ vanishes in the UV, which implies that it must lead in the IR.
	\item \textbf{$L^2>0$}\\
	In the non-axionic solutions this implies
	\be (2+z-\theta)(1+z-\theta)>0.\label{q10}\ee
	In the axionic solutions this implies
	\be (2+z-\theta)(1+z-\theta)(2V_0-k^2\Psi_0)>0.\label{q11}\ee
	\item \textbf{$q^2\geq 0,h^2\geq 0,k^2\geq 0$}\\
	This implies
	\be (2+z-\theta)(z-1)\geq 0\label{q12}\ee
	which is already true if the NEC holds.
\end{itemize}

Given that the above conditions are satisfied, the IR is located at $r=0$ if
\be \theta>2\label{q13}\ee
or at $r=\infty$ if
\be \theta<2.\label{q14}\ee

\subsubsection{The Axionic Solutions}

Specializing the above constraints for the class of axionic solutions we obtain constraints on the values of $\d,\g$
\begin{itemize}
	\item \textbf{The IR is located at $r=0$}\\
	\be \gamma(\delta-\gamma)>0\sp \delta^2+3\gamma^2-4\gamma\delta+1>0\label{q15}\ee
	\be 1+\gamma^2-\g\d>0\sp \gamma  \delta -\delta ^2+1\ge0\label{q16}\ee
	\item \textbf{The IR is located at $r=\infty$}
	\be \gamma(\delta-\gamma)<0\sp \delta^2+3\gamma^2-4\gamma\delta+1>0\label{q17}\ee
	\be 1+\gamma^2-\g\d>0\sp \gamma \delta -\delta^2+1\ge 0\label{q18}\ee
\end{itemize}

\begin{figure}[H]
	\centering
	\includegraphics[width=0.75\textwidth]{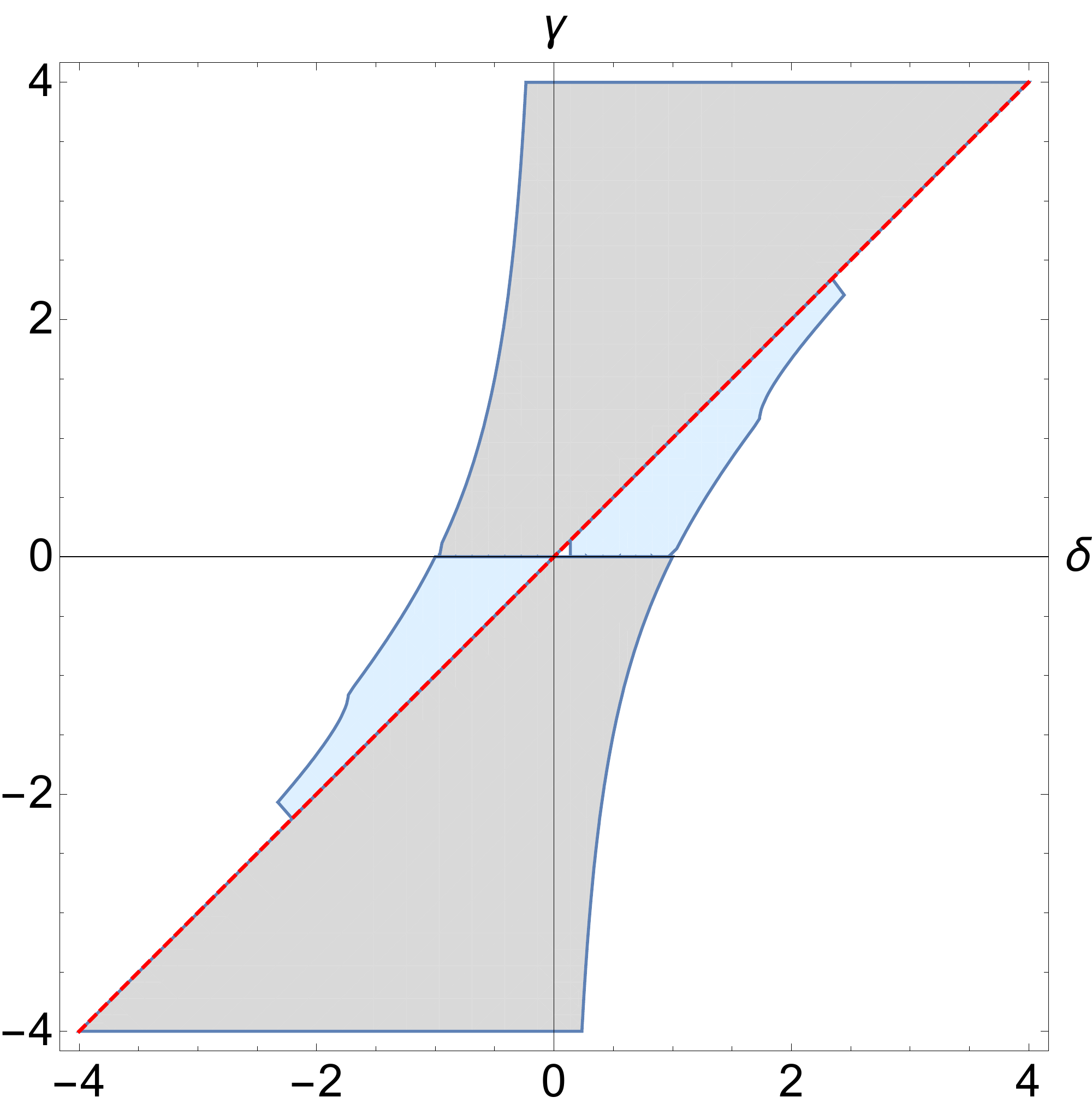}
	\caption{\small The region of validity of the axionic solution in the $\d-\g$ plane is presented above. The IR is located at $r\to_{IR}0$ within the light blue area and at $r\to_{IR}\infty$ in the light gray area. The red line corresponds to $\g=\d$. For values in the white area, either the null-energy condition does not hold or the IR is not well-defined.}
	\label{figIR1}
\end{figure}

%For small temperatures, the entropy scales as
%\be S\sim T^{2-\theta\over z}=T^{2{(\d-\g)^2\over 1+\g^2-\d^2}}.\ee

\subsubsection{The Case $\g=\d$}
Consider again the equations (\ref{ireq1}),(\ref{ireq2}) and set $\g=\d$ to obtain
\be L^2V_0r^{\alpha\g+\theta}={1\over 2}L^2k^2\Psi_0r^{2+\alpha\gamma}+(2+z-\theta)(1+z-\theta)\label{ireq3}\ee
\be \begin{split} V_0r^{\alpha\g+\theta}+{1\over 2}{h^2Z_0}r^{4+\alpha\beta-\theta}+{q^2\over 2Z_0}r^{4-\alpha\beta-\theta}-{hqW_0\over Z_0}r^{4 - \alpha\beta + \alpha\eta - \theta}\\+{h^2W_0^2\over 2Z_0}r^{4 - \alpha\beta + 2\alpha\eta - \theta}={1\over L^2}(2 + z - \theta) (2 z - \theta)\end{split}\label{q19}\ee
For $\theta=2$, from (\ref{dilsol0}) we obtain $\alpha=0$, hence the scalar does not run to infinity as we assumed. Nevertheless we can see what happens to the above equations:
\be V_0={1\over 2}k^2\Psi_0+r^{-2}{1\over L^2}z(z-1)\label{q20}\ee
\be  V_0+{1\over 2}{h^2Z_0}+{(q-hW_0)^2\over 2Z_0}=r^{-2}{1\over L^2}2z( z - 1)\label{q21}\ee
from which we obtain
\be z(z-1)=0\label{q22}\ee
\be k=0\sp q=0\sp h=0\sp V_0=0.\label{q23}\ee
The choices $z=1,z=0$ correspond to neutral Minkowski and Rindler spacetimes respectively with constant scalar.

\subsection{Solutions}

\subsubsection{Neutral solution\label{HVneutral}}

This neutral solution is obtained by setting $q=0,h=0$ and requiring that the axion term is subleading. This is solution 2.1.2 in \cite{GK2}.
\bear \begin{split} &ds^2=r^\theta(-r^{-2z}dt^2+L^2r^{-2}dr^2+r^{-2}dx_idx^i)\sp \theta={2\d^2\over \d^2-1}\sp z=1\\
	& L^2={(2 - \theta) (3 - \theta)\over V_0}\sp e^\phi=r^\alpha\sp \alpha={2\d\over 1-\d^2}\sp (\chi_x,\chi_y)=(kx,ky)\\
	&A=0\label{q24}
\end{split}
\eear
In this solution there is no magnetic field or charge density. The translation symmetry is restored in the IR, as the axion field is subleading.
The Lorentz symmetry is also restored $z=1$.

The constraints on the exponents are:
\be \delta^2<3\sp\;\; 1+\delta\g-\d^2<0.\label{q25}\ee
If $\delta^2<1$ the IR is located at $\infty$. If $\delta^2>1$ the IR is located at $0$.

\subsubsection{Axionic neutral solution\label{HVaxneutral}}

This neutral solution is obtained by setting $q=0,h=0$ and requiring that the axion participates to leading order. This is solution B.2 in \cite{Gouteraux:2014hca}.
\bear \begin{split} &ds^2=r^\theta(-r^{-2z}dt^2+L^2r^{-2}dr^2+r^{-2}dx_idx^i)\sp \theta=2{\delta/\gamma}\sp z={\delta^2-\gamma^2-1\over \gamma\delta-\gamma^2}\\
	& L^2={2(1 + z - \theta) (2 + z - \theta)\over 2V_0-k^2\Psi_0}\sp e^\phi=r^\alpha\sp \alpha=-2/\gamma\sp (\chi_x,\chi_y)=(kx,ky)\\
	&A=0\\
	& k^2={V_0\over\Psi_0}{2(z-1)\over 2z-\theta}
\end{split}\label{q26}
\eear

As in \ref{HVneutral} there is no charge density or magnetic field, but now the translational symmetry remains broken in the IR.
In this solution we have Lifshitz scaling ($z\ne 1$) caused by the axion fields. The Lorentz symmetry is restored if and only if the translational symmetry is restored ($k=0$). This also implies $\g=\d-1/\d$, which takes us to solution \ref{HVneutral}.

The constraints on the exponents are
\be \text{IR}\to 0:\;\;\;  \g(\d-\g)>0\sp 1+\g^2-\g\d>0\sp 1+\g\d-\d^2>0\sp 1+3\g^2+\d^2-4\g\d>0\label{q27}\ee
\be \text{IR}\to \infty:\;\;\;  \g(\d-\g)<0\sp 1+\g^2-\g\d>0\sp 1+\g\d-\d^2>0\sp 1+3\g^2+\d^2-4\g\d>0.\label{q28}\ee

\subsubsection{Dyonic solution\label{HVdyo}}
This solution is obtained when both the electric and magnetic terms are leading. The axion term is subleading in this solution. This is solution 6.1.4 in \cite{Angelinos:2014gea}.
\bear \begin{split} &ds^2=r^\theta(-r^{-2z}dt^2+L^2r^{-2}dr^2+r^{-2}dx_idx^i)\sp  \theta=4\sp z={3\d^2-4\over \d^2}  \\
	& L^2={( z - 3) (z - 2)\over V_0}\sp e^\phi=r^\alpha\sp \alpha=-4/\d\sp (\chi_x,\chi_y)=(kx,ky)\\
	&A=L{q\over Z_0(\zeta-z)}r^{\zeta-z}dt+hxdy\sp \zeta=2 \\
	& {h^2Z_0}+{q^2\over Z_0}=2(z-1)V_0
\end{split}\label{q29}
\eear

This solution is characterized by the coexistence of charge density and magnetic field in the IR, which can happen only for the value $\beta=0$. Changing the value of $\beta$ slightly, one lands in either the electric, \ref{HVel}, or magnetic solution, \ref{HVmag}.  The translation symmetry is restored in the IR, as the axion field is subleading.

The constraints on the exponents are
\be \beta=0\sp \delta^2<2\sp \d(\d-2\g)>0\sp \d\eta<0.\label{q30}\ee
The IR is located at $r\to 0$.

\subsubsection{Axionic dyonic solution}
This solution is obtained when both the electric and magnetic terms are leading, which implies $\beta=0$. The axion term participates in this solution to leading order. This is a new solution.
\bear \begin{split} &ds^2=r^\theta(-r^{-2z}dt^2+L^2r^{-2}dr^2+r^{-2}dx_idx^i)\sp  \theta=4\sp z={3\d^2-4\over \d^2}  \\
	& L^2={2( z - 3) (z - 2)\over 2V_0-k^2\Psi_0}\sp e^\phi=r^\alpha\sp \alpha=-4/\d\sp (\chi_x,\chi_y)=(kx,ky)\\
	&A=L{q\over Z_0(\zeta-z)}r^{\zeta-z}dt+hxdy\sp \zeta=2 \\
	&{h^2Z_0}+{q^2\over Z_0}+k^2\Psi_0(2z-\theta)=2(z-1)V_0
\end{split}\label{q31}
\eear

This is the axionic version of solution \ref{HVdyo}. Translation symmetry is broken in the IR. Changing the value of $\beta$ slightly, one lands in either the electric, \ref{HVaxel}, or magnetic solution, \ref{HVaxmag}.

The constraints on the exponents are
\be \beta=0\sp 2\gamma=\delta\sp \d^2<2\sp\d\eta<0.\label{q32}\ee
The IR is located at
$ r\to0$.

\subsubsection{Electric solution\label{HVel}}
This solution is obtained when the electric term is leading and the magnetic term is subleading. The axion term is subleading in this solution. This is solution 2.2.2 in \cite{GK2}.
\bear \begin{split} &ds^2=r^\theta(-r^{-2z}dt^2+L^2r^{-2}dr^2+r^{-2}dx_idx^i)\sp \theta={4\delta\over \d-\b}\sp z={3\delta^2-\b^2+2\b\d-4\over \d^2-\b^2} \\
	&L^2={(1 + z - \theta) (2 + z - \theta)\over V_0}\sp e^\phi=r^\alpha\sp \alpha={4\over \b-\d}\sp (\chi_x,\chi_y)=(kx,ky)\\
	&A=L{q\over Z_0(\zeta-z)}r^{\zeta-z}dt+hxdy\sp {q^2}=Z_0{2V_0(z-1)\over 1+z-\theta}\sp  \zeta=2{\d+\b\over\d-\b}
\end{split}\label{q33}
\eear

This solution is characterized by a finite charge density. The magnetic field is negligible in the IR and translational symmetry is restored, as the axion fields are subleading.

The constraints on the exponents are
\be  \label{q34}\begin{split} \text{IR}\to 0:\;\;\; &0<\d^2-\b^2<4\sp 4+3\b^2+2\b\d-\d^2>0\\ & \d^2+\b\d-2<0\sp 2+\b^2+\b\d>0
	\\&(2\g+\b-\d)(\b-\d)>0\sp \eta(\b-\d)>0\sp \b(\b-\d)>0\end{split}\ee
\be  \label{q35}\begin{split} \text{IR}\to \infty:\;\;\; &\d^2-\b^2<0\sp 4+3\b^2+2\b\d-\d^2>0\\ & \d^2+\b\d-2<0\sp 2+\b^2+\b\d>0
	\\&(2\g+\b-\d)(\b-\d)<0\sp \eta(\b-\d)<0\sp \b(\b-\d)<0\end{split}\ee

\subsubsection{Axionic electric solution\label{HVaxel}}
This solution is obtained when the electric term is leading and the magnetic term is subleading. The axion term participates in this solution to leading order. This is solution B.1 in \cite{Gouteraux:2014hca}.
\bear \begin{split} &ds^2=r^\theta(-r^{-2z}dt^2+L^2r^{-2}dr^2+r^{-2}dx_idx^i)\sp \theta={4\delta\over \d-\b}\sp z={3\delta^2-\b^2+2\b\d-4\over \d^2-\b^2} \\
	&L^2={2(1 + z - \theta) (2 + z - \theta)\over 2V_0-k^2\Psi_0}\sp e^\phi=r^\alpha\sp \alpha={4\over \b-\d}\sp (\chi_x,\chi_y)=(kx,ky)\\
	&A=L{q\over Z_0(\zeta-z)}r^{\zeta-z}dt+hxdy\sp {q^2}=Z_0{2V_0(z-1)-k^2\Psi_0(2z-\theta)\over 1+z-\theta}\sp  \zeta=2{\d+\b\over\d-\b}\\
	&
\end{split}\label{q36}
\eear

This is the axionic version of \ref{HVel}. The translational symmetry remains broken in the IR.

The constraints on the exponents are
\be  \label{q37}\begin{split} \text{IR}\to 0:\;\;\; & 2\gamma=\delta-\beta\sp\;\; 0<\d^2-\b^2<4\sp\;\; 4+3\b^2+2\b\d-\d^2>0\\ & \d^2+\b\d-2<0\sp\;\; \eta(\b-\d)>0\sp\;\; \b(\b-\d)>0\end{split}\ee
\be  \label{q38}\begin{split} \text{IR}\to \infty:\;\;\; &2\gamma=\delta-\beta\sp\;\; \d^2-\b^2<0\sp\;\; 4+3\b^2+2\b\d-\d^2>0\\ & \d^2+\b\d-2<0
	\sp \;\; \eta(\b-\d)<0\sp\;\; \b(\b-\d)<0\end{split}\ee

\subsubsection{Magnetic solution\label{HVmag}}
This solution is obtained when the magnetic term is leading and the electric term is subleading. The PQ term is subleading in the Einstein equations, but leading in the gauge field equation. The axion term is subleading in this solution. This is a new solution.
\bear \label{q39}\begin{split} &ds^2=r^\theta(-r^{-2z}dt^2+L^2r^{-2}dr^2+r^{-2}dx_idx^i)\sp \theta={4\delta\over \d+\b}\sp z={3\delta^2-\b^2-2\b\d-4\over \d^2-\b^2}\\
	& L^2={2(1 + z - \theta) (2 + z - \theta)\over 2V_0-k^2\Psi_0}\sp e^\phi=r^\alpha\sp \alpha=-{4\over\b+\d}\sp (\chi_x,\chi_y)=(kx,ky)\\
	&A=-L{hW_0\over Z_0(\xi-z)}r^{\xi-z}dt+hxdy\sp h^2={2V_0(z-1)\over Z_0 (1+z-\theta)}\sp \xi=2-{4(\eta-\b)\over\b+\d}
\end{split}
\eear

This solution is characterized by a finite magnetic field. The charge density is negligible in the IR and translational symmetry is restored, as the axion fields are subleading.

The constraints on the exponents are
\be  \label{q40}\begin{split} \text{IR}\to 0:\;\;\; &0<\d^2-\b^2<4\sp\;\; 4+3\b^2-2\b\d-\d^2>0\\ & \d^2-\b\d-2<0\sp\;\; 2+\b^2-\b\d>0
	\\&\eta(\b+\d)>0\sp\;\; (\b-\eta)(\b+\d)>0\sp\;\; (\b+\d-2\g)(\b+\d)>0\end{split}\ee
\be  \label{q41}\begin{split} \text{IR}\to \infty:\;\;\; &\d^2-\b^2<0\sp\;\; 4+3\b^2-2\b\d-\d^2>0\\ & \d^2-\b\d-2<0\sp\;\; 2+\b^2-\b\d>0
	\\&\eta(\b+\d)<0\sp\;\; (\b-\eta)(\b+\d)<0\sp\;\; (\b+\d-2\g)(\b+\d)<0\end{split}\ee

\subsubsection{Axionic magnetic solution\label{HVaxmag}}
This solution is obtained when the magnetic term is leading and the electric term is subleading. The PQ term is subleading in the Einstein equations, but leading in the gauge field equation. The axion term participates in this solution to leading order. This is a new solution.
\bear \begin{split} &ds^2=r^\theta(-r^{-2z}dt^2+L^2r^{-2}dr^2+r^{-2}dx_idx^i)\sp \theta={4\delta\over \d+\b}\sp z={3\delta^2-\b^2-2\b\d-4\over \d^2-\b^2}\\
	& L^2={2(1 + z - \theta) (2 + z - \theta)\over 2V_0-k^2\Psi_0}\sp e^\phi=r^\alpha\sp \alpha=-{4\over \b+\d}\sp (\chi_x,\chi_y)=(kx,ky)\\
	&A=-L{hW_0\over Z_0(\xi-z)}r^{\xi-z}dt+hxdy\sp h^2={2V_0(z-1)-k^2\Psi_0(2z-\theta)\over Z_0 (1+z-\theta)}\sp \xi=2-{4(\eta-\b)\over\b+\d}
\end{split}\label{q42}
\eear

This is the axionic version of \ref{HVmag}. Translational symmetry remains broken in the IR.

The constraints on the exponents are
\be  \label{q43}\begin{split} \text{IR}\to 0:\;\;\; &2\gamma=\beta+\delta\sp\;\; 0<\d^2-\b^2<4\sp\;\; 4+3\b^2-2\b\d-\d^2>0\\ & \d^2-\b\d-2<0\sp\;\; \eta(\b+\d)>0\sp\;\; (\b-\eta)(\b+\d)>0\end{split}\ee
\be  \label{q44}\begin{split} \text{IR}\to \infty:\;\;\; &2\gamma=\beta+\delta\sp\;\; \d^2-\b^2<0\sp\;\; 4+3\b^2-2\b\d-\d^2>0\\& \d^2-\b\d-2<0\sp\;\;
	\eta(\b+\d)<0\sp\;\; (\b-\eta)(\b+\d)<0\end{split}\ee

\subsubsection{Magnetic solution with intrinsic T-violation\label{HVmagT}}
This solution is obtained when the magnetic and PQ terms are leading. The axion term is subleading in this solution.
\bear \begin{split} &ds^2=r^\theta(-r^{-2z}dt^2+L^2r^{-2}dr^2+r^{-2}dx_idx^i)\sp \theta={4\delta\over \b+\d}\sp z={3\delta^2-\b^2-2\b\d-4\over \d^2-\b^2}\\
	& L^2={(1 + z - \theta) (2 + z - \theta)\over V_0}\sp e^\phi=r^\alpha\sp \alpha=-{4\over \b+\d}\sp (\chi_x,\chi_y)=(kx,ky)\\
	& A=-{hW_0\over Z_0(\xi-z)}r^{\xi-z}dt+hxdy\sp h^2={2Z_0V_0(z-1)\over (Z_0^2+W_0^2) (1+z-\theta)}\sp \xi=2 \\
	&\eta=\beta
\end{split}\label{q45}
\eear

This solution is characterized by a finite magnetic field. The charge density is negligible and translational symmetry is restored, as the axion fields are subleading. This solution can be obtained from \ref{HVmag} by setting $\eta=\beta$ and transforming
\be Z_0\to {Z_0^2+W_0^2\over Z_0}\sp W_0\to W_0{Z_0^2+W_0^2\over Z_0^2}.\label{q46}\ee

The constraints on the exponents are
\be  \label{q47}\begin{split} \text{IR}\to 0:\;\;\; &\eta=\beta\sp\;\; 0<\d^2-\b^2<4\sp\;\; 4+3\b^2-2\b\d-\d^2>0\\ & \d^2-\b\d-2<0\sp\;\; 2+\b^2-\b\d>0
	\\&(\b+\d-2\g)(\b+\d)>0\sp\;\; \b(\b+\d)>0\end{split}\ee
\be \label{q48} \begin{split} \text{IR}\to \infty:\;\;\; &\eta=\beta\sp\;\; \d^2-\b^2<0\sp\;\; 4+3\b^2-2\b\d-\d^2>0\\ & \d^2-\b\d-2<0\sp\;\; 2+\b^2-\b\d>0
	\\&(\b+\d-2\g)(\b+\d)<0\sp\;\; \b(\b+\d)<0\end{split}\ee

\subsubsection{Axionic magnetic solution with intrinsic T-violation}
This solution is obtained when the magnetic and PQ terms are leading. The axion term participates in this solution to leading order.
\bear \begin{split} &ds^2=r^\theta(-r^{-2z}dt^2+L^2r^{-2}dr^2+r^{-2}dx_idx^i)\sp \theta={4\delta\over \b+\d}\sp z={3\delta^2-\b^2-2\b\d-4\over \d^2-\b^2}\\
	& L^2={2(1 + z - \theta) (2 + z - \theta)\over 2V_0-k^2\Psi_0}\sp e^\phi=r^\alpha\sp \alpha=-{4\over\b+\d}\sp (\chi_x,\chi_y)=(kx,ky)\\
	& A=-{hW_0\over Z_0(\xi-z)}r^{\xi-z}dt+hxdy\sp h^2=Z_0{2V_0(z-1)-k^2\Psi_0(2z-\theta)\over (Z_0^2+W_0^2) (1+z-\theta)}\sp \xi=2 \\
	&\eta=\beta=2\gamma-\delta
\end{split}\label{q49}
\eear

This is the axionic version of \ref{HVmagT}. The translational symmetry is broken in the IR, since the axions are leading.  This solution can be obtained from \ref{HVaxmag} by setting $\eta=\beta$ and transforming
\be Z_0\to {Z_0^2+W_0^2\over Z_0}\sp W_0\to W_0{Z_0^2+W_0^2\over Z_0^2}.\label{q461}\ee

The constraints on the exponents are
\be  \label{q50}\begin{split} \text{IR}\to 0:\;\;\; &\eta=\beta\sp\;\; 2\gamma=\delta+\beta\sp\;\; 0<\d^2-\b^2<4\sp\;\; 4+3\b^2-2\b\d-\d^2>0\\ & \d^2-\b\d-2<0\sp\;\; \b(\b+\d)>0\end{split}\ee
\be  \label{q51}\begin{split} \text{IR}\to \infty:\;\;\; &\eta=\beta\sp\;\; 2\gamma=\delta+\beta\sp \;\; \d^2-\b^2<0\sp\;\; 4+3\b^2-2\b\d-\d^2>0\\ & \d^2-\b\d-2<0\sp\;\; \b(\b+\d)<0\end{split}\ee

\subsubsection{Solution with intrinsic T-violation\label{HVT}}
This solution is obtained when only the PQ term is leading. This is solution 6.2.2 in \cite{Angelinos:2014gea}.
\bear \begin{split} &ds^2=r^\theta(-r^{-2z}dt^2+L^2r^{-2}dr^2+r^{-2}dx_idx^i)\sp  \theta={4\delta\over \d+2\eta-\b}\sp \\
	& z={4+(\b-3\d-2\eta)(\b+\d-2\eta)\over (\b-\d-2\eta)(\b+\d-2\eta)}\sp (\chi_x,\chi_y)=(kx,ky)\\
	& L^2={(1 + z - \theta) (2 + z - \theta)\over V_0}\sp e^\phi=r^{\alpha}\sp \alpha={4\over \b-\d-2\eta}\\
	& A=-L{hW_0\over Z_0(\xi-z)}r^{\xi-z}dt+hxdy\sp {h^2}={Z_0\over W_0^2}{2V_0(z-1)\over 1+z-\theta}\sp \xi=2{\d+\b\over \d-\b+2\eta}
\end{split}\label{q52}
\eear

This solution is characterized by a finite magnetic field, which participates only through the PQ term. The charge density is negligible and translational symmetry is restored, as the axion field is subleading.

The constraints on the exponents are
\be  \begin{split} \text{IR}\to 0:\;\;\; &0<\d^2-(\b-2\eta)^2<4\sp\;\; 4+3(\b-2\eta)^2+2(\b-2\eta)\d-\d^2>0\\ & \d^2+(\b-2\eta)\d-2<0\sp\;\; 2+(\b-2\eta)^2+(\b-2\eta)\d>0
	\\&\eta(2\eta+\d-\b)>0\sp\;\; (\eta-\b)(2\eta+\d-\b)>0\\ &(2\g+\b-\d-2\eta)(\b-\d-2\eta)>0\end{split}\label{q53}\ee
\be  \begin{split} \text{IR}\to \infty:\;\;\; &\d^2-(\b-2\eta)^2<0\sp\;\; 4+3(\b-2\eta)^2+2(\b-2\eta)\d-\d^2>0\\ & \d^2+(\b-2\eta)\d-2<0\sp\;\; 2+(\b-2\eta)^2+(\b-2\eta)\d>0
	\\& \eta(2\eta+\d-\b)<0\sp\;\; (\eta-\b)(2\eta+\d-\b)<0\\ &(2\g+\b-\d-2\eta)(\b-\d-2\eta)<0\end{split}\label{q54}\ee

\subsubsection{Axionic solution with intrinsic T-violation\label{HVTax}}
This solution is obtained when only the PQ and axion terms are leading. This is a new solution.
\bear \begin{split} &ds^2=r^\theta(-r^{-2z}dt^2+L^2r^{-2}dr^2+r^{-2}dx_idx^i)\sp \theta={4\delta\over \d+2\eta-\b}\\ &z={4+(\b-3\d-2\eta)(\b+\d-2\eta)\over (\b-\d-2\eta)(\b+\d-2\eta)}\sp (\chi_x,\chi_y)=(kx,ky)\\
	& L^2={2(1 + z - \theta) (2 + z - \theta)\over 2V_0-k^2\Psi_0}\sp e^\phi=r^{\alpha}\sp \alpha={4\over \b-\d-2\eta}\\
	& A=-L{hW_0\over Z_0(\xi-z)}r^{\xi-z}dt+hxdy\sp {h^2}={Z_0\over W_0^2}{2V_0(z-1)-k^2\Psi_0(2z-\theta)\over 1+z-\theta}\\ &\xi=2{\d+\b\over \d-\b+2\eta}
\end{split}\label{q55}
\eear

This is the axionic version of \ref{HVT}. The translational symmetry remains broken in the IR.

The constraints on the exponents are
\be  \begin{split} \text{IR}\to 0:\;\;\; &2\gamma=2\eta-\beta+\delta\sp\;\; 0<\d^2-(\b-2\eta)^2<4\\ & 4+3(\b-2\eta)^2+2(\b-2\eta)\d-\d^2>0\sp\;\; \d^2+(\b-2\eta)\d-2<0
	\\&\eta(2\eta+\d-\b)>0\sp (\eta-\b)(2\eta+\d-\b)>0\end{split}\label{q56}\ee
\be  \begin{split} \text{IR}\to \infty:\;\;\; & 2\gamma=2\eta-\beta+\delta\sp\;\;\d^2-(\b-2\eta)^2<0\\ &4+3(\b-2\eta)^2+2(\b-2\eta)\d-\d^2>0\sp\;\; \d^2+(\b-2\eta)\d-2<0
	\\& \eta(2\eta+\d-\b)<0\sp (\eta-\b)(2\eta+\d-\b)<0\end{split}\label{q57}\ee

\subsection{Renormalization Group Stability of the Critical Solutions\label{HVpert}}

In this final subsection we study the RG stability of the solutions.
To do so, we must study small perturbations of these solutions with the same symmetry as before. Therefore, we perturb the solution as follows
\be D(r)=r^{\theta-2z}(1+D_1r^b)\sp B(r)=r^{\theta-2}(1+B_1r^b)\sp C(r)=r^{\theta-2}(1+C_1r^b)\label{q58}\ee
\be \phi(r)=\alpha \log r+\phi_1 r^b\sp A_t=r^{f}(A_0+A_1r^{f_1}),\label{q59}\ee
where $A_0,f$ indicate the coefficient and exponent in the leading order solution of $A_t$. We insert this ansatz into the equations of motion, \ref{eom2}, and, to linear order in the amplitudes, we obtain a $5\times 5$ homogeneous system of algebraic equations. Note that the above perturbation ansatz has a residual gauge freedom, stemming from the freedom to redefine the radial coordinate.

The perturbations are classified as relevant or irrelevant, depending on the sign of the exponents $b,f_1$. In particular, a perturbation is RG irrelevant if the IR is located at $r\to 0$ ($r\to\infty$) and the exponent of the perturbation is positive (negative). If the exponent is zero, it is marginal. Otherwise it is RG relevant.

The modes appear in pairs, with each pair summing to $2+z-\theta$.
In every solution there is a relevant mode $b=2+z-\theta$, which drives the solution to a finite temperature state, \cite{GK2}. Its conjugate mode is marginal ($b=0$) and can be absorbed by rescaling the coordinates. In solutions with $z\ne 1$ there is an additional pair of modes $b=b_\pm$. In the non-axionic solutions they satisfy,
\be b_++b_-=2+z-\theta\sp b_+b_-=-4{(z-1)(1+z-\theta)(2+z-\theta)\over 2z-2-\theta},\ee
while in the axionic solutions they are given by
\be \begin{split}&b_++b_-=2+z-\theta\\ &b_+b_-=-{(2+z-\theta)(1+z-\theta)\over 2z-2-\theta}\left(4(z-1)-(2-\theta){2k^2\Psi_0\over 2V_0-k^2\Psi_0}\right)\end{split}.\ee
The constraints in subsection \ref{HVIR} imply $b_+b_-<0$, hence one mode is positive and the other negative.

In the non-axionic solutions there is also a mode stemming from the axion term
\be b=2+\alpha\gamma,\ee
which becomes marginally relevant precisely when the axion term is leading in (\ref{ireq1}).

Turning to the gauge field perturbations, in every case we find a mode $f_1=-f$, which is a constant shift that can be gauged away.
In the neutral solutions (\ref{HVneutral}), (\ref{HVaxneutral}), turning on the magnetic field and charge density, there are two modes, with their corresponding amplitudes
\begin{subequations}
	\begin{align}
		& f_1^q=\zeta-z\sp A_1^q={qL\over Z_0(\zeta-z)}\\
		& f_1^h=\xi-z\sp A_1^h=-{hLW_0\over Z_0(\xi-z)}.
	\end{align}
\end{subequations}
These modes can be relevant and drive the solution to the corresponding electric or T-violating solution, depending on the values of the exponents $\beta,\eta$.

 In the solutions (\ref{HVdyo})-(\ref{HVaxel}) we find the following irrelevant mode and corresponding amplitude
\be f_1=\alpha\eta\sp A_1=-{hLW_0\over Z_0(\xi-z)}.\ee

In the solutions (\ref{HVmagT})-(\ref{HVTax}) we find the following irrelevant mode and amplitude
\be f_1=-\alpha\eta\sp A_1={qL\over Z_0(\zeta-z)}.\ee

\newpage

\section{Conclusions and Outlook\label{concl}}

In the present thesis we used a Holographic model in order to study charge transport and quantum criticality in strongly-coupled, large-$N$ quantum field theories at finite density, $q$, and in the presence of a magnetic field, $h$. The model was based upon the Einstein-Maxwell-Dilaton action, with the addition of the CP-violating PQ term (\ref{PQ}) and two axion fields (\ref{cc1}).

In Lorentz invariant systems with a magnetic field we confirmed that the DC conductivity is constrained to obey, \cite{hartnoll},
\be \sigma_{xx}=0\sp \sigma_{xy}={q\over h},\ee
while in the absence of the magnetic field it diverges.
The PQ term does not contribute, since it does not break any continuous symmetries.

Breaking the translational invariance with the axion fields we calculated finite DC conductivity, (\ref{cond}). Because of broken T-symmetry, a contribution analogous to the pair-creation in $\sigma_{xx}$ appears in the Hall conductivity, which is finite even at zero density and in the absence of the magnetic field.

The AC conductivity was found numerically in constant scalar black hole backgrounds. In the neutral case the longitudinal conductivity is finite and constant, \cite{Herzog:2007ij}, due to pair-creation, \cite{Karch:2007pd}. We found that the PQ term contributes an analogous constant term to the Hall conductivity, (\ref{ACneutral}).

The AC conductivity at finite charge density diverges at zero frequency, because of the translational symmetry. Including the axions, it is no longer divergent. There is a Drude peak around $\omega=0$ when the momentum relaxation is weak, (\ref{hydrcax}). As the relaxation parameter becomes larger, there are deviations from the Drude model. This has been documented as a coherent/incoherent metal transition \cite{DG},\cite{Kim:2014bza}. Meanwhile, there is a constant contribution to the Hall conductivity, as in the neutral case.

Turning on the magnetic field we find that the charge transport is dominated by the damped cyclotron resonance of \cite{Hartnoll:2007ih}, \cite{Hartnoll:2007ip}. The damping becomes stronger as the momentum relaxation scale increases. There are two kinds of contributions of the PQ term: a constant offset in the Hall conductivity and the charge density, $q$, always appearing shifted as $Q=q-hW_0$. The latter causes a system at finite density, $q$, and magnetic field, $h$, to behave effectively as a system at zero density\footnote{This happens precisely when the chemical potential is zero (see (\ref{chemp}))} when $q=hW_0$.

Finally, we classified the hyperscaling violating solutions of the EMD-PQ theory with axions. In the neutral case there is Lifshitz scaling ($z\ne 1$) if the axions are relevant in the IR solution, \cite{Gouteraux:2014hca}. In the rest of the cases, the presence of the axions does not affect the geometry, however the translational symmetry of the theory is broken in the IR.

As a final note, it would be interesting to calculate the electric conductivity in backgrounds in which the PQ coupling, $W$, has radial dependence, through the dilaton $\phi(r)$. In such cases there is mixing of the $x,y$ component of the gauge field (see (\ref{GFEP})), even in the absence of the magnetic field, and we would expect more interesting behaviors.

\section*{Acknowledgements}
I would like to thank Elias Kiritsis for helpful conversation, comments and suggestions during the course of this work.

\newpage
\appendix
\renewcommand{\theequation}{\thesection.\arabic{equation}}
\addcontentsline{toc}{section}{Appendix}
\section*{Appendix}

\section{AdS-RN Black Hole\label{rnapp}}

In this appendix we derive the simplest black hole solution to the action (\ref{actionax}).
We substitute the following AdS$_4$ black hole ansatz
\be ds^2={L^2\over r^2}\left(-f(r)dt^2+{dr^2\over f(r)}+dx_idx^i\right)\sp \phi(r)=\phi_0\ee
 into the equations (\ref{eom2}).
From the scalar equation (\ref{DEQ1}) we obtain constraints for the coupling functions
\be  \partial_\phi V=\partial_\phi\Psi=0\sp 2hQZ_0\partial_\phi W=(Q_0^2-h^2Z_0^2)\partial_\phi Z\ee
where
\be  Q_0=q-hW_0\sp W(\phi_0)=W_0\sp Z(\phi_0)=Z_0\sp \Psi(\phi_0)=\Psi_0.\ee
The solution for the gauge field, (\ref{GFEQ1}), is
\be A_t(r)=\mu+{Q_0\over Z_0}r\ee
and the blackening factor satisfies the equation
\be 2rf'-6f=k^2\Psi_0 r^2-L^2V_0+{Q_0^2+h^2Z_0^2\over L^2Z_0}r^4,\ee
which can be solved to obtain
\be f(r)=1-{1\over 2}k^2\Psi_0 r^2+{Q_0^2+h^2Z_0^2\over 4L^2Z_0}r^4-mr^3,\ee
where $m$ is the integration constant and we set
\be L^2=6/V(\phi_0)\ee
so that $f(0)=1$.

With appropriate choice of $m$, $f(r)$ has exactly two positive real roots $r_{\pm}$. Define
\be r_h=\min (r_+,r_-),\ee
then
\be f(r)=\left(1-{r\over r_h}\right)\left(1+{r\over r_h}+{r^2\over r_h^2}-{1\over 2}\Psi_0k^2r^2-{Q_0^2+h^2Z_0^2\over 4L^2Z_0}r^3r_h\right) \ee
and the temperature is
\be T=-{f'(r_h)\over 4\pi r_h}={1\over 4\pi r_h}\left(3-{1\over 2}\Psi_0k^2r_h^2-{Q_0^2+h^2Z_0^2\over 4L^2Z_0}r_h^4\right).\ee
We wrote the integration constant $m$ in terms of the outer horizon $r_h$
\be m=r_h^{-3}\left(1-{1\over 2}k^2\Psi_0r_h^2+{Q_0^2+h^2Z_0^2\over 4Z_0L^2}r_h^4\right).\ee

The relation between the chemical potential and charge density is found by $A_t(r_h)=0$, \cite{Koba}:
\be q=hW_0-Z_0{\mu\over r_h}.\label{chemp}\ee
In particular, this implies that in the absence of chemical potential ($\mu=0$), there is finite charge density, fixed in terms of the magnetic field
\be q=h W_0.\ee

\section{General Time-dependent Fluctuation Equations\label{gpert}}

In this thesis we used backgrounds of the form
\be \begin{split}ds^2=G(r)\left(-S(r)dt^2+{dr^2\over S(r)}\right)+C(r)(dx^2+dy^2)\\\ \phi=\phi(r)\sp  A_\m=(A_t(r),0,0,hx)\sp \chi_i=kx_i.\end{split}\ee
We remind that our action is (\ref{actionax}).

In sections \ref{hcond} and \ref{condax} we used the coordinate system in which $G=1$. In section \ref{ACcond} we used
\be G(r)=C(r)={L^2\over r^2}\sp S(r)=f(r).\ee

The fluctuation equations we used can be derived as special cases of the following perturbation ansatz:
\be \begin{split}& \delta A_i=\delta a_i (r,t)\sp \delta \chi_i=\delta k_i(r,t)\\
	& \delta g_{ti}=C(r)\delta z_i(r,t)\sp \delta g_{ri}=C(r)\delta g_i(r,t) \end{split}\ee

The linearized fluctuation equations are the following.

From the gauge field equations (\ref{GFEQ}):
\be \begin{split} &\left(ZS\delta a_x'+{Q}\delta z_x+h{SZ}\delta g_y\right)'-{Z\over S}\ddot{\delta a}_x+\dot{\delta a}_y W'-{hZ\over S}\dot{\delta z}_y-{Q}\dot{\delta g}_x=0\\
	&\left(ZS\delta a_y'+{Q}\delta z_y-h{SZ}\delta g_x\right)'-{Z\over S}\ddot{\delta a}_y-\dot{\delta a}_x W'+{hZ\over S}\dot{\delta z}_x-{Q}\dot{\delta g}_y=0 \end{split}\label{GFEP}\ee
from the $t-x,t-y$ components of the Einstein equation (\ref{EEQ}):
\be \begin{split}  & {G\over C}\left({ (C\delta z_x)'\over G}\right)'+K(r)\delta z_x+{QG\over C^2}\delta a_x'+h{GZ\over S C^2}\dot{\delta a}_y-\dot{\delta g}_x'\\
	&+\left({G'\over G}-{C'\over C}\right)\dot{\delta g}_x+h{GQ\over C^2}\delta g_y+{kG\Psi\over S}\dot{\delta k}_x=0\\
	& {G\over C}\left({(C\delta z_y)'\over G}\right)'+K(r)\delta z_y+{QG\over C^2}\delta a_y'-h{GZ\over S C^2}\dot{\delta a}_x-\dot{\delta g}_y'\\
	&+\left({G'\over G}-{C'\over C}\right)\dot{\delta g}_y-h{GQ\over C^2}\delta g_x+{kG\Psi\over SC}\dot{\delta k}_y=0\end{split}\label{EEQPt}\ee
where \be K(r)={G'C'\over GC}-{C''\over C}-h^2{GZ\over C^2S}-k^2{G\Psi\over CS}.\ee
From the $r-x,r-y$ components of the Einstein equations (\ref{EEQ}) we obtain:
\be \begin{split} &{hQ\over S}\delta z_y+hZ\delta a_y'+kC\Psi \delta k_x'=\delta g_x\left(h^2{Z}+k^2C\Psi\right)+{C^2\over GS}\ddot{\delta g}_x-{C^2\over GS}\left({\dot {\delta z}_x}\right)'-{Q\over S}\dot{\delta a}_x\\
	& -{hQ\over S}\delta z_x-hZ\delta a_x'+kC\Psi \delta k_y'=\delta g_y\left(h^2{Z}+k^2C\Psi\right)+{C^2\over GS}\ddot{\delta g}_y-{C^2\over GS}\left({\dot {\delta z}_y}\right)'-{Q\over S}\dot{\delta a}_y\label{EEQPr}\end{split}\ee
and finally from the axion equations (\ref{AEQ}):
\be
\begin{split}
	&\left(C\Psi S \delta k_i'-kC\Psi S \delta g_i\right)'+k{C\Psi\over S}\dot{\delta z}_i-{C\Psi\over S}\ddot{\delta k}_i=0\sp i=x,y.\label{AEQP}\\
\end{split}
\ee

\end{document}